%% file: main.tex
\documentclass{SciPost}

\hypersetup{
    colorlinks,
    linkcolor={red!50!black},
    citecolor={blue!50!black},
    urlcolor={blue!80!black}
}

\usepackage[bitstream-charter]{mathdesign}
\DeclareSymbolFont{usualmathcal}{OMS}{cmsy}{m}{n}
\DeclareSymbolFontAlphabet{\mathcal}{usualmathcal}

\fancypagestyle{SPstyle}{
\fancyhf{}
\lhead{\colorbox{scipostblue}{\bf \color{white} ~SciPost Physics Lecture Notes }}
\rhead{{\bf \color{scipostdeepblue} ~Submission }}

\fancyfoot[C]{\textbf{\thepage}}
}

\input{_preamb}

\input{_tikz_setup}

\begin{document}

\pagestyle{SPstyle}

\begin{center}{\Large \textbf{\color{scipostdeepblue}{
Les Houches Lecture Notes on Tensor Networks
}}}\end{center}

\begin{center}\textbf{
Bram Vancraeynest-De Cuiper\textsuperscript{1$\star$},
Weronika Wiesiolek\textsuperscript{2} and
Frank Verstraete\textsuperscript{1,2$\dagger$}
}\end{center}

\begin{center}
{\bf 1} Department of Physics and Astronomy, Ghent University, Krijgslaan 299, 9000 Gent, Belgium
\\
{\bf 2} Department of Applied Mathematics and Theoretical Physics, University of Cambridge,
Wilberforce Road, Cambridge, CB3 0WA, United Kingdom
\\[\baselineskip]
$\star$ \href{mailto:Bram.VancraeynestDeCuiper@UGent.be}{\small Bram.VancraeynestDeCuiper@UGent.be}\,,\quad
$\dagger$ \href{mailto:fv285@cam.ac.uk}{\small fv285@cam.ac.uk}
\end{center}

\section*{\color{scipostdeepblue}{Abstract}}
\textbf{\boldmath{Tensor networks provide a powerful new framework for classifying and simulating correlated and topological phases of quantum matter. Their central premise is that strongly correlated matter can best be understood by studying the underlying entanglement structure and its associated (generalised) symmetries. In essence, tensor networks provide a compressed, holographic description of the complicated vacuum fluctuations in strongly correlated systems, and as such they break down the infamous many-body exponential wall. These lecture notes provide a concise overview of the most important conceptual, computational and mathematical aspects of this theory.
}}

\vspace{\baselineskip}

\vspace{10pt}
\noindent\rule{\textwidth}{1pt}
\tableofcontents
\noindent\rule{\textwidth}{1pt}
\vspace{10pt}

\input{_Intro}
\input{_Lecture_1}
\newpage\input{_Lecture_2}
\newpage\input{_Lecture_3}
\newpage\input{_Lecture_4}
\newpage\input{_Lecture_5}

\section*{Acknowledgements}
We are grateful to the organisers Stephane Ouvry, Tomaž Prosen, Didina Serban and Masahito Yamazaki for making the 2025 Les Houches Summer School \emph{Exact Solvability and Quantum Information} such a wonderful event. Moreover we thank Dante Bosgoed, Lander Burgelman, Boris De Vos, Jutho Haegeman, Laurens Lootens, Anton Martin, Zohar Nussinov, Alex Turzillo and Vic Vander Linden for constructive feedback on the manuscript.

\paragraph{Funding information}
We acknowledge funding from the UKRI
grant EP/Z003342/1,  BOF-GOA (Grant No. BOF23/GOA/021), EOS (Grant
No. 40007526), IBOF (Grant No. IBOF23/064), and the Research Foundation Flanders (FWO) through a doctoral fellowship No. 11O2423N awarded to BVDC.

\begin{appendix}
\numberwithin{equation}{section}

\newpage\input{_App_Cohomology}
\newpage\input{_App_Cat}

\end{appendix}
\bibliography{refs.bib}

\end{document}

%% file: _preamb.tex
\usepackage{float}
\usepackage{cleveref}
\usepackage[dvipsnames,table]{xcolor}
\usepackage[linesnumbered,ruled,vlined]{algorithm2e}

\usepackage{physics}
\usepackage{bm}

\newcommand{\mc}{\mathcal}
\newcommand{\mbb}{\mathbb}
\newcommand{\msf}{\mathsf}
\newcommand{\sss}{\scriptstyle}
\newcommand{\ssss}{\scriptscriptstyle}
\newcommand{\rU}{{\rm U}}
\newcommand{\Vect}{\mathsf{Vec}}
\newcommand{\Rep}{\mathsf{Rep}}
\newcommand{\Fib}{\mathsf{Fib}}
\newcommand{\Fun}{\mathsf{Fun}}
\newcommand{\cat}{\lhd}
\newcommand{\act}{\rhd}
\newcommand{\mF}{{}^{\act\!} F}
\newcommand{\nF}{{}^{\cat\!} F}
\newcommand{\dF}{{}^{\circ\!} F}
\newcommand{\rF}{{}^{\bowtie\!} F}
\newcommand{\upup}[1]{\raisebox{0.5pt}[0pt]{$\sss{#1}$}}
\newcommand{\dodo}[1]{\raisebox{-1.2pt}[0pt]{$\sss{#1}$}}
\newcommand{\CC}[6]{\Big[{}^{\upup{#1}}_{\dodo{#4}}\, {}^{\upup{#2}}_{\dodo{#5}} \big| {}^{\upup{#3}}_{\dodo{#6}}\Big]}

\newcommand{\RCR}[3]{{\textcolor{BlueViolet}{#1}\textcolor{BrickRed}{#2}\textcolor{BlueViolet}{#3}}}
\newcommand{\RDR}[3]{{\textcolor{BlueViolet}{#1}\textcolor{black}{#2}\textcolor{BlueViolet}{#3}}}

\newcommand{\ZtZt}{\mbb Z_2\times\mbb Z_2}

\renewcommand{\vec}[1]{{\bf #1}}

\usepackage{graphicx}
\usepackage{tikz}
\usetikzlibrary{shapes.geometric,arrows,matrix,calc,scopes,decorations.markings,intersections,arrows.meta,decorations.pathreplacing,angles,quotes,patterns,decorations.pathmorphing}

%% file: _tikz_setup.tex
\tikzstyle{obj1}=[line width = 0.8pt, line join=round, line cap=round]
\tikzstyle{obj2}=[line width = 0.8pt, line join=round, line cap=round, color=BrickRed]
\tikzstyle{obj3}=[line width = 0.8pt, line join=round, line cap=round, color=BlueViolet]
\tikzstyle{mpstensor}=[draw=black, line width = 0.7pt, fill=BlueViolet!14, circle, inner sep=3pt]
\tikzstyle{mpotensor}=[draw=black, line width=.7pt, fill=gray!20, rectangle, inner sep=3.8pt, rounded corners=2pt]
\tikzstyle{matrix}=[draw=black, line width = 0.7pt, fill=black!10, circle, inner sep=1.6pt]
\tikzstyle{mor}=[draw=none, fill=gray!22]

\tikzset{singlet/.style={decorate, decoration=snake}}

\tikzset{->1-/.style={decoration={
			markings,
			mark=at position #1 with {\arrow{Triangle[fill=black, width=3.5pt, length=3.5pt]}}},postaction={decorate}}}
\tikzset{-<1-/.style={decoration={
			markings,
			mark=at position #1 with {\arrowreversed{Triangle[fill=black, width=3.5pt, length=3.5pt]}}},postaction={decorate}}}
			
\tikzset{->2-/.style={decoration={
			markings,
			mark=at position #1 with {\arrow{Triangle[fill=BrickRed, width=3.5pt, length=3.5pt]}}},postaction={decorate}}}
\tikzset{->3-/.style={decoration={
			markings,
			mark=at position #1 with {\arrow{Triangle[fill=BlueViolet!80, width=3.5pt, length=3.5pt]}}},postaction={decorate}}}

\tikzset{
    mask point/.style={ transform shape, sloped, 
    minimum width=20pt, minimum height=4pt, inner sep=0pt, ultra thin, font=\tiny}
}
\tikzset{
    clip even odd rule/.code={\pgfseteorule},
    invclip/.style={clip,insert path=[clip even odd rule]{
    [reset cm](-\maxdimen,-\maxdimen)rectangle(\maxdimen,\maxdimen)
}}}
\newcommand\clipIntersection[1]{
    (#1.north west) -- (#1.north east) -- (#1.south east) -- (#1.south west) -- (#1.north west)
}
\newcommand\clipAroundNode[2]{
    \begin{pgfscope}
        \begin{scope}[overlay]
            \path[invclip] \clipIntersection{#1};#2
        \end{scope}
    \end{pgfscope}
}
\newcommand{\middleCoo}[6]{
    \path (#1) --coordinate[pos=0.5](#3) (#2);
    \path (#1) --coordinate[pos={0.5-\sp}](#4) (#2);
    \path (#1) --coordinate[pos={0.5+\sp}](#5) (#2);
}
\newcommand{\barycentre}[8]{
    \middleCoo{#1}{#2}{A}{}{}{}{#8};
    \middleCoo{#2}{#3}{B}{}{}{}{#8};
    \middleCoo{#1}{#3}{C}{}{}{}{#8};
    \path[name path=PA#3] (A) -- (#3);
    \path[name path=PB#1] (B) -- (#1);
    \path[name path=PC#2] (C) -- (#2);
    \path [name intersections={of=PA#3 and PB#1,by={#4}}];
    \path (#1) --coordinate[pos=0.84](#5) (#4);
    \path (#2) --coordinate[pos=0.84](#6) (#4);
    \path (#3) --coordinate[pos=0.84](#7) (#4);
}
\newcommand{\barycentreSq}[9]{
    \middleCoo{#1}{#2}{A1}{A2}{A3}{};
    \middleCoo{#3}{#4}{B1}{B2}{B3}{};
    \middleCoo{#1}{#3}{C1}{C2}{C3}{};
    \middleCoo{#2}{#4}{D1}{D2}{D3}{};
    \middleCoo{A1}{B1}{#5}{}{};
    \path[name path=PA2B2] (A2) -- (B2);
    \path[name path=PA3B3] (A3) -- (B3);
    \path[name path=PC2D2] (C2) -- (D2);
    \path[name path=PC3D3] (C3) -- (D3);       
    \path [name intersections={of=PA2B2 and PC2D2,by={#6}}];
    \path [name intersections={of=PA2B2 and PC3D3,by={#8}}];
    \path [name intersections={of=PA3B3 and PC2D2,by={#7}}];
    \path [name intersections={of=PA3B3 and PC3D3,by={#9}}];
}

%% file: _Intro.tex
\section*{Introductory remarks}
We are currently witnessing a paradigm shift in the way we understand and do quantum many-body physics. The \emph{quantum many-body problem} has been the central problem in theoretical physics and quantum chemistry for the best of the last century, and its associated exponential wall has precluded progress in describing strongly interacting phases of matter. During the last twenty years however, there has been remarkable progress in simulating and classifying those strongly correlated systems using the theory of \emph{entanglement}, as epitomised by the concept of \emph{tensor networks}.

In essence, tensor networks circumvent the exponential quantum many-body wall by providing a \emph{holographic} description of the many-body wave functions. Instead of describing the wave function in an exponentially large Hilbert space, tensor networks model how entanglement is routed between different parts of the system. Underlying the success of tensor networks lies the fact that all the information needed to calculate any observable of a low-energy wave function can be compressed into a local tensor with so-called \emph{entanglement degrees of freedom} making up a \emph{virtual} Hilbert space. The reason that this can be done is related to the existence of \emph{area laws} for the entanglement entropy: tensor networks are concise and precise representations of the manifold of all states satisfying such an area law. By projecting the Hamiltonian defined in the exponentially large physical Hilbert space on this low-dimensional manifold, it has become possible to construct numerical algorithms to simulate large-scale quantum many-body systems. DMRG (\emph{Density Matrix Renormalisation Group}), MPS (\emph{Matrix Product State}) and PEPS (\emph{Projected Entangled-Pair State}) algorithms are revolutionising the way we simulate state-of-the-art experiments in atomic physics, current-day quantum computers and the low-energy physics of quantum many-body systems.

But there is more: since the pioneering insights of Landau and Anderson, it is known that many-body physics is all about \emph{symmetries} and \emph{symmetry breaking}. It turns out that the global, local and topological symmetries of the full system are reflected in the symmetries of the local tensors of the tensor network, which provide an extremely effective compression of the full many-body wave function. As a consequence, the classification of gapped \emph{phases of matter}, and in particular \emph{topological phases}, is reduced to the problem of classifying the irreducible ways in which tensors transform under a set of relevant symmetries. As those symmetries act on the virtual (entanglement) degrees of freedom, there is no reason why those symmetries should form a group, and more general associative structures emerge that can be described by \emph{fusion categories}. Indeed, tensor networks provide an explicit way of representing those \emph{generalised symmetries} (and the dualities between them) in terms of MPO (\emph{Matrix Product Operator}) algebras and their higher-dimensional analogues, PEPOs (\emph{projected entangled-pair operators}). Several of the most active research areas in mathematical, condensed matter, statistical and high-energy physics are precisely dealing with the consequences of this insight: a holographic description of strongly correlated and topological phases of matter is the only one which reveals its inner structure.

\bigskip\noindent These lecture notes are based on five lectures given by Frank Verstraete at the Les Houches Summer School of Physics in August 2025 on \emph{Exact Solvability and Quantum Information}, covering a selection of elementary computational and theoretical results of tensor networks. They do not aim to provide a comprehensive review of these topics. Rather, they are meant as an invitation for further study and we can refer to a vast literature of existing reviews and lecture notes:~\cite{Verstraete2008,Schollwoeck2010,Bridgeman2016,Vanderstraeten2019,Cirac2020,Xiang2023,Banuls2022,Verstraete2023,Chen2025}.

%% file: _Lecture_1.tex
\section{Lecture I: Motivation \& the manifold of MPS}
\subsection{The double-exponential wall \& the illusion of Hilbert space}\label{sec:Illusion}
Before we dive into these lectures, it is important to recognise the true challenge posed by the quantum many-body problem. As one learns in every course on quantum mechanics, the dimension of the Hilbert scales exponentially in the number of constituent particles. This is in fact not so different from the classical case where the number of possible strings of classical bits also scales exponentially in the number of bits. However, since quantum mechanics allows for \emph{superpositions} of states and consequently \emph{entanglement}, the natural question to ask is not by how many basis states the Hilbert space is spanned, but rather how many of the states in the Hilbert space can actually be explored by some physical mechanism originating from local (few-body) interactions, and in particular with  a quantum computer.

To be more precise and concrete, let us consider the following question raised in ref.~\cite{Poulin2011}. Within the Hilbert space of $N$ qubits, which portion of the exponentially large many-body Hilbert space can be reached by running a quantum computer starting from a reference state for a time scaling polynomially in system size? To estimate this, we consider so-called \emph{$\varepsilon$-balls} of states around points $\ket{\psi}$ in the Hilbert space, i.e. we consider states $\ket{\varphi}$ for which
\begin{equation}
    |\braket{\psi}{\varphi}| \gtrsim 1 - \varepsilon.
\end{equation}
A simple geometric argument shows that such $\varepsilon$-balls occupy a fraction of the total volume of the Hilbert space which scales according to $\mc O\big(\varepsilon^{2^N}\big)$, i.e. a \emph{double-exponentially} (!) small fraction of the total Hilbert space! Due to this scaling, it would take any physical system (including a quantum computer) a time that is at least exponential in the system size to have overlap with a specific $\varepsilon$-ball. We conclude that the vast majority of states is unphysical and is completely irrelevant for the description of physical systems: Hilbert space is a (convenient) illusion, as states with many particles that take exponentially long to create cannot be created by any physically reasonable process. 

\bigskip\noindent This dramatic result points to a paradigm shift in our understanding of quantum many-body systems, namely that physical wave functions are those which have low complexity. More specifically, we expect ground states of quantum many-body systems to live on a physically relevant low-complexity \emph{manifold} of states, which for all practical purposes we will want to parametrise with a \emph{small} number of parameters, meaning scaling polynomially in the number of particles. Roughly speaking, a manifold is a space which locally resembles flat Euclidean space. This means that to every point of the manifold one can associate a linear \emph{tangent space} whose dimension is the dimension of the manifold.
One of the (many) great insights of Dirac was that one can now actually do physics on such a manifold without ever leaving it by interpreting the manifold as a variational set of states. Let us denote a general variational manifold of states as
\begin{equation}
    \mc M = \big\{\ket{\Psi(\vec z)}|\vec z \in \mbb C^n\big\}.
\end{equation}
We will write $T_\vec z \mc M$ for the linear tangent space at any point $\vec z$. A general state in this tangent space is then of the form
\begin{equation}
    \sum_i x^i \pdv{}{z^i} \ket{\Psi(\vec z)},
\end{equation}
for complex coefficients $x^i$. Suppose now that we want to compute the time evolution of a certain state $\ket{\Psi(\vec z)}$ subject to some many-body Hamiltonian $H$:
\begin{equation}
    i\pdv{}{t} \ket{\Psi(\vec z)} = H\ket{\Psi(\vec z)}.
\end{equation}
Then $i\pdv{}{t} \ket{\Psi(\vec z)}=i\sum_i\dot z^i\pdv{}{z^i} \ket{\Psi(\vec z)}$ can be thought of as a tangent vector in $T_\vec z \mc M$. However, for a generic Hamiltonian, $H\ket{\Psi(\vec z)}$ is a vector in the ambient Hilbert space in which $\mc M$ is embedded that points \emph{out of} the tangent plane, and thus doesn't belong to the variational class anymore. Dirac's proposal was to define an action on this manifold, but for our purposes it is convenient to think about this as the orthogonal projection of $H\ket{\psi(\vec z)}$ on the tangent space of the manifold $T_\vec z \mc M$. Geometrically, one can think of this projection as minimising the distance
\begin{equation}
    \min_{\{\dot z^i\}_i} \norm{i\sum_i\dot z^i\pdv{}{z^i} \ket{\Psi(\vec z)} - H \ket{\Psi(\vec z)}}.
\end{equation}
Denoting the orthogonal projector at a point $\vec z$ by $\mc P_{T_\vec z}$, we can depict its action on $H\ket{\psi(\vec z)}$ thus as:
\begin{equation}
    %
	\input{figs/TangentPlane.tex}%
 \,\, .
\end{equation}
The resulting \emph{time-dependent variational principle} (TDVP) equations
\begin{equation}
    i\pdv{}{t} \ket{\Psi(\vec z)} = \mc P_{T_\vec z}H\ket{\Psi(\vec z)},
\end{equation}
then describe a collection of highly non-linear (due to the appearance of the tangent space projector) differential equations of the generic many-body state on the manifold $\mc M$. Importantly, it can be shown that the manifold $\mc M$ is \emph{symplectic}. This means in particular that the expectation value of the energy, generators of symmetries and other constants of motion are exactly conserved under evolution by the TDVP equations.

\bigskip\noindent This manifold approach to the many-body problem has been applied in many fields with exceptional success. In fact, it is merely a small exaggeration to state that all progress in tackling the quantum many-body problem since the discovery of the Schr\"odinger equation has been in finding such good manifolds of states in which the relevant physics takes place. \emph{Matrix product states} (MPS) and its higher dimensional generalisations such as two-dimensional \emph{projected entangled-pair states} (PEPS) form such classes, and we will introduce them shortly. But before doing so, let us take some time to discuss some other, perhaps to some readers more familiar, variational classes of states from this manifold point of view.
\paragraph{Free fermions \& Hartree-Fock}
Probably the most well-known variational manifold of many fermions is that of \emph{Slater determinants}~\cite{Slater1929}. A Slater determinant parametrises an $N$-fermion wave function in terms of $N$ single-particle orbitals $\psi_i,i=1,...,N$ as a determinant of the form:
\begin{equation}
    \Psi({\bf x}_1,\dots,{\bf x}_N) =
    \begin{vmatrix}
    \psi_1({\bf x}_1) & \dots & \psi_N({\bf x}_1) \\
    \psi_1({\bf x}_2) & \dots & \psi_N({\bf x}_2) \\
    \vdots & \ddots & \vdots \\
    \psi_1({\bf x}_N) & \dots & \psi_N({\bf x}_N)
    \end{vmatrix}.
\end{equation}
By construction, Slater determinants are manifestly anti-symmetric under the exchange of any two fermions, in correspondence with the Pauli exclusion principle. Moreover, note that the number of parameters needed to represent Slater determinants scales \emph{linearly} in the number of electrons and Slater determinants therefore only span an exponentially small submanifold in the Hilbert space. The \emph{Hartree-Fock} method for approximately solving the Schr\"odinger equation of $N$ fermions is then a self-consistent variational method over this class of states: approximations of both the time-dependent and time-independent Schrödinger equations are obtained by applying the TDVP principle to this manifold.  The success of this method can hardly be overstated. It has allowed us to simulate molecules and materials with great accuracy which would otherwise be completely intractable with exact methods. In fact, Hartree-Fock allowed us to understand the structure of the periodic table of elements and of band structures from first principles~\cite{Slater1935,Hartree1947}.
\paragraph{Field coherent states of interacting bosons} The bosonic counterpart of Slater determinants also exists in the form of the manifold of \emph{field coherent states}. For the bosonic creation operators\footnote{In this paragraph we will exceptionally denote operators with hats to avoid confusion with scalar functions.} $\hat \psi^\dagger(\vec x)$, one has for any function $\varphi (\vec x):\mbb R^3\rightarrow\mbb C$ the coherent state
\begin{equation}
    \ket{\Psi[\varphi]} = \exp(\int d^3\vec x \varphi(\vec x)\hat\psi^\dagger(\vec x)) \ket{\Omega},
\end{equation}
where $\ket{\Omega}$ is the bosonic vacuum characterised by $\hat\psi(\vec x)\ket{\Omega}=0$, $\forall\vec x$. Recall that coherent states are exact eigenstates of the annihilation operators, i.e. $\hat\psi(\vec x)\ket{\Psi[\varphi]} = \varphi(\vec x)\ket{\Psi[\varphi]}$. Let us now apply the TDVP to the Hamiltonian
\begin{equation}
    H = \int d^3\vec x \hat \psi^\dagger(\vec x) \bigg( -\frac{1}{2m}\nabla^2 + V(\vec x)\bigg)\hat\psi(\vec x) + \frac{U}{2} \hat\psi^\dagger(\vec x)\hat\psi^\dagger(\vec x)\hat\psi(\vec x)\hat\psi(\vec x),
\end{equation}
describing a dilute boson gas prone to an external potential $V$ and a contact interaction. One then finds that the TDVP equations give rise to  the  (time-dependent) \emph{Gross-Pitaevskii} equation:
\begin{equation}
    i\pdv{}{t}\varphi(\vec x) = -\frac{\nabla^2}{2m}\varphi(\vec x) + V(\vec x)\varphi(\vec x) + U|\varphi(\vec x)|^2\varphi(\vec x).
\end{equation}
Despite the simplicity of this derivation, the Gross–Pitaevskii has been employed with great success to describe the behaviour of Bose-Einstein condensates~\cite{GP}.

\bigskip\noindent We highlighted here only a few of the most common and successful methods to define and parametrise variational manifolds in the exponentially large Hilbert space. Besides these we have omitted the plethora of post-Hartree-Fock methods such as \emph{coupled cluster methods}~\cite{CC}.
\subsection{Quantum spin systems: monogamy \& frustration}
A very useful and insightful idealisation of a generic quantum many-body system or of a quantum field theory is provided by quantum lattice systems. In this framework, the lattice is a discretisation of space that hosts quantum degrees of freedom on its vertices, and in some cases on the edges. From the point of view of quantum field theory, this is the best controlled way of circumventing UV-divergences~\cite{Kogut1979,Kogut1982}. From the point of view of condensed matter and atomic physics, lattice systems provide effective (compressed) descriptions of the relevant degrees of freedom. 

In these lecture notes,  we will be concerned with the low-energy physics of quantum many-body lattice models in one or two spatial dimensions. Restricting to the low-energy sector is often permitted due the observation that for most realistic materials the Fermi temperature is of the order $10^4K$, much larger than the typical temperature scale of an experiment (even room temperature is very cold if you consider systems of electrons). But also in quantum field theory the ground state and its low lying excitations are typically the states of interest when computing for example cross sections of scattering experiments. Typically, the Hilbert spaces under consideration are a tensor product of qudits which we will denote by $\mc H=(\mbb C^d)^N\equiv\bigotimes_{i=1}^N \mbb C^d$, and unless stated otherwise we will consider the thermodynamic limit $N\rightarrow \infty$, while neglecting subtleties regarding order of limits. Note however that the dimension of the corresponding Hilbert space is still exponential as a function of the system size, so lattice systems also suffer from the same exponential wall.

Obtaining a good approximation to the ground state of such a model is generically very non-trivial due to \emph{frustration}. In this context, the term frustration refers to the inaptitude of the ground state to simultaneously minimise the energy of all Hamiltonian terms. As an example, consider for instance the paradigmatic spin 1/2 antiferromagnetic Heisenberg chain in $(1+1)d$:
\begin{equation}\label{eq:Heisenberg}
    H = \sum_{\expval{\msf i,\msf j}} \vec{S}_\msf i \cdot \vec{S}_\msf j.
\end{equation}
If we would consider only two spins, the ground state would be the singlet state with density matrix $\rho=\ketbra{\phi}$ where $\ket{\phi}=\frac{1}{\sqrt{2}}(\ket{\uparrow\downarrow} - \ket{\downarrow\uparrow})$.
In the case of spin chains with $N$ spins, the total energy would clearly be minimised if all reduced density matrices $\rho_{\msf i,\msf i+1}$ of the ground state are equal to this singlet -- but this is clearly not possible: the \emph{monogamy of entanglement} dictates that such a state cannot exist~\cite{Coffman1999,Osborne2006}. Indeed, there exists no three-partite state $\rho_{ABC}\geq 0$, such that $\Tr_C(\rho_{ABC}) = \Tr_B(\rho_{ABC})=\ketbra{\phi}$. As such, monogamy implies frustration.

Monogamy of entanglement can also be used to explain why mean field theory becomes increasingly exact when the number of spatial dimensions increases. Consider instead of the one-dimensional chain a lattice with a large coordination number and again a Hamiltonian which is the sum of antiferromagnetic Heisenberg interactions between all neighbouring spins. Intuitively, in that case, for any given spin its entanglement with other spins then needs to be distributed over so many neighbours that the ground state necessarily becomes very close to a seperable state $\otimes_i |\psi_i\rangle$ and mean field theory becomes exact. Though presented in a hand waving fashion here, this argument can be made precise by invoking the \emph{quantum de Finetti theorem}~\cite{Raggio1989}. This theorem characterises multi-partite density matrices that are invariant under permutations of the different parties, and plays an important role in proving the security of quantum cryptographic protocols.

From this discussion we have deduced the fact that the ground state cannot be such that all its $\rho_{\msf i, \msf i+1}$ are spin singlets. Now we could rather look for a state whose marginals are, in a sense, \emph{as close as possible} to the spin singlet. To this end, one could for example try the \emph{resonating valence bond state} as an ansatz to the ground state:
\begin{equation}
    \ket{\Psi} = \ket{\Psi_+} + \ket{\Psi_-}, \quad {\rm where} \quad
    \ket{\Psi_\pm} = \prod_\msf i \big( \ket{\uparrow\downarrow}_{2\msf i,2\msf i\pm 1} - \ket{\downarrow\uparrow}_{2\msf i,2\msf i\pm 1} \big).
\end{equation}
Graphically one could depict this state as
\begin{equation}
    %
	\input{figs/RVB.tex}%
,
\end{equation}
where the wiggly lines stand for spin singlets between neighbouring sites. In fact, $\ket{\Psi_\pm}$ span the two-dimensional ground state subspace of the model introduced by Majumdar and Ghosh in ref.~\cite{Majumdar1969}. Note that this state has all the symmetries we could expect: it is ${\rm SU}(2)$ symmetric and has translational, time-reversal and reflection invariance. This is important as ground states (or more precisely, ground state subspaces) should inherit all the symmetries of the Hamiltonian.  More importantly, it has some non-vanishing entanglement, and we would therefore expect the energy of this state to be much closer to the actual ground state energy than the mean field approximation. Nevertheless, the energy of this state turns out to be still a fairly bad approximation of the true ground state energy of the Heisenberg model as it overestimates the energy quite a lot.

At this point it should be clear that writing down entangled states with the correct symmetries that can serve as ansatz for the ground state of our favourite many-body quantum Hamiltonian, is very hard. However, the valence bond state suggest one particular way of doing this. Inspired by this Majumdar-Ghosh model, Affleck, Kennedy, Tasaki and Lieb (AKLT) proposed in 1987 a state constructed of pairs of entangled spin 1/2 particles which are projected onto a spin 1 particle on each lattice site~\cite{Affleck1987}. Their construction goes as follows. One starts from a chain of spin 1/2 particles where every other pair of neighbouring particles is in a spin singlet. On every site $\msf i$ the rightmost particle is then taken together with the leftmost particle on the next site $\msf i+1$ to form a \emph{physical} site. The two \emph{virtual} spin 1/2 particles that make up the physical site are then projected onto their spin 1 subspace. A graphical depiction of this state is more enlightening:
\begin{equation}
    \ket{\Psi_{\rm AKLT}} = %
	\input{figs/AkLT.tex}%
 \, ,
\end{equation}
where
\begin{equation}
    %
	\input{figs/AKlT_proj.tex}%
.
\end{equation}
As shown by the aforementioned authors, this state is the ground state of a Hamiltonian which is very closely related to the spin 1 Heisenberg model, and is believed to belong to the same (symmetry-protected) phase (discussed further in lecture~\cref{sec:PhasesOfMatter}). Since two neighbouring sites exactly involve one spin singlet, the 4 constituent spin 1/2 particles which make up these physical degrees of freedom can not be in a coupled spin 2 state. As such, this state is annihilated by the projector on the spin 2 subspace of two spin 1's. Therefore, this state is the ground state of the AKLT Hamiltonian
\begin{equation}\label{eq:AKLTHam}
    H = \sum_\msf i \vec{S}_\msf i \cdot \vec{S}_{\msf i+1} + \frac{1}{3} \big( \vec{S}_\msf i \cdot \vec {S}_{\msf i+1} \big)^2,
\end{equation}
where the local two-body term can be recognised as the spin 2 projector, up to a shift and a prefactor.
\subsection{Tensor networks \& area laws}\label{sec:AreaLaw}
A few years after AKLT, in 1992, Fannes, Nachtergaele and Werner~\cite{Fannes1990} introduced the class of one-dimensional \emph{finitely correlated states} which generalises the construction of the AKLT state using ideas of quantum Markov chains.  The name \emph{finitely correlated states} stems from  the fact that these states provably have exponentially decaying connected correlation functions, as will be demonstrated below in \cref{sec:NormalObs}. They also proved that all those states are ground states of gapped frustration-free Hamiltonians. The modern point of view, as first advocated in ref.~\cite{Delgado2004}, is to construct those states by starting from a chain of maximally entangled \emph{virtual} qudits between neighbouring sites that are projected onto the \emph{physical} spin on each site by a linear map. This linear map is often referred to as  \emph{projector} (even though it doesn't have to be an idempotent linear map), which is the origin of the name \emph{projected entangled-pair states}, although this name is now typically used for the higher dimensional generalisations. These projectors contain parameters which play the role of the \emph{variational} degrees of freedom of that state. This class of states coincides with the class of (infinite and translational invariant) uniform \emph{matrix product states} (MPS), and this is the name that is most often used to characterise them. Let us make this more explicit. The construction of MPS thus starts from considering maximally entangled states
\begin{equation}
    \ket{I}_\msf i = \sum_{\alpha=1}^\chi \ket{\alpha\alpha}_{\msf i, \msf i+1}
\end{equation}
between each pair of neighbouring sites $\msf i,\msf i+1$, $\msf i\in\mbb Z$. The integer $\chi$ is called the \emph{bond dimension} and directly controls the number of variational parameters as we will see. We then consider on every physical site of the chain following linear map
\begin{equation}
    A = \sum_{i,\alpha,\beta} A_{\alpha\beta}^i \ket{i}\otimes\bra{\alpha}\otimes\bra{\beta} \in \mbb C^{d\chi^2},
\end{equation}
on which we act on the entangled pairs, so that the state corresponding to the tensor $A$ is given by:
\begin{equation}
    \ket{\Psi[A]} = \big(\bigotimes_{\msf i\in\mbb Z} A\big) \cdot \big(\bigotimes_{\msf i\in\mbb Z} \ket{I}_{\msf i}\big).
\end{equation}
We will say that $A$ \emph{generates} the MPS $\ket{\Psi[A]}$. Expanding now the product in this expression yields the more familiar expression of \emph{uniform} matrix product states, directly in the thermodynamic limit:
\begin{equation}\label{eq:MPS}
    \ket{\Psi[A]} = \sum_{\{i_j\}} \big(\cdots A^{i_j}A^{i_{j+1}}\cdots \big) \ket{\dots, i_j,i_{j+1},\dots}.
\end{equation}
This expression for the wave function explains the name \emph{matrix product} states, as the wave function amplitudes are computed as the products of the matrices $\{A^i\}_i$. In the uniform case the number of variational parameters is $d\chi^2$. Notice however that the dimension of the manifold of states spanned by uniform MPSs is strictly smaller than $d\chi^2$ since for any non-singular matrix $X$ a transformation of the kind
\begin{equation}
    A^i \mapsto XA^iX^{-1}, \quad \forall i=1,2,\dots,d,
\end{equation}
called a \emph{gauge transformation}, leaves the state $\ket{\Psi[A]}$ invariant. A more precise characterisation of this manifold is relegated to the next section.

Note that we could also have chosen our tensors (projectors) different at every site, leading to a generic non-uniform MPS with bond dimensions that can vary over the different edges of the lattice. This way we can easily deal with finite spin chains with open, periodic or twisted boundary conditions.  Remark that, for the uniform case we did not specify the boundary conditions at infinity. For generic matrix product states, which are \emph{injective} (defined below), the choice of these boundary conditions is irrelevant. Throughout these lecture notes we will focus almost exclusively on matrix product states in the thermodynamic limit for which this injectivity property holds. For completeness let us display here generic MPSs on finite spin chains. A generic MPS on an open chain can be written as:
\begin{equation}\label{eq:OpenMPS}
    \ket{\Psi[A]} = \sum_{\{i_j\}} \vec v_L^\dagger A^{i_1}[1]A^{i_2}[2]\cdots A^{i_N}[N]\vec v_R \ \ket{i_1,i_2,\dots,i_N}.
\end{equation}
where the vectors $\vec v_{L/R}$ encode the boundary conditions, and the integers between square brackets index the physical sites. A translation invariant state on a periodic chain with $N$ sites on the other hand takes the form
\begin{equation}
    \ket{\Psi[A]} = \sum_{\{i_j\}} {\rm Tr}\big(A^{i_1}A^{i_2}\cdots A^{i_N}\big) \ket{i_1,i_2,\dots,i_N}.
\end{equation}

\bigskip\noindent As is customary when dealing with tensor networks, we will use a graphical language in which a single MPS tensor is depicted as:
\begin{equation}
    %
	\input{figs/MPSTensor.tex}%
 = \ A_{\alpha\beta}^i, \qquad \substack{i=1,2,\dots,d\\ \alpha,\beta=1,2,\dots,\chi}
\end{equation}
and where a leg connecting two tensors indicates a contraction over the corresponding indices. With these conventions in mind, we can depict the MPS \eqref{eq:MPS} as following diagram:
\begin{equation}
    \ket{\Psi[A]} = %
	\input{figs/MPS.tex}%
 \ .
\end{equation}
Often, we will not explicitly indicate the MPS tensor in the figures when they can be deduced from the context. 

\bigskip\noindent Now that we have introduced the variational class of matrix product states, let us argue why these states approximate well the ground states of local gapped Hamiltonians~\cite{Verstraete2006a}. Physically, this follows from the \emph{area law} of entanglement entropy satisfied by gapped ground states. In generic dimensions, an area law for the entanglement entropy means that the bipartite entanglement entropy between a region $\mc A$ and its complement $\bar{\mc A}$, defined as the Von Neumann entropy of the reduced matrix $\rho_\mc A={\rm Tr}_{\bar{\mc A}}(\rho)$, i.e. $S=-{\rm Tr}(\rho_\mc A\log(\rho_\mc A))$, does not scale with the size, or \emph{volume}, of the region $\mc A$ but rather by the size of its boundary $\partial\!\mc A$, or thus its \emph{area}. In the one-dimensional setting the boundary of a connected subregion of the chain is just a constant, and in this setting the area law was proven in 2007 by Hastings~\cite{Hastings2007}. As we will now demonstrate MPS inherently satisfy an area law.

To this end, we make use of the graphical calculus of MPS. Herein, let us calculate the reduced density matrix of a uniform MPS when tracing out the left half of the system. This can be depicted as
\begin{equation}
    \rho_R = {\rm Tr}_L\big(\ketbra{\Psi[A]}\big) = %
	\input{figs/MPSReducedDensity_1.tex}%
\,\,,
\end{equation}
where the dotted vertical line represents the entanglement cut between the left ($L$) and right ($R$) subsystem. An object which plays a pivotal role in the computation of the reduced density matrix, but also appears in e.g. the computation of correlation functions (see \cref{sec:NormalObs}), is the \emph{transfer matrix}
\begin{equation}
    \mbb T_A = \sum_i A^i\otimes \bar A^i,
\end{equation}
which we have indicated as well in the above figure by the dotted box. Let us write its spectral decomposition\footnote{More generally, Jordan blocks could be needed, but this is not relevant for the following discussion.} as
\begin{equation}\label{eq:TransferSpec}
    \mbb T_A = \sum_{i=1}^{\chi^2} \lambda_i \ketbra{\rho_i}{\lambda_i},
\end{equation}
where the eigenvalues are ordered according to decreasing absolute value. The eigenvectors are clearly invariant subspaces of the completely positive linear maps $\$_L(X)=\sum_i A^{i \dagger} X A^i$ and $\$_R(X)=\sum_i A^i X A^{i \dagger}$. If those maps are ergodic, the quantum Perron-Frobenius theorem dictates that the largest eigenvalue in magnitude will be unique, real and positive. Furthermore, the corresponding left and right eigenvectors  $\ket{\rho_1}\equiv\ket{\rho}$ and $\ket{\lambda_1}\equiv\ket{\Lambda}$, when reshaped as matrices (they have 2 indices), are positive definite. This property of having a unique dominant eigenvector is equivalent to the notion of injectivity mentioned above. Another equivalent characterisation of injectivity is as follows: an MPS is injective if and only if there exists a finite $L$, called the \emph{injectivity length}, such that the collection of matrices $\{A^{i_1}A^{i_2}\dots A^{i_L}\}$ is a basis for all $\chi\times\chi$ matrices. This is also precisely the property needed to guarantee that the MPS is the \emph{unique} ground state of its gapped parent Hamiltonian. For a generic MPS, this will always be the case: the set of non-injective MPSs is of measure zero, as this is only possible when the matrices $A^i$ have common invariant subspaces. 

To ensure that the state is properly normalised to one we will always normalise the MPS tensor $A$ such that the corresponding eigenvalue $\lambda_1$ is equal to one (see also \cref{sec:NormalObs}). From these considerations, it follows that we can express the reduced density matrix as follows:
\begin{equation}\label{eq:ReducedDensity}
    \rho_R = %
	\input{figs/MPSReducedDensity_2.tex}%
.
\end{equation}
As $\rho_R$ is the product of a $d^L\times\chi$ and a $\chi\times d^L$ matrix with $L$ the total number of sites on the right hand side, we conclude that the rank of the reduced density matrix $\rho_R$ is upper bounded by $\chi$. The entanglement entropy is hence bounded by $S \leq \log \chi$, as this is the maximal entropy for a state of rank $\chi$. The theorems in refs.~\cite{Verstraete2006a,Hastings2007} ensure that the converse is also true: whenever a state satisfies an area law for the  entanglement entropy, there exists an MPS that will be a faithful approximation of that state.  In the next lecture we will study the entanglement properties of general matrix product states in more detail.
\subsection{Matrix product operators}\label{sec:MPO}
Throughout these lecture notes we will also be making use of tensor network representations of linear operators. Uniform \emph{matrix product operators}~\cite{Verstraete2004c,Pirvu2010} are written as
\begin{equation}
    \sum_{\{i_k,j_k\}} \big(\cdots M^{i_k,j_k}M^{i_{k+1},j_{k+1}}\cdots \big) \ketbra{\dots, i_k,i_{k+1},\dots}{\dots, j_k,j_{k+1},\dots},
\end{equation}
and graphically depicted by
\begin{equation}
    %
	\input{figs/MPO.tex}%
\,\, .
\end{equation}
In analogy with MPS, the generalisation to open boundary conditions, and the non-uniform case is straightforward.

MPOs have since long been used extensively in the field of statistical physics, where MPOs of a particular type are known as transfer matrices (see \cite{Haegeman2016} for a unifying discussion). Similarly to the case of MPS, we call an MPO injective when the matrices $\{M^{ij}\}_{i,j}$ have no common invariant subspaces and span the entire matrix algebra. 

Non-injective MPOs are however also used extensively, for example to encode Hamiltonians of quantum spin chains. Let us e.g. consider the Hamiltonian for the transverse field Ising model with open boundary conditions:
\begin{equation}\label{eq:TFIM}
    H = -\sum_{\msf i=1}^{L-1} \sigma^Z_{\msf i}\sigma^Z_{\msf i+1} - \lambda\sum_{\msf i=1}^L \sigma^X_{\msf i}.
\end{equation}
This Hamiltonian can be represented as an MPO with site-independent tensors\footnote{In this notation, the element $M_{\alpha\beta}$ is an operator acting on the physical Hilbert space.} $M$ of bond dimension $3$:
\begin{equation}
    M_{\alpha\beta} = \begin{pmatrix}
    \mbb 1 & -\sigma^Z & -\lambda\sigma ^X\\
    0 & 0 & \sigma^Z\\
    0 & 0 & \mbb 1
    \end{pmatrix}_{\alpha\beta},
\end{equation}
and boundary vectors 
\begin{equation}
\vec{v}_L=\begin{pmatrix}1\\0\\0\end{pmatrix},\qquad
\vec{v}_R=\begin{pmatrix}0\\0\\1\end{pmatrix}.
\end{equation}

Note that since we are dealing with a non-injective MPO here (there are invariant subspaces), the boundary vectors must be defined. 
This MPO is one of the generators of the Onsager algebra.  
Its exponential, $\exp(itH)$, cannot be written anymore as an MPO with a bond dimension independent of the system size -- unless $\lambda=0$, then (exercise) it can be written as an injective MPO with $\chi=2$.

\subsection{The manifold of MPS}\label{sec:MPSmanifold}
Now that we have introduced matrix product states and operators through the lens of the entangled-pair construction, let us investigate in more detail the manifold of uniform matrix product states~\cite{Haegeman2013a,Haegeman2014b,Vanderstraeten2019}. To this end we consider all uniform MPSs with a fixed bond dimension $\chi$, $\ket{\Psi[A]}$. As a first observation, note that the these states don't span a linear subspace of the full Hilbert space since the sum of two MPSs is typically not an MPS of the same bond dimension. Instead, they live on a surface whose local coordinates are given by the variational parameters of the state, i.e. $A^i_{\alpha\beta}$. Recall from above that to every point of the manifold we can associate a linear tangent space. For the case at hand, the tangent vectors are of the form
\begin{equation}
    \ket{\Psi[B,A]} = \sum_I B^I\pdv{}{A^I}\ket{\Psi[A]},
\end{equation}
where here and below $I$ collectively denotes the physical and virtual indices of the MPS tensors.
Graphically, such a tangent vector boils down to
\begin{equation}\label{eq:TangentVector}
    \ket{\Psi[B,A]} = \sum_\msf n %
	\input{figs/TangentVector.tex}%
 \,\, ,
\end{equation}
where the sum over $\msf n$ is a sum over all the positions of the $B$ tensor in the infinite chain.

At any given point $\ket{\Psi[A]}$ of this manifold the overlap of two tangent vectors $\ket{\Psi[B,A]}$ and $\ket{\Psi[B',A]}$ provides a metric or \emph{Gram matrix} $G$ via
\begin{equation}
    \braket{\Psi[\bar B',\bar A]}{\Psi[B,A]} = \sum_{I,J} \bar B'^I \underbrace{\braket{\pdv{}{\bar A^I}\Psi[\bar A]}{\pdv{}{A^J}\Psi[A]}{}}_{G(A,\bar A)_{I,J}} B^J.
\end{equation}
Naively one would now assume that the dimension of the manifold is equal to the number of parameters in the tensor $B$, viz. $d\chi^2$. But note that gauge transformations of the form $B^i \rightarrow B^i + XA^i - A^iX$ leave the state $\ket{\Psi[B,A]}$ invariant so that $d\chi^2$ overcounts the dimension of the MPS manifold. A more careful argument which is made precise in ref.~\cite{Vanderstraeten2019} demonstrates that only $(d-1)\chi^2$ of those parameters can be chosen independently. Provided that $A_L^i$ is in the so called \emph{left canonical gauge} (see \cref{sec:NormalObs}), a faithful parametrisation of $B$ is then $B^i=V^iX$, where the $V^i$ spans the null space of $(A_L)_{\alpha\beta}^i$ (interpreted as matrix from $\beta$ to $\alpha,i$) and $X$ encodes the independent degrees of freedom of $B$. With this parametrisation of $B$, the metric is locally Euclidean, i.e. the Gram matrix is equal to the identity. From the computational point of view this parametrisation is useful as it vastly simplifies the implementation of algorithms and improves their stability.

One such algorithm is the TDVP applied to the class of MPS. Recall from \cref{sec:Illusion} that TDVP aims to approximately solve the time-dependent Schr\"odinger equation
\begin{equation}
    i\pdv{t}\ket{\Psi[A]} = H \ket{\Psi[A]}
\end{equation}
by projecting the right-hand side of this equation onto the tangent plane:
\begin{equation}\label{eq:TDVP}
    i\pdv{t}\ket{\Psi[A(t)]} = \mc P_{A(t)}H \ket{\Psi[A(t)]},
\end{equation}
where $\mc P_{A(t)}$ is the projector on the tangent plane. A further manipulation of the TDVP equation \cref{eq:TDVP} reveals that the TDVP is equivalent to a (semi-)classical set of equations of motions governed by a \emph{Poisson bracket} on the MPS manifold. The famous Density Matrix Renormalisation Group (DMRG) algorithm \cite{White1992} -- discussed below in \cref{sec:DMRG} -- and all its variants can be understood as specific splitting methods for solving this differential equation~\cite{Haegeman2016b}.  Recall that a Poisson bracket is an bilinear, antisymmetric function on phase space which satisfies the Jacobi identity. If we define the Poisson bracket for arbitrary functions $f$ and $g$ that depend on $A$ and $\bar A$ as follows:
\begin{equation}
    \{f,g\} = - \frac{\partial f}{\partial A_I} (G^{-1})^{IJ} \frac{\partial g}{\partial \bar A_J} + \frac{\partial g}{\partial A_I} (G^{-1})^{IJ} \frac{\partial f}{\partial \bar A_J},
\end{equation}
and write $f(A,\bar A)=\expval{F}{\Psi[A]}$, it can be shown that the time evolution of the expectation value $f$ is governed by
\begin{equation}
    \partial_t f = i\{f,h\}.
\end{equation}
This has two significant consequences. For one, it follows from $\{h,h\}=0$ that the energy expectation value is exactly preserved under time evolution. This was anticipated above and implies that the manifold is symplectic. On the other hand, for any operator $K$ which commutes with the Hamiltonian, and thus constitutes a symmetry generator, as well as with the tangent space projector $\mc P_A$, it can be shown that $\{k,h\}=0$ and hence $\partial_t k=0$. In particular, this is the case for symmetry generators which are tensor products of local on-site unitaries.

Besides TDVP it should be mentioned that the manifold picture has also provided the necessary framework for ground state optimisation directly in the thermodynamic limit, the so-called \emph{variational uniform MPS} (VUMPS) algorithm, as well as the construction of \emph{quasiparticle excitations}. These topics are covered in \cref{sec:DMRG_VUMPS} and \cref{sec:Excitations} respectively, but first we now delve deeper into the calculus of MPS.
\subsection{Synopsis of lecture I} The main message of lecture I is the fact that the full many-body Hilbert space is an illusion: to describe relevant many-body states such as ground states of quantum spin systems, it is sufficient to consider a low-dimensional manifold of states which captures all relevant physics. For the particular case of gapped quantum spin chains, the monogamy of quantum entanglement implies the existence of an area law for the entanglement entropy. This implies that the manifold of MPS is in one-to-one correspondence with all states satisfying an area law. Instead of focusing on the full Hilbert space, we can hence project the Hamiltonian on this MPS manifold using the prescription of the time-dependent variational principle; all tensor network algorithms can be understood within this framework.

%% file: figs/AKLT_proj.tex
\begin{tikzpicture}[baseline=(current bounding box.center)]
	\def\l{1.4};
    \def\h{0.8};

    \draw[fill=black!8, draw=black, line width=.7pt] (0,0) ellipse (16pt and 8pt); 
    \node[anchor=west] at (\l/2,0) {$=\ketbra{+}{\uparrow\uparrow} + \ket{0}{\frac{\bra{\uparrow\downarrow}+\bra{\downarrow\uparrow}}{\sqrt{2}}} + \ketbra{+}{\downarrow\downarrow} $};
\end{tikzpicture}

%% file: figs/MPO.tex
\begin{tikzpicture}[baseline={([yshift=-.5ex]current bounding box.center)}]
	\def\l{1.6};
    \def\h{.65};

	\draw[obj2] (-.5*\l,0) -- (2.5*\l,0);
	
	\foreach \x in {0,1,2}{
		\draw[obj1] (\x*\l,\h) -- (\x*\l,-\h);
		\node[mpotensor] at (\x*\l,0) {};
	}
\end{tikzpicture}

%% file: _Lecture_2.tex
\section{Lecture II: The calculus of MPS \& algorithms}
In this section we will introduce the calculus of matrix product states. In particular this constitutes the computation of observables, which is simplified greatly by introducing appropriate \emph{normal forms} for the MPS tensors. With these tools at hand we will focus on the renowned \emph{density matrix renormalization group} (DMRG) algorithm for ground state optimisation on finite chains, as well as the \emph{variational uniform matrix product state} (VUMPS) algorithm, which works directly in the thermodynamic limit. We then study quasiparticle excitations on top of the ground state, which are parametrised as the \emph{quasiparticle ansatz}. Let us however kick off with some more examples of matrix product states.
\subsection{Examples, injectivity \& the parent Hamiltonian}\label{sec:Examples}
\paragraph{Product states} Any product state $\ket{\Psi}=\ket{\psi}^{\otimes N}$, $\ket{\psi}=\sum_i \psi_i\ket{i}$ can be written as a bond dimension 1 MPS of the form
\begin{equation}
    A^i_{11} = \psi_i.
\end{equation}
\paragraph{AKLT} The paradigmatic example of a non-trivial matrix product state encountered above in \cref{sec:AreaLaw} is the Affleck-Kennedy-Lieb-Tasaki (AKLT) state. This state is generated by the MPS matrices:
\begin{equation}\label{eq:AKLTState}
    A^i = \sqrt{\frac{1}{3}}\sigma^i, \quad i=X,Y,Z,
\end{equation}
where the $\sigma^i$ are the Pauli matrices, and the prefactor ensures that the state is normalised in the thermodynamic limit.

On periodic boundary conditions, this state is the unique gapped ground state of the spin 1 Hamiltonian \eqref{eq:AKLTHam}. As will be explained below in \cref{sec:SPT1d}, this Hamiltonian has a fourfold ground state degeneracy on open chains, characterised by gapless edge modes. This is a hallmark of its \emph{symmetry-protected topological} (SPT) order.
\paragraph{GHZ} Another example of a state which can be expressed as a bond dimension 2 MPS is the so called Greenberger–Horne–Zeilinger (GHZ) state \cite{Greenberger1989}:
\begin{equation}
    \ket{\Psi} = \frac{1}{\sqrt{2}} \big( \ket{0}^{\otimes N} + \ket{1}^{\otimes N} \big).
\end{equation}
Its MPS representation reads:
\begin{equation}
    A^i =
    \begin{cases}
        \frac{1}{\sqrt{2}}\begin{pmatrix} 1 & 0 \\ 0 & 0\end{pmatrix}, \quad i=0 \\
        \frac{1}{\sqrt{2}}\begin{pmatrix} 0 & 0 \\ 0 & 1\end{pmatrix}, \quad i=1.
    \end{cases}
\end{equation}
Note that the GHZ state may be viewed as exhibiting $\mbb Z_2$ symmetry breaking. This is reflected in the fact that the MPS matrices are block diagonal, and the blocks are swapped under the action of $\prod_\msf i \sigma^X_\msf i$. Clearly, this MPS representation of the GHZ-state is non-injective.

This has to be contrasted with the AKLT state, whose MPS matrices span the entire $\mbb C^{2\times 2}$ matrix algebra, therefore cannot be brought in block diagonal form, and thus define an injective MPS. Note that the injectivity length of the AKLT state is $2$, which can be checked by an immediate computation.

It holds generically that every uniform MPS can be brought\footnote{After getting rid of off-diagonal blocks, as reviewed in e.g. ref.~\cite{Cirac2020}.} in a \emph{block injective} or a \emph{standard} form whereby all MPS matrices are of the form~\cite{PerezGarcia2006,Cadarso2012}
\begin{equation}
    A^i = \bigoplus_{k=1}^r \mu_k A_k^i,
\end{equation}
and where all $\{A^i_k\}_k$, $k=1,...,r$ are injective, and sometimes called \emph{injective blocks}. It should be noted that this notion of injectivity should be revisited in the case of \emph{fermionic} MPS, which is the topic of \cref{sec:Fermions}.

Interestingly, every MPS can be regarded as a ground state of a corresponding \emph{parent Hamiltonian}, thereby placing the study of MPS and their associated parent Hamiltonians on the same footing. This parent Hamiltonian is gapped, local and frustration-free, meaning that the energy of each term is minimised by the MPS. Its local terms are the projectors on the null space of the local reduced density matrix on the corresponding contiguous region of spins~\cite{PerezGarcia2006,Cirac2020}. For the GHZ state for example, the local parent Hamiltonian term is given by $\frac{1}{2}(\mbb 1 - \sigma^Z_{\msf i}\sigma^Z_{\msf i+1})$. When an MPS is injective, it can be shown that it is the unique ground state of its parent Hamiltonian. More interestingly, when the parent Hamiltonian commutes with an on-site representation of a symmetry group $G$, as in the case of the GHZ state, the action of $G$ permutes (a minimal irreducible subset of) the different injective blocks. In that case, one finds a maximal subgroup $H\subseteq G$ whose action leaves a particular reference block invariant~\cite{Schuch2011}. The blocks are then labelled by \emph{cosets} in $G/H$.\footnote{Recall that a (left) coset $gH$ is the set $gH=\{gh|h\in H\}$. The set of all cosets is denoted $G/H$ and partitions the group $G$.}

Below, in lecture~\ref{sec:PhasesOfMatter} and lecture~\ref{sec:GenSym}, we will re-encounter the parent Hamiltonian as a tool to classify \emph{quantum phases of matter}, which -- roughly speaking -- are equivalence classes of Hamiltonians which can be adiabatically transmuted into each other.

\subsection{Normal forms, entanglement \& observables}\label{sec:NormalObs}
As mentioned in \cref{sec:AreaLaw}, the set of matrices $\{A^i\}$ is not a unique representation of a given MPS. This is because matrix product states are invariant under reparametrisations of the form
\begin{equation}
    A^i \mapsto XA^iX^{-1}, \quad \forall i=1,2,\dots,d,
\end{equation}
for any non-singular matrix $X$, since the matrices $X$ and $X^{-1}$ cancel each other on the bonds connecting neighbouring MPS matrices. This redundancy is often called \emph{gauge} freedom of the MPS description. It often turns out that this gauge freedom can be heavily exploited to simplify certain computations and improve the numerical stability of MPS algorithms.

By appropriate gauge transformations, the MPS matrices can be brought in the so-called \emph{left} or \emph{right canonical forms}, i.e. respectively defined via the conditions:
\begin{align}
    \sum_{i} A^{i \dagger} A^i &= \mbb 1_\chi \label{eq:MPSLeftCanonical}, \\
    \sum_{i} A^i A^{i \dagger} &= \mbb 1_\chi. \label{eq:MPSRightCanonical}
\end{align}
Diagrammatically, these conditions can thus be written as follows:
\begin{align}
    %
	\input{figs/MPSRightCanonical.tex}%
 \quad &= \quad %
	\input{figs/MPSRightIdentity.tex}%
\,\, , \\
    %
	\input{figs/MPSLeftCanonical.tex}%
 \quad &= \quad %
	\input{figs/MPSLeftIdentity.tex}%
\,\, .
\end{align}
Note that the conditions \cref{eq:MPSLeftCanonical,eq:MPSRightCanonical} don't fix the MPS matrices completely as we can still conjugate the matrices with \emph{unitary} gauge matrices. In what follows we will  write $A_L$ and $A_R$ for the left and right canonical forms of the MPS tensor $A$. Let us stress again that $A$, $A_L$ and $A_R$ all generate the same state. The gauge transformation or \emph{pivot matrix} that transmutes the left canonical form into the right canonical form is often denoted by $C$ and is thus characterised by $A_L^iC=CA_R^i$, $\forall i$. For future reference let us also introduce the notation $A_C^i =A_L^iC=CA_R^i$. An MPS is said to be in \emph{mixed canonical form} if it can be written as:
\begin{equation}\label{eq:MPSMixedGauge}
    \ket{\Psi[A]} = %
	\input{figs/MPSMixedGauge.tex}%
\,\,.
\end{equation}
As this depiction suggests, the matrix $C$ contains information about the entanglement between the subsystems to the left and right of $C$. Let us make this more precise.

Consider a Hilbert space that admits a tensor product decomposition $\mc H=\mc H_A\otimes \mc H_B$, with $d_A=\dim(\mc H_A), d_B=\dim(\mc H_B)$. In virtue of the \emph{singular value decomposition} (SVD), any state in $\mc H$ can be written in the form
\begin{equation}\label{eq:SchmidtState}
    \ket{\Psi} = \sum_k s_k \ket{\psi_k^A}\otimes\ket{\psi_k^B},
\end{equation}
where $\ket{\psi_k^A}$ and $\ket{\psi_k^B}$ are orthonormal basis vectors for $\mc H_A$ and $\mc H_B$ respectively, and the real numbers $s_k$ for $k=1,2,...,\min(d_A,d_B)$ are called the \emph{singular values} or \emph{Schmidt numbers} and constitute the \emph{entanglement spectrum} of the state. By tracing out either subsystem we can write the reduced density matrices as
\begin{align}
    \rho_A &= {\rm Tr}_B(\ketbra{\Psi}) = \sum_{k=1}^{d_A} s_k^2 \ket{\psi_k^A}\bra{\psi_k^A}, \\
    \rho_B &= {\rm Tr}_A(\ketbra{\Psi}) = \sum_{k=1}^{d_B} s_k^2 \ket{\psi_k^B}\bra{\psi_k^B}.
\end{align}
A direct comparison with the MPS \eqref{eq:MPSMixedGauge} reveals that the Schmidt spectrum of the state $\ket{\Psi[A]}$ coincides with the singular values of $C$. The MPS $\ket{\Psi[A]}$ can eventually be brought in the form \eqref{eq:SchmidtState} via the residual unitary gauge freedom $A_L^i\mapsto UA_L^iU^\dagger $, $A_R^i\mapsto V^\dagger A_R^iV$, with $U$ and $V$ obtained from the SVD $C=USV^\dagger$.

\bigskip\noindent One of the key advantages of MPS is the efficient computation of expectation values of local observables and their correlation functions. \emph{Efficient} means here with a computational cost that scales linear in the distance between the local operators. As an illustration, let us focus on the two-point correlation function of local observables $O_1$ and $O_2$ at sites $\msf i$ and $\msf j$. Due to translation invariance this expectation value depends only on the distance $L=|\msf j - \msf i|$. Defining expectation values of observables as usual:
\begin{equation}
    \expval{O_1[\msf i]O_2[\msf j]}_{\Psi[A]} := \mel{\Psi[A]}{O_1[\msf i]O_2[\msf j]}{\Psi[A]},
\end{equation}
this boils down to the contraction
\begin{equation}
    \expval{O_1[\msf i]O_2[\msf j]}_{\Psi[A]} \, = %
	\input{figs/MPSCorrelator_1.tex}%
.
\end{equation}
By assuming injectivity and thus again that the transfer matrix has a unique largest eigenvalue $1$ with corresponding left and right eigenvectors $\Lambda$ and $\rho$, we are able to rewrite this contraction as
\begin{equation}
    %
	\input{figs/MPSCorrelator_2.tex}%
\,\, ,
\end{equation}
which demonstrates that this correlator can be computed with a cost that scales as $\mc O(L\chi^3)$. Indeed, the contraction of this tensor network proceeds from left to right, and by choosing the contraction order in a judicious way, the main cost is the multiplication of the $\chi\times\chi$ matrices (see ref.~\cite{Pfeifer2014} for an algorithm that identifies the optimal contraction order of a generic tensor network).
As particular cases of this expression, we have the expectation value of a local operator by taking $O_1\equiv O$ and $O_2\equiv\mbb 1$:
\begin{equation}
    %
	\input{figs/MPSExpVal.tex}%
\,\, ,
\end{equation}
as well as the norm squared of the MPS
\begin{equation}
    \braket{\Psi[A]} := %
	\input{figs/MPSNorm.tex}%
\,\, .
\end{equation}
If we now define the \emph{connected} two-point correlation function as customary:
\begin{equation}
    C(O_1[\msf i],O_2[\msf j]) := \expval{O_1[\msf i]O_2[\msf j]}_{\Psi[A]} - \expval{O_1[\msf i]}_{\Psi[A]} \expval{O_2[\msf j]}_{\Psi[A]},
\end{equation}
one can show, by once more making use of the spectral decomposition of the transfer matrix \cref{eq:TransferSpec} (assuming it exists), that the dominant contribution to this correlation function decays \emph{exponentially} in $L$, more precisely:
\begin{equation}
    C(O_1[\msf i],O_2[\msf j]) \sim \lambda_2^{L-1} \sim \exp(-L/\xi),
\end{equation}
if we define the \emph{correlation length} $\xi$ as $\xi = - 1/\log(|\lambda_2|)$. This exponential decay of correlation functions is a witness of the fact that MPS describe well the ground state physics of \emph{gapped} Hamiltonians, and can thus not reproduce the algebraic decay of correlations present in \emph{critical systems}. Nevertheless, by making use of the theory of \emph{entanglement scaling}, the topic of the next section, precise approximations to critical correlation functions can be achieved.
\subsection{Scaling theory}\label{sec:scaling}
Aside from gapped ground state physics, MPS methods are also a powerful tool to compute properties of \emph{critical} systems. Using uniform methods which work directly in the thermodynamic limit, i.e. on the infinite chain (see Sec. \ref{sec:DMRG_VUMPS}), it is possible to extract critical exponents and the central charge describing the universality class or CFT of a continuous phase transition. There is a large body of existing research about the topic, originally introduced in refs.~\cite{Nishino1996,Tagliacozzo2007,Pollmann2009,Pirvu2012}. Here we will follow the approach developed in the papers \cite{Zauner2015,Rams2018,Vanhecke2019}.

Close to a second-order critical point, we expect correlation functions to be of the Ornstein–Zernike form, it is a product of a power law contribution multiplied with an exponential decaying contribution. This can be understood from the K{\"a}ll\`en--Lehmann representation of correlation functions as a linear combination of a continuum of exponentials \cite{Zauner2015}. To express the continuum directly using MPS, one would need an infinite bond dimension MPS. Using finite bond dimension MPS is equivalent to a discretisation of this continuum spectrum. We can quantify the `discreteness', or degree of approximation, by the gaps in the transfer matrix spectrum. Let $\lambda_i$ as before be the eigenvalues of the transfer matrix, with $\lambda_1=1$, and define $\epsilon_i = - \ln\lambda_i$. Take a linear combination of the top $n$ spectral variables $\epsilon_i$, excluding the first one:
\begin{equation}
    \delta := \sum_{i=2}^{n} c_i \epsilon_i  \quad \text{with} \quad \sum_{i=2}^n c_i = 0.
\end{equation}
In refs.~\cite{Zauner2015,Rams2018}, it was demonstrated that the $\epsilon_i$ have a meaning of momenta, and that the difference between consecutive values of $\epsilon_i$ therefore introduces an inverse length scale $\delta$ which goes to zero if $\chi\rightarrow\infty$, both in the gapped and in the gapless case. In analogy to the finite-size scaling hypothesis of Cardy \cite{Cardy}, where the length scale is the system size, it is possible to use this inverse length scale $\delta$ to formulate a \textit{scaling hypothesis} for any order parameter $m$ of a second-order phase transition as follows:
\begin{equation}
    m(t,\delta) = s^{-\beta/\nu} \, m\!\left( s^{1/\nu} t,\, s\delta \right).
\end{equation}
An analogous scaling hypothesis for the correlation length is:
\begin{equation}
    \delta \xi(t,\delta) = \xi\!\left( \delta^{-1/\nu} t,\, 1 \right)
= \tilde{\xi}\!\left( \delta^{-1/\nu} t \right).
\end{equation}
From the forms above, one can also deduce the scaling hypothesis for the entanglement entropy. Using the CFT form for the entanglement entropy $S$ \cite{Calabrese2004}, we know that $\text{exp}(\frac{6}{c}S)$ scales as length, so the scaling hypothesis for $S$ is:
\begin{equation}
    \exp\!\left( \frac{6}{c} S(t,\delta) \right)
= s \exp\!\left( \frac{6}{c} S\!\left( s^{-1/\nu} t,\, s\delta \right) \right),
\end{equation}
where $c$ is the central charge. Performing simulations using uniform MPS methods with various bond dimensions $\chi$ then allows for fitting $m, \xi, c$ using standard optimisation methods. The whole point of this is the fact that many different simulations for many different parameters of the Hamiltonian close to the critical point and for different bond dimensions can be related to each other, and that the correct values of the critical point and critical exponents will lead to a data collapse of all the ensuing curves. We refer to ref.~\cite{Vanhecke2019} for details. Working with a finite bond dimension is like opening a gap in the system, and its effect is indistinguishable to adding a relevant perturbation to a critical theory. This way, the finite entanglement scaling theory yields the state-of-the-art way of simulating critical theories -- the fact that MPS are always gapped is not a bug, it is a feature!
\subsection{Variational MPS algorithms}
\label{sec:DMRG_VUMPS}
\subsubsection{Density matrix renormalisation group}\label{sec:DMRG}
The \emph{density matrix renormalisation group} (DMRG) algorithm, originally due to Steven White \cite{White1992}, is a variational method for finding the ground state of \emph{local one-dimensional Hamiltonians}. It has been applied with tremendous success to a plethora of models and is still the current state of the art for ground state optimisation in (1+1)d for strongly-correlated spin chains~\cite{Schollwoeck2010,Verstraete2023}. 

Although DMRG was originally derived and explained in terms of the renormalisation group, its modern interpretation is that of a variational method over the class of matrix product states, where the optimum is iteratively found by using an alternating least squares method~\cite{Verstraete2004a}. DMRG minimises the expectation value
\begin{equation}\label{eq:min}
    \underset{\{A\}}{\rm min}\left(\frac{\expval{H}{\Psi[A]}}{\braket{\Psi[A]}}\right),
\end{equation}
over all finite MPS of a maximum (bounded) bond dimension $\chi_{\rm max}$. The larger you choose $\chi_{\rm max}$, the better the approximation will be. Even with modest computational resources, one can easily achieve bond dimensions $\chi\sim\mc O(10^3)$ by making use of sparse eigensolvers such as a Lanczos method to solve the effective eigenvalue problem (see below), or even $\mc O(10^5)$ when symmetries are exploited as well~\cite{Sanz2009,Singh2009,Weichselbaum2012,Singh2013,Lootens2024,Devos2025}.

Assuming that the Hamiltonian is written in MPO form (see \cref{sec:MPO}), \cref{eq:min} boils down to:
\begin{equation}
    \underset{\{A\}}{\rm min}\,\, \frac{%
	\input{figs/DMRG_H.tex}%
}{%
	\input{figs/DMRG_norm.tex}%
}\,\,.
\end{equation}
The algorithm is initiated by providing an initial guess for the ground state, typically simply a random MPS, and proceeds with a series of local sequential optimisations of all tensors, sweeping back and forth through the chain, until convergence is reached.

Fixing a site $\msf i$ and bringing all tensors to its left in left canonical form and vice versa for the tensors to its right, the update of the tensor $A[\msf i]$ boils down to replacing it by the eigenvector corresponding to the smallest real eigenvalue of the \emph{effective Hamiltonian}
\begin{equation}
    %
	\input{figs/DMRG_Heff1.tex}%
\,\,.
\end{equation}
In this figure, $L$ and $R$ stand for the (site-dependent) \emph{environments} obtained by contracting all MPS and MPO tensors to the left, respectively right, of site $\msf i$.

As such, this version of the algorithm updates each site individually and is referred to as \emph{one-site DMRG}. The downside of this approach is that there is no convenient way of increasing the bond dimension during the sweeps. This is particularly problematic when symmetries are imposed, since in that case the algorithm can not explore distinct symmetry charges on the virtual level. This shortcoming is remedied by a \emph{two-site} version of the algorithm.

This version of the algorithm proceeds by blocking two sites in every step of every sweep, and subsequently updating the blocked tensor $A'[\msf i,\msf i+1]$ as the minimal eigenvector of the effective two-site Hamiltonian 
\begin{equation}
    %
	\input{figs/DMRG_Heff2.tex}%
\,\, .
\end{equation}
A singular value decomposition is then used to split $A'$ into $A[\msf i],A[\msf i+1]$. It is in this step that the bond dimension can be dynamically controlled by truncating the SVD so as to only keep the largest singular values. The high-level pseudocode of DMRG can be found below.

\begin{algorithm}[H]
\caption{Two-site DMRG}
 Initialize the MPS tensors $A[\msf i]$ (usually to a random state) \\
\For{$k = 1$ to \texttt{num\_sweeps}}{
    \tcp{Left-to-right sweep}
    \For{$\msf i = 1$ to $N-2$}{
         Optimise and update pair of sites $\msf i, \msf i+1$ \\
         Update left environment at site $\msf i+1$ \\
    }

    \tcp{Right-to-left sweep}
    \For{$\msf i = N-1$ to $2$}{
         Optimise and update pair of sites $\msf i, \msf i+1$ \\
         Update right environment at site $\msf i+1$ \\
    }
}
\Return $\{A[\msf i]\}$
\end{algorithm}
\subsubsection{Variational uniform matrix product state}\label{sec:VUMPS}
The \emph{variational uniform matrix product state} (VUMPS) algorithm is an infinite-chain equivalent of DMRG, and was introduced in ref.~~\cite{Zauner-Stauber2018a} (see  ref.~\cite{Vanderstraeten2019} for a pedagogical review, and \cite{McCulloch2008} for an alternative method).  It works directly in the thermodynamic limit, which makes it particularly useful for the simulation of critical systems and extracting their scaling laws. Of course, in the infinite limit the total ground state energy is unbounded, so that VUMPS rather minimises the \emph{energy density}.

Before introducing the algorithm, let us first fix some notation. Recall the central gauge MPS $A_C$ and bond tensor $C$ introduced in \cref{sec:NormalObs}, alongside the left and right gauges $A_L, A_R$. We then define the effective single-site Hamiltonian, $H_{\rm eff}$, as well as the effective environment for the bond tensor $C$, $H_C$, as:
\begin{equation}
\label{eq:VUMPSEnvironments}
    H_{\text{eff}} := %
	\input{figs/VUMPS_EffectiveHamiltonian.tex}%
\,\,, \qquad
    H_C := %
	\input{figs/VUMPS_EffectiveEnvironment.tex}%
\,\,.
\end{equation}
Here, the left and right environments $L,R$ are the fixed points of the transfer matrices $\mbb T_L^{[H]}, \mbb T_R^{[H]}$ respectively:
\begin{equation}
    \mbb T_L^{[H]} := %
	\input{figs/VUMPS_TransferMatrixHamiltonianLeft.tex}%
\,\, , \qquad \mbb T_R^{[H]} := %
	\input{figs/VUMPS_TransferMatrixHamiltonianRight.tex}%
\,\,.
\end{equation}
They can be calculated by solving a system of linear equations.

As derived in ref.~\cite{Zauner-Stauber2018a}, the variational minimum of the ground state energy density in the manifold of uniform MPS is characterised by a simultaneous solution to
\begin{equation}
\begin{split}
    &H_{\rm eff}(A_C) \propto A_C ,\\
    &H_C(C) \propto C, \\
    &A^i_C = A_L^iC = CA_R^i.
\end{split}
\end{equation}
The last equation is included so as to avoid having to take $A_L=A_CC^{-1}$, $A_R=C^{-1}A_C$, which is potentially ill-defined if $C$ is close to being singular. Given an initial guess for the ground state, the VUMPS algorithm iteratively approaches this solution until convergence is reached. In each step, the current values of $A_C$ and $C$ are updated by replacing them by the eigenvectors with minimal real eigenvalue of the current $H_{\rm eff}$ and $H_C$. Subsequently, approximate $\tilde A_L$ and $\tilde A_R$ are found from $\tilde A_C$ and $\tilde C$ by invoking the singular value decomposition for $A_CC$ and $CA_C$:
\begin{equation}\label{eq:approxALAR}
\begin{split}
    \tilde A_L &= U_LV_L^\dagger, \quad U_L\Sigma_LV_L^\dagger =\tilde A_C\tilde C,\\
    \tilde A_R &= U_RV_R^\dagger,  \quad \tilde C^\dagger \tilde A_C = U_R\Sigma_R V_R^\dagger.
\end{split}
\end{equation}
As such, the 2-norms
\begin{equation}
\begin{split}
    \epsilon_L &= ||\tilde A_C - \tilde A_L\tilde C||_2, \\
    \epsilon_R &= ||\tilde A_C - \tilde C\tilde A_R||_2, \\
\end{split}
\end{equation}
are minimised.

The VUMPS pseudocode for an MPO Hamiltonian can be found below.

\begin{algorithm}[H]
\caption{VUMPS}
    Initialise $\{A_L,A_R,C\}$ (usually to a random state)\\
    Initialise current precision $\epsilon_{\rm cur}$\\
    \While{$\epsilon_{\rm cur}>\epsilon_{\rm tol}$}{
        Calculate $\mbb T_L^{[H]}$ from $A_L, H$ \\
        Calculate $\mbb T_R^{[H]}$ from $A_R, H$ \\
        Calculate the left environment $L$ from $\mbb T_R^{[H]}$ \\
        Calculate the right environment $R$ from $\mbb T_L^{[H]}$ \\
        Construct the effective infinite Hamiltonian $H_{\text{eff}}$ from $L, R, H$ \\
        Construct the effective infinite environment $H_C$ from $L, R$ \\
        Find $A_{C}$, the leading eigenvector of $H_{\text{eff}}$ \\
        Find $C$, the leading eigenvector of $E_{\text{eff}}$ \\
        Calculate $A_L, A_R$ from $A_{C}, C$ as in \cref{eq:approxALAR}\\
        Set $\epsilon_{\rm cur}=\max(\epsilon_L,\epsilon_R)$
    }
    \Return $\{A_L,A_R,C\}$
\end{algorithm}
\subsection{Quasiparticle excitations}\label{sec:Excitations}
Up until this point we have mainly focussed on the ground state, and algorithms to variationally approximate it, both in the finite (\cref{sec:DMRG}) as well as in the infinite (\cref{sec:VUMPS}) setting. A comprehensive understanding of a quantum many-body system should also explain its low-lying excitations. Typically, the spectrum of a gapped quantum many-body Hamiltonian consists of a few isolated branches which have an interpretation in terms of localised \emph{quasiparticle excitations}, as well as a tower of continuum \emph{scattering states} which can be constructed by essentially adding up quasiparticle states. These quasiparticles show up in dynamical correlation functions -- which can be probed in e.g. inelastic neutron scattering experiments -- but in the case of critical systems also encode universal properties of their effective continuum description\footnote{Remember that a finite bond dimension gaps out the excitations when approximating critical ground states using MPS.}.

Matrix product state techniques can be used to capture these quasiparticle excitations and their scattering states by means of the so-called \emph{quasiparticle ansatz}, first introduced in ref.~\cite{Ostlund1995} and subsequently generalised in ref.~\cite{Haegeman2011b} to single-particle excitations in one-dimensional chains and to two-particle states in ref.~\cite{Vanderstraeten2013}. Here, we assume a translationally invariant local and gapped Hamiltonian $H$. For the time being we moreover assume a unique ground state to exclude domain-wall excitations, which will be discussed below. We denote an MPS approximation of the ground state by $\ket{\Psi[A]}$. The ansatz for a quasiparticle with momentum $k$ then takes the form
\begin{equation}\label{eq:QP}
    \ket{\Psi[A,B],k}= \sum_\msf n e^{ik\msf n} %
	\input{figs/TangentVector.tex}%
\,\,.
\end{equation}
That is, the ansatz replaces one of the ground state tensors by a tensor $B$ which contains all variational parameters of the ansatz, and we take a momentum superposition over the position of $B$ in the infinite chain. As such, it can be thought of as a \emph{boosted} version of a tangent vector \cref{eq:TangentVector}, allowing one to exploit the full arsenal of techniques developed for the tangent space. Also, it is interesting to note that the quasiparticle ansatz can be interpreted as a generalisation of the Feynman-Bijl ansatz, or \emph{single mode approximation} developed to describe collective excitations in liquid Helium~\cite{Bijl1941,Feynman1953,Feynman1954}. Note that due to the finite bond dimension, we should not think about $\ket{\Psi[A,B],k}$ as describing a localised particle on the location of the tensor $B$, but rather as a lump of increased energy density with respect to the uniform vacuum. Given $A$, the optimisation of $B$ boils down to a generalised eigenvalue problem for an effective Hamiltonian, similar in spirit to \cref{sec:VUMPS}. A detailed exposition of the algorithm is beyond the scope of these notes, but let us remark that in it, the use of the efficient parametrisation of $B$, very similar to the one described in \cref{sec:MPSmanifold}, is crucial.

It is worth emphasising that the ground state encodes information about the excited states. Indeed, for a generic self-adjoint operator, one of its eigenvectors typically does not reveal anything about the others. However, the locality of the Hamiltonian is the distinguishing feature that ensures that \cref{eq:QP} provides a \emph{good} ansatz for quasiparticle excitations. Indeed, in an application of the famous Lieb-Robinson bounds~\cite{Lieb1961,Nachtergaele2006,Hastings2010}, it was demonstrated in ref.~\cite{Haegeman2013b} that every (exact) eigenstate $\ket{\Psi_{k,\alpha}}$ can be approximated arbitrary well via acting with a momentum $k$ superposition of a localised (in space) operator on the ground state. The ansatz \cref{eq:QP} should then be recognised as the extreme case where the MPS ground state is acted upon by said operator whose support is truncated to one site.

Interestingly, in the presence of an internal symmetry the excitations are organised in degenerate multiplets transforming in an (irreducible) representation of the symmetry group. This is not compatible with the quasiparticle approach that we presented so far, as in that case the energy window around an excitation energy $\Delta$, $[\Delta-\delta,\Delta+\delta]$, contains more than one excitation for any value of $\delta$. To remedy this, one can modify the ansatz by replacing the operator $\widetilde O$ by a multiplet of operators $\widetilde O_m$, which in the MPS picture boils down to specifying the charge of the $B$ tensor~\cite{Zauner-Stauber2018b}.

In the case of symmetry breaking and thus a degenerate ground state subspace, the excitations with lowest energy are not of the form \cref{eq:QP}. Rather, they are \emph{topological domain wall excitations} that form an interface between different ground states. In this context, \emph{topological} refers to the fact that these excitations are created from the ground state by an extensive operator, namely a symmetry operator acting on the half-infinite chain terminated at the location of the domain wall. This can be accounted for by adapting the ansatz so as to encode different ground state tensors on either side of the $B$ tensor. All computational MPS techniques work equally well in this case, but care has to be taken to contract the right environments with each other. 

\begin{figure}[htb]
    \centering
    \includegraphics[width=0.7\linewidth]{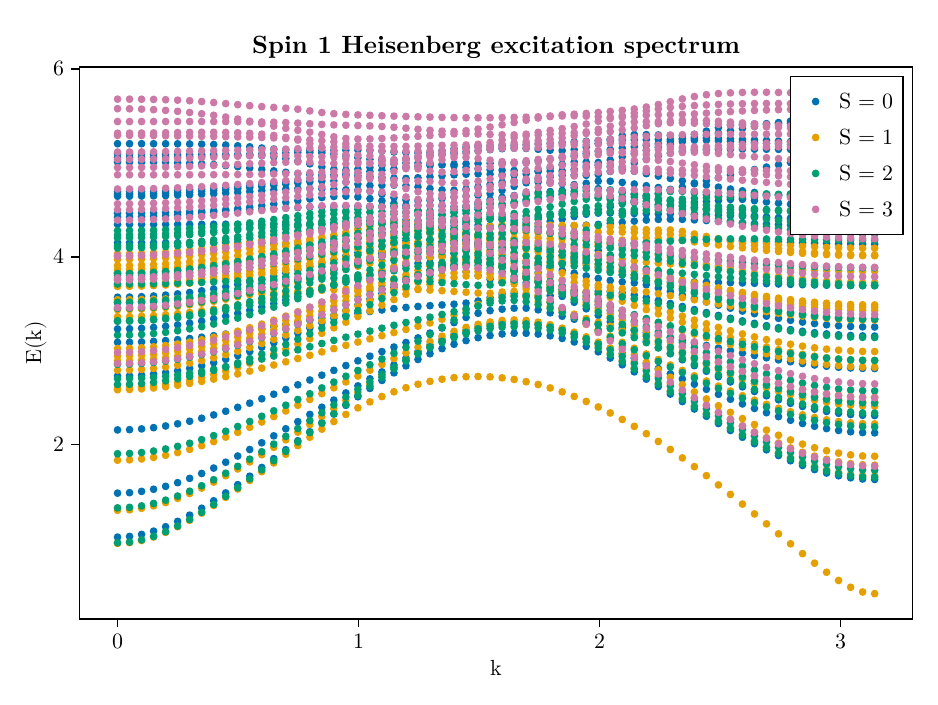}
    \caption{Excitation spectrum of the Heisenberg spin 1 antiferromagnet obtained with the one-site quasisparticle ansatz at bond dimension $48$. Excitations are labelled by their ${\rm SU}(2)$ spin. Note in particular that the \emph{Haldane} gap is situated at momentum $\pi$. Created with MPSKit.jl~\cite{VanDamme}.}
    \label{fig:HeisenbergExcitations}
\end{figure}

In \cref{fig:HeisenbergExcitations} we depict the low-energy excitation spectrum of the spin 1 antiferromagnetic Heisenberg model. It was obtained by explicitly enforcing the ${\rm SU}(2)$ symmetry, and we labelled the excitations by their spin. The isolated $S=1$ branch is clearly visible and reaches its minimum $\Delta\approx 0.41$ at $k=\pi$, famously known as the \emph{Haldane gap}~\cite{Haldane1982,Haldane1983}. Astonishingly, this energy gap can be computed using the quasiparticle ansatz with a precision of ca. 10 digits, even with very modest computational resources.
\subsection{Synopsis of lecture II} Lecture II has been dealing with the computational aspects of matrix product states - how to define normal forms, how to minimise the energy on the manifold of MPS using alternating least squares methods (DMRG), how to extend those methods to infinite system sizes (VUMPS), and how to define excitations by projecting the full Hamiltonian on the tangent plane of the manifold of MPS. Furthermore, we touched upon the topic of entanglement scaling - through which critical systems can be studied with unprecedented precision. 

The fact that - in all those applications - the computational cost only scales polynomially in both the system size and the bond dimension is a clear witness of the fact that tensor network methods effectively manage to break down the exponential many-body wall - and hence has the potential of resolving central problems in physics and chemistry.

%% file: figs/MPSRightIdentity.tex
\begin{tikzpicture}[baseline=(current bounding box.center)]
	\def\l{1.2};
    \def\h{1.3};
    
    \draw[obj1, rounded corners=3pt] (0,0) -- (-\l/2,0) -- (-\l/2,\h) -- (0,\h);
\end{tikzpicture}

%% file: figs/MPSLeftIdentity.tex
\begin{tikzpicture}[baseline=(current bounding box.center)]
	\def\l{1.2};
    \def\h{1.3};
    
    \draw[obj1, rounded corners=3pt] (0,0) -- (\l/2,0) -- (\l/2,\h) -- (0,\h);
\end{tikzpicture}

%% file: figs/MPSMixedgauge.tex
\begin{tikzpicture}[baseline=(current bounding box.center)]
	\def\l{1.2};
    \def\h{.65};

    \draw[obj1] (-\l/2,0) -- (9*\l/2,0);
    \foreach \i/\j in {0/L,1/L,3/R,4/R}{
        \draw[obj1] (\l*\i,0) --+ (0,\h);
        \node[mpstensor, inner sep=4.4pt] at (\l*\i,0) {};
        \node at (\l*\i,0) {$\sss A_\j$};
    }
    \node[matrix] at (2*\l,0) {$\sss C$};
\end{tikzpicture}

%% file: figs/MPSNorm.tex
\begin{tikzpicture}[baseline=(current bounding box.center)]
	\def\l{1.2};
    \def\h{1.3};
    
    \draw[obj1,rounded corners=3pt] (-\l/2,-\h) -- (-\l/2,0)-- (0,0) -- (0,-\h) -- cycle;

    \node[matrix, inner sep=4pt] at (-\l/2,-\h/2) {};
    \node[matrix, inner sep=4pt] at (0,-\h/2) {};
    \node at (-\l/2,-\h/2) {$\sss \Lambda$};
    \node at (0,-\h/2) {$\sss \rho$};
\end{tikzpicture}

%% file: _Lecture_3.tex
\section{Lecture III: Phases of matter}
\label{sec:PhasesOfMatter}
\subsection{Gapped phases of matter, entanglement \& symmetries}
A fundamental challenge in quantum many-body physics is to determine the \emph{phase diagram} of a given model. Two models belong to the same phase if there exists a a smooth path of gapped, bounded-strength, local Hamiltonians interpolating between them. From a pragmatic point of view, this is equivalent to the existence of a \emph{sublinear-depth quantum circuit} that maps their ground states into each other~\cite{Bravyi2006,Hastings2010}. Distinct phases are then separated by phase transitions, which in the case of a second-order phase transition -- as encountered above in \cref{sec:scaling} -- is typically described by a conformal field theory~\cite{DiFrancesco1997}. For a model with two parameters $h_{1,2}$ a fictitious phase diagram is sketched below.

The \emph{Landau paradigm} of phase transitions states that every phase of matter is in one-to-one correspondence with a symmetry breaking pattern~\cite{Landau1937}. The most famous (classical) example is of course the phase transition of water to ice, in which continuous Euclidean symmetry is broken to one of its 230 subgroups, also known as the \emph{space groups}. Moreover, the paradigm states that this breaking of symmetry can be diagnosed via a local order parameter that is charged under the symmetry group. Vice versa, \emph{symmetric} phases can be identified through \emph{disorder operators}, as argued by Kadanoff and Ceva~\cite{Kadanoff1970}. The Landau paradigm has since its conception been generalised in a number of ways.

For one, it has since been realised that after breaking its symmetry down to a subgroup, a quantum system can realise its residual symmetry in distinct ways on its entanglement degrees of freedom. These distinguishable entanglement patterns correspond to distinct phases if we require that the adiabatic paths between Hamiltonians commute with the symmetry. Such \emph{symmetry-protected topological} phases are for this reason naturally described and classified in terms of tensor networks~\cite{Chen2010,Chen2011,Schuch2011,GarreRubio2022}.

The goal of this lecture will be to sketch the classification of one-dimensional quantum phases of matter in the presence of a group symmetry using the framework of MPS. As remarked in \cref{sec:NormalObs}, symmetry breaking in MPS is manifested by the existence of distinct injective blocks in the ground state MPS. These are labelled by cosets in $G/H$ for a subgroup $H\subseteq G$ and are permuted under the group action. The challenge will thus be to classify the \emph{symmetric phases} realised on each of these injective blocks. To this end, we will first derive the \emph{fundamental theorem} of MPS which states that an MPS is invariant under a symmetry if and only if its local tensors transform trivially under that symmetry. This theorem exemplifies the mantra that tensor networks encode \emph{global} properties of states in the \emph{local} tensors. This will bring us in the realm of \emph{projective representations} whose corresponding cohomology classification provides a \emph{topological index} that classifies distinct symmetric phases in the sense that two injective MPSs belong to the same phase if and only if their topological index is the same. Notably, one-dimensional bosonic systems all belong to the same phase if no symmetry is imposed. For fermions the story is more intricate and non-trivial phases can exist due to the always present fermionic parity symmetry; this will be the topic of \cref{sec:Fermions}.

Let us also foreshadow our objectives of subsequent lectures. With the discovery of the fractional quantum Hall effect and other two-dimensional \emph{topological phases of matter}, it also became clear that even when a system does not exhibit any symmetry, it can still belong to a non-trivial quantum phase of matter. These phases of matter can be characterised by a ground state degeneracy which depends on the topology of the underlying space, gapped \emph{anyonic excitations} and in some cases chiral edge modes. They are also not detected by any local order parameter. As we will discuss in the next lecture, these phases owe their stability to the entanglement in these systems being \emph{long range}. In particular, in the case of \emph{non-chiral topological order}, the entanglement patterns are also responsible for the ground state degeneracy, as it has been appreciated that entanglement degrees of freedom exhibit string-like `symmetries' which are described by a \emph{fusion category}. We will encounter a particularly simple class of topological orders below in \cref{sec:QuantumDoubles}, and a microscopic construction of the generic non-chiral case in \cref{sec:SN}. The existence of these \emph{generalised symmetries} on the entanglement degrees of freedom implies in particular that the boundary Hamiltonians of these topological orders, describing the dynamics of their edge modes, inherit also these symmetries. As such, we are inclined to consider a Landau paradigm for these generalised symmetries as well, and this will be one of the topics of the last lecture.
\begin{equation}
    %
	\input{figs/PhaseDiagram.tex}%

\end{equation}
\subsection{Fundamental theorem of MPS}\label{sec:FundamentalTheorem}
As anticipated above, the \emph{fundamental theorem} of MPS \cite{Perez2008} is crucial in the classification of one-dimensional symmetric gapped phases of matter and many other applications of MPS. The incarnation of the theorem that we will apply is as follows. Two uniform \emph{injective} matrix product states represent the same state on any system with periodic boundary conditions, independent of the length, if and only if the MPS tensors are related by a non-singular gauge transformation on the virtual level:
\begin{equation}\label{eq:FundTh}
\boxed{
    \ket{\Psi[A]} \sim \ket{\Psi[B]} \iff A^i = e^{i\chi} XB^iX^{-1}, \quad\forall i
    }.
\end{equation}
Note the we assumed, without loss of generality, that both MPSs have the same bond dimension. When both MPSs are in canonical form, $X$ is unitary.

As an application of some of the key concepts introduced earlier in these lectures, we provide a proof here.

Note first that the $\impliedby$ direction is trivial since all appearances of $X$ and $X^{-1}$ cancel on the virtual level. To prove the $\implies$ direction, let us assume without loss of generality that $\ket{\Psi[A]}$ is in right canonical form, while $\ket{\Psi[B]}$ is in left canonical form. Consider now the following mixed transfer matrix:
\begin{equation}
    \mbb T_{A,B} = %
	\input{figs/MPSMixedTransfer.tex}%
 \,\, .
\end{equation}
Since $|\braket{\Psi[A]}{\Psi[B]}|=1$, the leading eigenvalue of the mixed transfer has unit norm~\cite{Cirac2016}. We will denote its corresponding right eigenvector by $X$, so that $\mbb T_{A,B}X = e^{i\phi} X$. Let us now consider following contraction
\begin{equation}
    e^{-i\phi}\,%
	\input{figs/FundTh_1.tex}%
\,.
\end{equation}
First note that on the one hand this expression is equal to $\Tr(X^\dagger X)$ via $\mbb T_{A,B}X = e^{i\phi} X$. On the other hand, interpreting this expression as the inner product of two vectors with respect to the diagonal cut and applying the Cauchy-Schwarz inequality, we obtain that
\begin{equation}
    \Tr(X^\dagger X) = e^{-i\phi} %
	\input{figs/FundTh_1.tex}%
 \leq \left|\, %
	\input{figs/FundTh_2.tex}%
 \,\right|^{1/2} \times \hspace{10pt} \left|\,%
	\input{figs/FundTh_3.tex}%
 \,\right|^{1/2} = \Tr(X^\dagger X),
\end{equation}
where we use that $A$ and $B$ are in right and left canonical form. So clearly the equality holds, which implies that $A^iX = e^{i\chi}XB^i$, or since $X$ is full rank thus \cref{eq:FundTh}.
\subsection{SPT classification of one-dimensional bosonic matter}\label{sec:SPT1d}
Let us consider a gapped Hamiltonian $H$ with a unique ground state, and assume it commutes with a symmetry $G$: $[U(g)^{\otimes L},H]=0$, for a representation $\{U(g),g\in G\}$. As $H$ is gapped and has a unique ground state, the ground state must inherit that symmetry. It is guaranteed that this ground state is faithfully represented by an injective MPS. That MPS $\ket{\Psi[A]}$ must therefore transform trivially under $G$, i.e.:
\begin{equation}
    U(g)^{\otimes N} \ket{\Psi[A]} \simeq \ket{\Psi[A]}, \quad \forall g \in G.
\end{equation}
A direct application of the fundamental theorem \cref{eq:FundTh} reveals that the MPS tensor itself has to transform trivially under the symmetry group, namely:
\begin{equation}\label{eq:gMPS}
    \forall g \in G,\, \exists X_g,\chi_g: \, \sum_j U(g)_{ij} A^j = e^{i\chi_g} X_gA^iX_g^\dagger
\end{equation}
where it should be noted that the gauge matrices and phases themselves depend on the group element $g$. Let us now consider acting successively with the symmetry operators labelled $g$ and $h$ on the state. We can do this in two different ways: either by acting with $U(gh)$ directly and using the fundamental theorem once or acting sequentially with $U(g)$ and $U(h)$ and applying the fundamental theorem twice. Equating these two results in:
\begin{equation}
    e^{i\chi_{gh}}X_{gh} A^i X_{gh}^\dagger = e^{i(\chi_g+\chi_h)}X_gX_h A^i X_h^\dagger X_g^\dagger,\quad\forall i,g,h.
\end{equation}
As $A^i$ is injective and hence spans the full matrix algebra, this is only possible if both $\chi_{gh}=\chi_g +\allowbreak \chi_h$ and $ X_gX_h = \psi(g,h) X_{gh}$ for $\rU(1)$ phases $\psi(g,h)$. The former means that $\{e^{i\chi_g},g\in G\}$ constitutes a one-dimensional representation of $G$. Note that after blocking a finite number of $M$ sites, we obtain an enlarged unit cell for which the phase $e^{iM\chi_g}=1$ disappears. Therefore we could say that this phase is \emph{unstable against blocking} and we will not take it further into account for phase classification. On the other hand, we have that the set of matrices $\{X_g,g\in G\}$ obeying
\begin{equation}\label{eq:ProjRep}
    X_gX_h = \psi(g,h) X_{gh}, \quad \forall g,h\in G,
\end{equation}
for some phases $\psi(g,h)\in\rU(1)$ forms a so-called \emph{$\psi$-projective representation} of $G$. Projective representations generalise ordinary \emph{linear} representations by allowing for the projective phase $\psi(g,h)$ in their multiplication. Due to the associativity of the matrix multiplication the phases $\psi(g,h)$ have to satisfy
\begin{equation}\label{eq:2cocycle}
    \psi(g,h)\psi(gh,k) = \psi(g,hk)\psi(h,k), \quad \forall g,h,k\in G,
\end{equation}
a consistency equation commonly known as the \emph{second cocycle equation}. This condition has trivial solutions of the form $\psi(g,h)=\frac{\phi_{gh}}{\phi_g \phi_h}$ for any function $\phi:G\rightarrow\rU(1)$. In the context of group cohomology such a trivial cocycle is called a \emph{coboundary}. In order to classify phases we will want to consider solutions of the cocycle equation \emph{modulo} coboundaries. Indeed, from the fundamental theorem it is clear that we can freely redefine $X_g\mapsto \phi_gX_g$ for any function $\phi$. To this end, note that both the cocycles and the coboundaries form abelian groups, which are denoted by $Z^2(G,\rU(1))$ and $B^2(G,\rU(1))$ respectively. Since clearly $B^2(G,\rU(1))$ is a normal subgroup of $Z^2(G,\rU(1))$, we can consider the quotient $H^2(G,\rU(1)):=Z^2(G,\rU(1))/B^2(G,\rU(1))$, the \emph{second cohomology group} of $G$ over $\rU(1)$.

Note that the cocycle equation \cref{eq:2cocycle} -- after taking the logarithm of both sides, and collecting the unknown phases $\psi(g,h)$ in a vector $\bm{\psi}$ -- can be written as a linear system of equations, satisfied modulo $2\pi$:
\begin{equation}
    \Omega \bm{\psi} = \vec{0}, \mod 2\pi.
\end{equation}
Computing the second cohomology group and corresponding representative cocycles can then be done by bringing $\Omega$ in \emph{Smith normal form}, a matrix decomposition of matrices with integer coefficients of the form $\Omega=P\Lambda R$, where both $P$ and $R$ are unimodular integer matrices, and $\Lambda$ is a diagonal matrix. A detailed explanation is relegated to \cref{sec:GroupCohomology}. The main observation is that the second cohomology group is always finite and of the form
\begin{equation}\label{eq:H2}
    H^2(G,\rU(1)) \simeq \mbb Z_{d_1}\times\mbb Z_{d_2}\times\dots\times\mbb Z_{d_r},
\end{equation}
where the $d_i$ are the diagonal entries of $\Lambda$. The group structure of $H^2(G,\rU(1))$ is physically realised by \emph{stacking}  two $G$-symmetric chains and considering the diagonal subgroup of $G\times G$ as the $G$ symmetry of the whole.

One can prove that if two MPSs transform under a $G$ symmetry with projective representations corresponding to distinct cohomology classes in $H^2(G,\rU(1))$, these MPSs must represent ground states in different phases~\cite{Schuch2011}. In other words, there is no symmetric adiabatic path one can take to transform one into the other without crossing a phase transition. In that case, the distinct cohomology classes are sometimes said to form a \emph{topological obstruction} to do so, while a representative cocycle is sometimes called a \emph{topological index} characterising that phase. If we were to allow paths that break the symmetry, then all MPSs would again belong to the same trivial phase. In particular, the number of symmetric phases is always finite due to \cref{eq:H2}. The proof is beyond the scope of these lecture notes, but let us remark that the argument relies heavily on the parent Hamiltonian construction of MPS (see \cref{sec:NormalObs}).

\bigskip\noindent An important property of any projective representation is that its matrix dimension is necessarily strictly bigger than one, when it corresponds to a non-trivial cohomology class. As an example, consider the smallest group with a non-trivial second cohomology group, namely $\mbb Z_2\times \mbb Z_2$, $H^2(\mbb Z_2\times \mbb Z_2,\rU(1))\simeq \mbb Z_2$. If we denote its group elements by tuples $(a,b), a,b=0,1$, a non-trivial projective representation is given by the Pauli matrices as follows: $X_{(a,b)}=(\sigma^X)^a(\sigma^Z)^b$, and this representation cannot be further reduced. This fact has an important consequence for the entanglement of an MPS in a non-trivial SPT phase, and also guarantees the existence of gapless symmetry-protected edge modes~\cite{Pollmann2009}.

\begin{figure}[htb]
    \centering
    \includegraphics[width=0.7\linewidth]{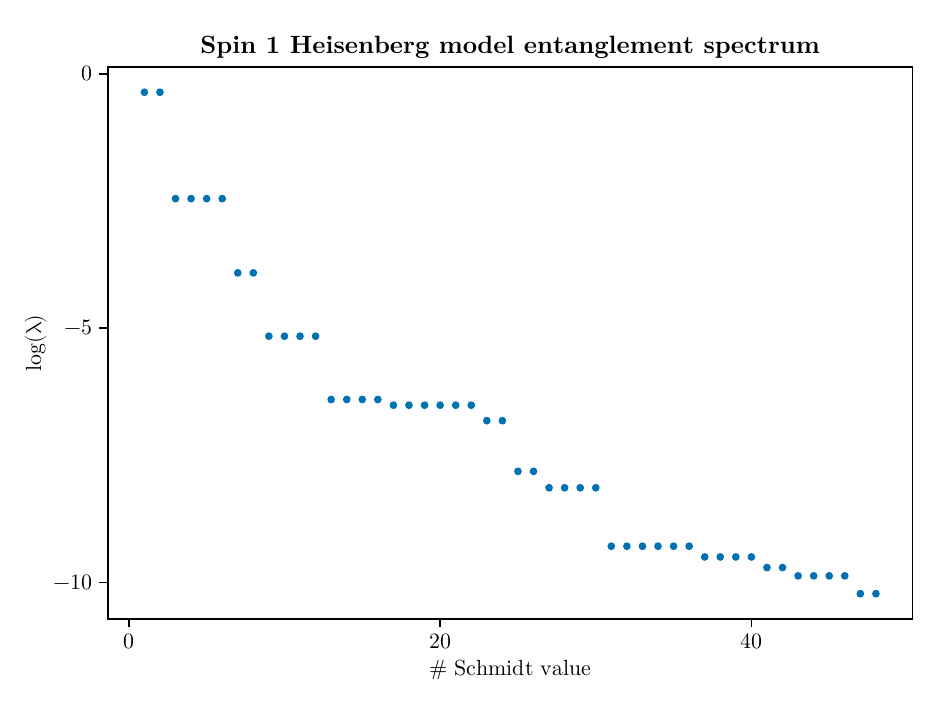}
    \caption{The entanglement spectrum of the spin 1 Heisenberg model ground state in the thermodynamic limit with bond dimension 48. Created with MPSKit.jl~\cite{VanDamme}.}
    \label{fig:HeisenbergEntanglement}
\end{figure}

Indeed, it turns out that the entanglement spectrum (see \cref{sec:NormalObs}) is characterised by a degeneracy which is at least the dimension of the smallest irreducible projective representation in that class. This is beautifully illustrated in the spin 1 antiferromagnetic Heisenberg model. In \cref{fig:HeisenbergEntanglement} the first 48 Schmidt values are plotted, and the spectrum organises itself visibly in multiplets which are at least twofold degenerate. In this case, the protecting symmetry is ${\rm SO}(3)$ whose second cohomology group indeed has a non-trivial element, $H^2({\rm SO}(3),\rU(1))\simeq \mbb Z_2$. All irreps of ${\rm SO}(3)$ have an odd dimension, so these degeneracies clearly originate from the half-integer irreps of ${\rm SU}(2)$, which indeed form projective representations of ${\rm SO}(3)$. In fact, the phase is also protected by a $\mbb Z_2\times \mbb Z_2$ subgroup of ${\rm SO}(3)$, so that we can even perturb the Hamiltonian with terms which break part of the ${\rm SO}(3)$ symmetry as long as they still commute with $\mbb Z_2\times \mbb Z_2$, without leaving the SPT phase.

As anticipated in \cref{sec:Examples}, another hallmark of SPT phases is the presence of symmetry-protected gapless \emph{edge modes}, when they are considered on open chains. A feature of matrix product states (and their higher-dimensional generalisations, see \cref{sec:PEPS}) is that a tensor product basis for these edge modes is provided by the entanglement degrees of freedom on the boundary.  Indeed, suppose we have a uniform MPS $\ket{\Psi[A]}$ of bond dimension $\chi$. We can consider then following $\chi^2$ states on an open chain of $N$ sites:
\begin{equation}
    \ket{\Psi[A]} = \sum_{\{i_j\}} (A^{i_1}A^{i_2}\cdots A^{i_N})_{\alpha\beta} \ \ket{i_1,i_2,\dots,i_N}.
\end{equation}
All states $\ket{\Psi[A],\alpha,\beta}$, $\forall \alpha,\beta=1,\dots,\chi$ have the same energy for the parent Hamiltonian, so that seemingly we have $\chi^2$ degenerate edge modes. However, by perturbing the Hamiltonian, some of these edge modes can be gapped out. Nevertheless, as long as we restrict to symmetry-respecting perturbations, the edge modes will transform under the projective representation on the virtual level according to:
\begin{equation}
    U(g)^{\otimes N}\ket{\Psi[A],\alpha,\beta} = \sum_{\alpha',\beta'} (X_g)_{\alpha\alpha'} (\bar X_g)_{\beta\beta'}\,\,\ket{\Psi[A],\alpha',\beta'},
\end{equation}
and hence, at least a two-fold degeneracy of the edge modes persists.

\bigskip\noindent This concludes the classification of \emph{bosonic} $G$-symmetric gapped one-dimensional phases. Provided a symmetry group $G$, gapped phases are in one-to-one correspondence with a subgroup $H\subseteq G$ which represents the \emph{unbroken} symmetry in the ground state subspace, and a cohomology class $[\psi]\in H^2(H,\rU(1))$ representing the SPT phase of the surviving $H$ symmetry.

As an aside we note here that such pairs $(H,[\psi])$ also characterise the (indecomposable) \emph{module categories} over $\Vect_G$, the \emph{fusion category} of $G$-graded vector spaces (see also \cref{sec:cat}). There is thus a one-to-one correspondence between $\Vect_G$-module categories and gapped $G$-symmetric phases. Of course this observation is completely superfluous for the simple case of group symmetries, but in the last lecture we will argue that for generic fusion category symmetries this correspondence continues to hold.
\subsection{Spacetime symmetries}
The above discussion was focused on internal symmetries of the Hamiltonian, but applies with some slight modifications to spacetime symmetries as well. When dealing with these symmetries, one has to address two issues which are not present in the internal case. First, note that some spacetime symmetries such as translation symmetry and spatial reflection act in a non-local way. On the other hand, \emph{orientation-reversing} symmetries such as time-reversal or reflection symmetry act in an antiunitary way and thus depend on the chosen basis. The latter has strong consequences for the cohomological classification of SPT phases, as we will see below. Nevertheless, MPS provide a suitable setting in which the classification of symmetric phases protected by spatial symmetries can be derived explicitly~\cite{Pollmann2009,Vancraeynest2022}.

Let us illustrate how a similar reasoning as above leads to the classification of phases protected by time-reversal symmetry. In general, time-reversal symmetry acts as $UK$ where $U$ is a unitary operation, $(\sigma^Y)^{\otimes N}$ in the case of spin 1/2's as demonstrated by Wigner~\cite{Wigner1960}, and $K$ is complex conjugation. We assume here for simplicity that $U=\mbb 1$. Acting with complex conjugation on an injective MPS $\ket{\Psi[A]}$ and applying the fundamental theorem results in
\begin{equation}
    \bar A^i = e^{i\chi}QA^iQ^{-1}.
\end{equation}
Applying time-reversal symmetry then a second time and exploiting injectivity of the MPS shows that:
\begin{equation}\label{eq:TRProjective}
    Q\bar Q = \pm \mathbb 1.
\end{equation}
We thus obtain a non-trivial topological index corresponding to the sign, giving rise to a $\mbb Z_2$ classification of time-reversal SPT phases. As an illustration consider the AKLT state~\cref{eq:AKLTState}. It is easily checked that in that case $Q=\sigma^Y$, $Q\bar{Q}=-\mathbb 1$ so that the AKLT state thus belongs to a non-trivial SPT phase w.r.t. time-reversal symmetry.

Note that the $Q$ in \cref{eq:TRProjective} cannot be interpreted as a non-trivial projective representation of $\mbb Z_2$ due to the complex conjugation on $Q$. Consequently, the phase $\pm 1$ in \cref{eq:TRProjective} also cannot be interpreted as a non-trivial cocycle for $H^2(\mbb Z_2,\rU(1))$, which is in fact trivial. We therefore need a generalised notion of projective representations and cocycles to account for symmetry groups where some group elements act in an antiunitary fashion. It is easy to verify that when a generic symmetry group $G$ has antiunitary elements, the gauge matrices $X_g$ multiply according to
\begin{equation}
    X_g\beta_g(X_h) = \psi(g,h) X_{gh},
\end{equation}
where $\beta_g(\bullet)$ acts as complex conjugation when $g$ is an antiunitary element and as the identity otherwise. Clearly, this expression is equivalent to \cref{eq:ProjRep} when all group elements act unitarily and to \cref{eq:TRProjective} when the symmetry group is $\mbb Z_2$ time-reversal. It is straightforward to verify as well that the consistency condition on $\psi(g,h)$ then reads
\begin{equation}
    \psi(g,h)\psi(gh,k) = \psi(g,hk)\beta_g(\psi(h,k)), \quad \forall g,h,k\in G.
\end{equation}
Accordingly, also the notion of coboundaries has to be slightly adapted. Putting things together, this means that in case of antiunitary symmetries, the cohomology classification of SPT phases has to be adapted in a way so as to account for the non-trivial group action $\beta_g(\bullet)$. The corresponding cohomology group is then denoted $H^2(G,\rU(1)^\beta)$, and is generically different from $H^2(G,\rU(1))$. Importantly, this group can be computed with the same method as outlined above and detailed in \cref{sec:GroupCohomology}. In correspondence with the example of the AKLT state, we have indeed that $H^2(\mbb Z_2,\rU(1)^\beta)=\mbb Z_2$.
\subsection{Fermionic MPS \& symmetric phases}\label{sec:Fermions}
Up until now we have focused solely on \emph{bosonic} matrix product states. The goal of this section is to sketch the extension of the MPS framework to \emph{fermionic systems} and highlight some of the technical complications that arise, thereby closely following ref.~\cite{Bultinck2016}, see also ref.~\cite{Kapustin2016,Mortier2024}.

\bigskip\noindent Crucial is the notion of a \emph{super vector space}, which is a vector space that admits a direct sum decomposition of the form
\begin{equation}
    V = V^0 \oplus V^1,
\end{equation}
where vectors in $V^0$, respectively $V^1$, are said to possess a \emph{fermion even}, respectively \emph{odd}, parity. Without loss of generality we assume that basis vectors $\ket{i}$ have definite parity, which we will denote by $|i| = 0,1$. A super vector space is moreover endowed with a \emph{graded tensor product} behaving as follows under a re-arrangement of the tensor product:
\begin{equation}
    \ket{i_1} \otimes_\mathfrak{g} \ket{i_2} \mapsto (-1)^{|i_1||i_2|} \ket{i_2} \otimes_\mathfrak{g} \ket{i_1}.
\end{equation}
This property is the mathematical manifestation of fermionic particle exchange statistics. The \emph{supertrace} is then defined as
\begin{equation}
    \mc C (\ket{i}\otimes_\mathfrak g \bra{j}) = (-1)^{|i||j|} \delta_{i,j}.
\end{equation}

A uniform fermionic MPS (fMPS) is then constructed from an \emph{even-parity} tensor
\begin{equation}
    A = \sum_{i,\alpha,\beta}A_{\alpha\beta}^i \ket{\alpha} \otimes_\mathfrak{g} \ket{i} \otimes_\mathfrak{g} \bra{\beta},
\end{equation}
i.e. $|i| + |\alpha| + |\beta| = 0 \mod 2$ for all indices. Depending on whether the resulting state has even or odd fermionic parity the construction is slightly different.
\paragraph{Even-parity states}
An even-parity state is constructed from the tensor $A$ as follows
\begin{equation}\label{eq:EvenfMPS}
    \ket{\Psi}_e = \mc C_v\big(A\otimes_\mathfrak g A \otimes_\mathfrak g\dots \otimes_\mathfrak g A\big).
\end{equation}
In this expression, $\mc C_v$ stands for the supertrace over the virtual indices. We can now rewrite this fMPS in a more explicit form. Note that the entries of a generic even-parity tensor are of the form
\begin{equation}
    A^i =
    \begin{pmatrix}
        B^i & 0 \\
        0 & C^i
    \end{pmatrix}, \,\, |i| = 0, \quad
    A^i =
    \begin{pmatrix}
        0 & D^i \\
        F^i & 0
    \end{pmatrix}, \,\, |i| = 1.
\end{equation}
Defining the \emph{parity matrix}
\begin{equation}
    \mc P = \begin{pmatrix}\mbb 1 & 0\\ 0 & -\mbb 1\end{pmatrix},
\end{equation}
and carefully working out the expression \cref{eq:EvenfMPS} yields then:
\begin{equation}
    \ket{\Psi[A]}_e = \sum_{\{i_j\}} {\rm Tr}\big(\mc PA^{i_1}A^{i_2}\dots A^{i_N}\big) \ket{i_1,i_2,\dots,i_N}.
\end{equation}
\paragraph{Odd-parity states}
The construction of odd-parity states proceeds analogously to the construction of even states, but requires the insertion of an additional matrix in the supertrace:
\begin{equation}
    \ket{\Psi}_o = \mc C_v\big(Y\otimes_\mathfrak gA\otimes_\mathfrak g A \otimes_\mathfrak g\dots \otimes_\mathfrak g A\big).
\end{equation}
The requirement to introduce the matrix $Y$ ultimately stems from the fact that accessing odd-parity states necessitates twisted boundary conditions. We will encounter similar considerations in lecture~\ref{sec:GenSym}.

Similar manipulations as in the previous case allow us to rewrite the odd-parity fMPS as
\begin{equation}
    \ket{\Psi[A]}_o = \sum_{\{i_j\}} {\rm Tr}\big(YA^{i_1}A^{i_2}\dots A^{i_N}\big) \ket{i_1,i_2,\dots,i_N}.
\end{equation}
Imposing then translation invariance reveals that $Y$ can be chosen to be of the form
\begin{equation}\label{eq:fMPSY}
    Y =
    \begin{pmatrix}
        0 & -\mbb 1 \\ \mbb 1 & 0
    \end{pmatrix},
\end{equation}
and in this basis the MPS matrices read:
\begin{equation}
    A^i =
    \begin{pmatrix}
        B^i & 0 \\
        0 & B^i
    \end{pmatrix}, \,\, |i| = 0, \quad
    A^i =
    \begin{pmatrix}
        0 & B^i \\
        -B^i & 0
    \end{pmatrix}, \,\, |i| = 1.
\end{equation}
Contrary to the even case, this fMPS is specified by one block $B^i$.

Note that since the odd fMPS matrices commute with $Y$, the MPS tensors have a common invariant subspace. Nevertheless, contrary to the bosonic case, the odd-parity states $\ket{\Psi[A]}_o$ define an irreducible matrix product state when the $B^i$ span the full $\mbb C^{\chi/2\times \chi/2}$ matrix algebra. We refer to appendix D of \cite{Piroli2020} for a detailed technical discussion and proofs on canonical forms of fermionic graded MPS. Physically, the underlying reason can be traced back to the superselection rule for fermionic parity: there exist no superpositions of odd and even-parity states.

A notable example of an odd fMPS state is the ground state of the Kitaev chain of spinless fermions whose Hamiltonian is a simple sum of pairing terms~\cite{Kitaev2000}:
\begin{equation}
    H = -i\sum_{\msf i=1}^{N-1} \gamma_{2\msf i}\gamma_{2\msf i+1},
\end{equation}
for Majorana modes $\gamma_\msf i=\gamma^\dagger_\msf i$, $\{\gamma_\msf i, \gamma_\msf j\}=2\delta_{\msf i,\msf j}$. The translational invariant fMPS ground state is found to have odd parity and its MPS matrices read
\begin{equation}\label{eq:KitaevTensor}
    A^0 =
    \begin{pmatrix}
        1 & 0 \\
        0 & 1
    \end{pmatrix}, \quad
    A^1 =
    \begin{pmatrix}
        0 & 1 \\
        -1 & 0
    \end{pmatrix}.
\end{equation}
This model is noteworthy for exhibiting Majorana edge modes. Indeed, on open boundary conditions the state takes the form
\begin{equation}
    \sum_{\{i_j\}} \big(A^{i_1}A^{i_2}\dots A^{i_N}\big)_{\alpha\beta} \ket{i_1,i_2,\dots,i_N}.
\end{equation}
It is then straightforward to check that due to the structure of the tensors, choosing the boundary conditions $\alpha=\beta=0$ or $\alpha=\beta=1$ yields one and the same ground state. Similarly, both boundary conditions $\alpha=0,\beta=1$ and $\alpha=1,\beta=0$ yield a second ground state, resulting eventually in a two-fold ground state degeneracy, which cannot be lifted by adding an extra local interaction to the Hamiltonian.

\bigskip\noindent Similar to the bosonic case, one can extract from fermionic matrix product states certain discrete topological invariants when global internal or spacetime symmetries are imposed. An illustrious example is the case of interacting time-reversal symmetric fermionic SPT phases. In that case there are three distinct topological indices valued in $\{0,1\}$, one of which is related to whether the fMPS has Majorana edge modes, and two of which are related to the local transformation properties of the fMPS tensors under time-reversal. As carefully demonstrated, in ref.~\cite{Bultinck2016} it can be shown that under stacking of the corresponding fMPSs, the $\mbb Z_8$ classification of Fidkowski and Kitaev is recovered~\cite{Fidkowski2010}.
\subsection{Synopsis of lecture III} Lecture III explored the manifold of MPS when additional symmetries are imposed on the tensors. The main workhorse of symmetric tensor networks is the \emph{fundamental theorem of MPS}, which states that any \emph{global} symmetry must be reflected in the symmetries of the \emph{local} MPS tensors. Interestingly, those \emph{virtual} symmetries can be represented \emph{projectively}. Any symmetry-preserving adiabatic path between symmetric MPSs which transform according to different projective representations is guaranteed to have a gapless point. As a consequence, all bosonic gapped phases of matter of one-dimensional quantum spin chains with a Hamiltonian that is $G$-symmetric are in one to one correspondence with all possible subgroups $H\subseteq G$ and its possible projective representations labelled by the second group cohomology $H^2(H,\rU(1))$. The latter classify the so-called \emph{SPT phases}. If no symmetry is considered, then all MPSs can adiabatically converted into each other, and therefore the phases are only protected when the symmetry is imposed. We showed that this classification can be generalised to the case of spacetime symmetry groups. Interestingly, when superselection rules are considered such as the parity superselection rule for fermions, genuine topological phases of matter emerge (i.e. a phase that does not need a symmetry to be protected).

%% file: figs/MPSMixedTransfer.tex
\begin{tikzpicture}[baseline=(current bounding box.center)]
	\def\l{1.8};
    \def\h{1.3};

    \draw[obj1] (-\l/2,0) -- (\l/2,0);
    \draw[obj1] (-\l/2,\h) -- (\l/2,\h);
    
    \draw[obj1] (0,0) -- (0,\h);
    
    \node[mpstensor, inner sep=4pt] at (0,0) {};
    \node[mpstensor, inner sep=4pt] at (0,\h) {};
    \node at (0,0) {$\sss A$};
    \node at (0,\h) {$\sss \bar B$};
\end{tikzpicture}

%% file: _Lecture_4.tex
\section{Lecture IV: MPO algebras: from the algebraic Bethe ansatz to G-injective PEPS}
Let us now turn our attention towards integrable models, one of the central topics of this school. It turns out that tensor networks provide a very concise way of rephrasing the \emph{algebraic Bethe ansatz} in terms of matrix product states and operators. In particular, we will see that the \emph{Yang-Baxter equation}, which lies at the heart of its integrability, can be understood as a consequence of the fundamental theorem (\cref{sec:FundamentalTheorem}). Although tensor network techniques are not teaching us anything fundamentally new about the Bethe ansatz, the insights of this lecture will be instrumental in the MPO representation theory of \emph{generalised} or \emph{categorical symmetries} described by \emph{fusion categories}, which is the topic of the final lecture.

After the discussion on the algebraic Bethe ansatz, we will focus in \cref{sec:PEPS} on the two-dimensional generalisation of MPS, namely \emph{projected entangled-pair states}. We will conjecture a generalisation of the fundamental theorem, and discuss the role of MPO symmetries in the tensor network representation of two-dimensional SPT and topological phases, as well as their edge modes.
\subsection{Magnon states}\label{sec:magnon}
Tensor networks can be used as ansatze for analytic solutions of Hamiltonians, as was exemplified for the AKLT mode \cref{eq:AKLTState}. The example of interest in this lecture however, is a famous family of exactly solvable models, called \emph{integrable systems}. An integrable quantum system is roughly speaking a system with an extensive number of conserved charges -- local operators commuting with the Hamiltonian -- to which they owe their exact solvability. Integrable systems can be solved with the ansatz wave function called the \emph{Bethe ansatz} (see e.g. \cite{Gaudin2014}). There are two formulations of this approach, called the \emph{coordinate Bethe ansatz} and \emph{algebraic Bethe ansatz}, which produce the same set of solutions. Both formulations can be written down using tensor networks.

For the sake of concreteness, we consider the spin $1/2$ XXZ Hamiltonian 
\begin{equation}
    H_{XXZ}=\sum_{\msf i=}^L \sigma^X_{\msf i}\sigma^X_{\msf i+1}+\sigma^Y_\msf i\sigma^Y_{\msf i+1}+\Delta \sigma^Z_\msf i\sigma^Z_{\msf i+1}
\end{equation}
on periodic boundary conditions, i.e. $L+1\equiv 1$. The \emph{magnon states} used as a basis when solving the Bethe equations can readily be written down as an MPS. For instance, the \emph{periodic} boundary conditions 1-magnon state can be written down as an \emph{open} boundary conditions MPS (see \cref{eq:OpenMPS}) with left boundary vector $\vec v_L^\dagger = \bra{0}$ and right boundary vector $\vec v_R = \ket{1}$, and MPS tensor:
\begin{equation}
\begin{split}
A^0(k) = 
\begin{pmatrix}
    e^{ik} & 0 \\
    0 & 1
\end{pmatrix}, \quad
A^1(k) = \begin{pmatrix}
    0 & e^{ik} \\
    0 & 0
\end{pmatrix} \\
\implies \ket{\Psi_1[A]} = \sum_{\msf n=1}^L e^{ik\msf n} \ket{0_0 ...0_{\msf n-1} 1_\msf n 0_{\msf n+1}...0_{L}}
\end{split}.
\label{AAAA}
\end{equation}
Note that this is precisely how excited states were constructed as boosted tangent vectors on the manifold of MPS in \cref{sec:Excitations}. The general n-magnon state can be can be found in refs.~\cite{Alcaraz2004,Murg2012,Haegeman2016}, and involves an MPS with a bond dimension that scales exponentially in the number of particles in the system.
\subsection{The Yang-Baxter equation as the fundamental theorem of MPS}\label{sec:YangBaxterFromMPO}
There is however a completely different way of deriving the exact eigenstates in terms of the algebraic Bethe ansatz (ABA). Before the formulation of both approaches in terms of tensor networks, it was not clear how those two approaches were related to each other; it is now clear that both approaches are related by a non-trivial gauge transformation \cite{Katsura2010}, as has to be because of the fundamental theorem of MPS. In the algebraic Bethe ansatz, all eigenstates of the Hamiltonian can be constructed using the multiplication of MPOs that depend on a continuous \emph{spectral parameter} $\mu_i$:
\begin{equation}\label{eq:BetheAnsatzMPS}
    %
	\input{figs/YB_ABA.tex}%
.
\end{equation}
Here, all MPO tensors are of the form 
\begin{equation}
    A_{\alpha\beta}^{ij}(\mu) =\mu\delta_{\alpha\beta}\delta_{ij}+\delta_{\alpha i}\delta_{\beta j}= \mu %
	\input{figs/YB_A_1.tex}%
 + %
	\input{figs/YB_A_2.tex}%
.
    \label{AA}
\end{equation}
It can easily be checked that every MPO adds one magnon to the system, as the left boundary is $\langle 0|$ and the right boundary $|1\rangle$. The coefficients $\mu_i$ are functions of the magnons' momenta, and can be determined by solving the so-called \emph{Bethe equations}. A crucial ingredient of the ABA is the fact that all those MPOs commute with each other. 

The underlying logic is as follows. We start from a transfer matrix of a two-dimensional statistical mechanical mechanical model that can be written in the form of an injective MPO $T(\mu)$ determined by the same tensor $A(\mu)$ as in equation~\ref{AA}, but this time with periodic boundary conditions. It turns out that this system will be integrable if and only if all those MPOs commute with each other independent of their spectral parameter:
\begin{equation}
        \forall \lambda, \mu: [T(\lambda), T(\mu)] = 0.
\end{equation}
If this is the case, we can invoke the fundamental theorem of MPS \cref{eq:FundTh} to prove the existence of a gauge transform $R$ that transforms those products into each other~\cite{Haegeman2016}. Indeed, note that we can merge together two legs of an MPO to form a single leg, so that the fundamental theorem of MPS can be used for rank-4 MPOs. Concretely, there is an intertwiner tensor $R(\lambda, \mu)$ obeying the equation
\begin{equation}\label{eq:YB_intertwiner}
    %
	\input{figs/YB_intertwiner_1.tex}%
 \,=\, %
	\input{figs/YB_intertwiner_2.tex}%
\,\,.
\end{equation}
This $R$-tensor is a local implementation of the global symmetries given by the commuting MPOs. 

Now let us take three copies of the $R$-tensor. Due to the associativity of the MPO algebra, these $R$-tensors can be fused together in two different ways, giving rise to the following equation:
\begin{equation}
    %
	\input{figs/YB_1.tex}%
 \,=\, %
	\input{figs/YB_2.tex}%
\,\,.
\end{equation}
This is precisely the Yang–Baxter equation -- hence the fundamental theorem of MPS  reproduces this  result in a very concise way. Given any integrable transfer matrix, the corresponding intertwiners have therefore to satisfy the Yang-Baxter equations. It can be checked that the tensor in \cref{AA} is precisely one of those solutions.  A constructive way of finding integrable models is henceforth to turn the tables around: start with finding a solution of the Yang-Baxter equations (which is typically very hard!), leading to explicit representations of the tensors $R(\lambda,\mu)$. As a next step, solve for the transfer matrix $T$ as a solution of \cref{eq:YB_intertwiner}. What is  surprising is the fact that this equation is equivalent to the one of the Yang-Baxter equation by identifying the  MPO tensors to be $A_{\alpha \beta}^{ij}(\mu) := R_{j \beta}^{i \alpha}(\mu, 0)$. This so-called \emph{fundamental representation} of the Yang-Baxter algebra hence allows to construct a one-parameter family of commuting transfer matrices. Note that this is not the only family of MPOs solving \cref{eq:YB_intertwiner}. One can construct different representations of the same algebra -- later on, we will encounter an analogous situation where we will construct different representations of the same symmetric system.  As a last step, it is now possible to construct the magnon MPOs of \cref{eq:BetheAnsatzMPS} with the same tensors $A$, but with different boundary conditions. 

The key principle and advantage of the tensor network formulation is thus as follows: \emph{global} conservation laws and commutation relations can be written down as equations involving \emph{local} degrees of freedom, i.e. MPS and MPO tensors, which are usually easier to solve.

Given the MPO formulation, it is also possible to recover the integrable Heisenberg Hamiltonian. The transfer matrix of the 6-vertex model with all weights equal to each other, the corresponding integrable statistical mechanical model, has exactly the form of~\cref{AA}. By construction, all MPOs $T(\mu)$ with those tensors $A(\mu)$ will commute with each other, so the derivative of this MPO is defined univocally. The Heisenberg Hamiltonian is indeed the logarithmic derivative of this transfer matrix; we leave the proof as a (very interesting) exercise (see \cite{Murg2012} for an elaborate discussion):
\begin{equation}
    H = T^{-1}(0) \frac{{\rm d}T(\mu)}{{\rm d}\mu}\vert_{\mu=0} = \sum_{\msf i}\quad%
	\input{figs/YB_Ham.tex}%
\,.
\end{equation}

In conclusion, both Bethe ansatz formulations (coordinate and algebraic) have a natural representation in terms of tensor networks. Tensor networks and Bethe ansatz clearly  share the same DNA, but tensor networks turn out to be more generally applicable -- they work equally fine for non-integrable systems and higher dimensions. 
\subsection{Projected entangled-pair states}\label{sec:PEPS}
The natural generalisation of matrix product states to two dimensions are the so-called \emph{projected entangled-pair states} (PEPS)~\cite{Verstraete2004b}. Similar to the one-dimensional case, a translation invariant PEPS is constructed from a single tensor, which in the case of the square lattice has one physical and four virtual indices:
\begin{equation}
    A = \sum_{\substack{i,\alpha,\beta\\\gamma,\delta}} A^i_{\alpha\beta\gamma\delta} \ket{i}\otimes\bra{\alpha}\otimes\bra{\beta}\otimes\bra{\gamma}\otimes\bra{\delta} \in \mbb C^{d\chi^4}.
\end{equation}
Such a tensor is then depicted by
\begin{equation}
    %
	\input{figs/PEPSTensor.tex}%
 = \ A^i_{\alpha\beta\gamma\delta}, \qquad \substack{i=1,2,\dots,d\\ \alpha,\beta,\gamma,\delta=1,2,\dots,\chi},
\end{equation}
so that the full PEPS wave function can be represented by
\begin{equation}
    \ket{\Psi[A]} = %
	\input{figs/PEPS.tex}%
\,\, .
\end{equation}
Note that PEPS naturally satisfy an area law since the entanglement entropy of a contiguous region of spins scales linearly with the number of bonds on the boundary of the region, and its perimeter. Therefore PEPS are expected to form a good variational class for low-energy states of local gapped Hamiltonians, which have since long been conjectured to satisfy an area law.  Conversely, as for MPS, it can be shown that every PEPS is the exact ground state of a local frustration free gapped parent Hamiltonian. It is also obvious how to generalise this construction to open boundary conditions, arbitrary lattices or to higher dimensions. Similarly, one can generalise MPOs to the two-dimensional setting. Such operators are referred to as \emph{projected entangled-pair operators} (PEPOs).

Considerable effort has been made to generalise DMRG and VUMPS algorithms to the case of PEPS. Although it is in principle straightforward to generalise the manifold picture to the case of PEPS, the actual computational scaling with the bond dimension is much worse than in the 1D case. Nevertheless those PEPS algorithms are  providing state of the art variational results for strongly correlated lattice systems in 2D such as for the Hubbard model, and it is a very active research field to come up with better algorithms. We refer to refs.~\cite{Verstraete2008,Vanderstraeten2016,Corboz2016,Liao2019,Vanderstraeten2022} for state of the art results. In these lectures, we will focus on the theory of PEPS. 

Contrary to the case of matrix product states, there exists no completely general fundamental theorem yet that provides necessary and sufficient conditions for two distinct PEPS tensors $A$ and $B$ to generate the same state for all system sizes. However, it is clear that a sufficient condition for this to hold is the existence of a matrix product operator that \emph{intertwines} the PEPS tensors according to
\begin{equation}
    %
	\input{figs/PEPSPullingThrough_1.tex}%
 \,\,=\,\,
    %
	\input{figs/PEPSPullingThrough_2.tex}%
\,\,,
\end{equation}
and all other ways of pulling the MPO through, such as:
\begin{equation}
    %
	\input{figs/PEPSPullingThrough_3.tex}%
 \,\,=\,\,
    %
	\input{figs/PEPSPullingThrough_4.tex}%
\,\,.
\end{equation}
In the absence of a proof, we call this the fundamental \emph{conjecture} of PEPS. Assuming periodic boundary conditions, the symmetry action is then implemented by starting from the PEPS $\ket{\Psi[A]}$, nucleating a small bubble of the intertwining MPO, pulling it through the entire lattice and annihilating it with itself so as to end up with the PEPS $\ket{\Psi[B]}$. As we will see in the remainder of these lectures, large classes of PEPS implement their symmetries exactly according to this prescription. In some cases, this MPO is just a tensor product (bond dimension $1$), so that the virtual symmetry simply acts as a tensor product on the virtual legs.

In the following sections, we will discuss some particularly interesting PEPS wave functions which represent topological phases of matter. The fact that topological wave functions can easily be constructed using the tensor network formalism is clearly one of the main reasons that they are studied so extensively. Additionally, those tensor network representations identify the long-range entanglement with (MPO) symmetries acting on the virtual degrees of freedom. This is the reason that this fundamental conjecture is so relevant: topological order is characterised by non-trivial MPO symmetries.
\subsection{Two-dimensional SPTs: a case study}\label{sec:SPT2d}
A paradigmatic example of a PEPS with non-trivial MPO symmetry is the so-called \emph{CZX model} introduced by Chen, Liu and Wen in ref.~\cite{Chen2011} (see~\cite{Buerschaper2013,Williamson2016,Molnar2017} for its tensor network representation). Their model is an example of a non-trivial SPT protected by a $\mbb Z_2$ symmetry which acts in an on-site manner $U_g^{\otimes N}$ on the physical level of the PEPS. At the virtual level of the PEPS tensors, this symmetry is implemented via a bond dimension 2 MPO as:
\begin{equation}\label{eq:PEPS_CZX}
    %
	\input{figs/PEPS_CZX_1.tex}%
 = 
    %
	\input{figs/PEPS_CZX_2.tex}%
\,\, ,
\end{equation}
for $g\in\{0,1\}$. The MPOs indexed by $g$ form a non-local representation of $\mbb Z_2$, which graphically can be represented by
\begin{equation}\label{eq:GroupMPO}
    %
	\input{figs/MPOGroupStacked.tex}%
 =\,
    %
	\input{figs/MPOGroupgh.tex}%
\,\, .
\end{equation}
The multiplication \cref{eq:GroupMPO} can be implemented locally by means of an intertwining tensor satisfying\footnote{The MPO for $g=0$ is not injective, but can not be decomposed in injective blocks. This can be explained via the underlying algebraic structure, which is not semisimple~\cite{Liu2025,FloridoLlinas2025}. As such, the intertwiners cannot be found directly by invoking the fundamental theorem.}~\cite{Chen2011,Williamson2016,Liu2025}:
\begin{equation}\label{eq:GroupIntertwiner}
    %
	\input{figs/GroupMPOIntertwiner_1.tex}%
 = %
	\input{figs/GroupMPOIntertwiner_2.tex}%
\,\, .
\end{equation}
Recall that in \cref{sec:SPT1d} we argued that in the case of one-dimensional SPT phases, there exists a topological index that takes values in the second cohomology group $H^2(G,\rU(1))$. This index is fully captured by the multiplication rules of the virtual matrices which form a projective representation of the symmetry. It is now natural to wonder what the corresponding topological index for two-dimensional SPT phases is and how it is encoded in the MPO algebra. As it turns out, if you consider the multiplication of three MPOs representing the $\mbb Z_2$ symmetry, then the fusion tensors defined above can be contracted in two different ways. These two ways have to be compatible, so that there must exist a phase $\omega$ depending on three group elements satisfying
\begin{equation}\label{eq:GroupMPOFusion}
    %
	\input{figs/GroupMPOPentagon_1.tex}%

    = \omega(g,h,k) \,\,
    %
	\input{figs/GroupMPOPentagon_2.tex}%
\,\, .
\end{equation}
Associativity of the matrix multiplication reveals that $\omega$ has to satisfy
\begin{equation}
    \omega(g,h,k)\omega(g,hk,l)\omega(h,k,l) = \omega(gh,k,l)\omega(g,h,kl).
\end{equation}
This condition is known as the \emph{3-cocycle equation} (see also \cref{sec:GroupCohomology}). Note that we can freely multiply the fusion tensors with a phase $\varphi(g,h)$ and that under this redefinition the 3-cocycle changes according to the prescription
\begin{equation}\label{eq:3cocycle}
    \omega(g,h,k) \mapsto \omega(g,h,k) \frac{\varphi(g,h)\varphi(gh,k)}{\varphi(g,hk)\varphi(h,k)}.
\end{equation}
Modding out such redefinitions, we can conclude that the index $\omega$ takes values in the third cohomology group $H^3(G,\rU(1))$. This group can be computed by the methods laid out in \cref{sec:GroupCohomology}, and is in particular also finite. For the case of $G=\mbb Z_2$ considered above, we have that $H^3(\mbb Z_2,\rU(1))\simeq \mbb Z_2$ so that there exists exactly one non-trivial SPT phase protected by $\mbb Z_2$ symmetry. A representative 3-cocycle for the corresponding non-trivial class reads
\begin{equation}
    \omega(g,h,k) = (-1)^{ghk},
\end{equation}
which can be immediately deduced from \cref{eq:GroupIntertwiner} for the fusion tensors of the CZX model. Just as in the case of quantum spin chains, two PEPS wave functions with different virtual symmetries/3-cocycles cannot be converted into each other via symmetry-preserving adiabatic paths without closing the gap.
\subsection{Edge Hamiltonians \& anomalies}
A noteworthy feature of PEPS emerges when they are defined on an open patch $\mc A$. In that case, the uncontracted bonds define a Hilbert space of dimension $\chi^{|\partial \mc A|}$ where $|\partial\! \mc A|$ is the number of links intersecting the boundary. Graphically:
\begin{equation}
    %
	\input{figs/PEPS_open.tex}%
\,\,.
\end{equation}
These edge degrees of freedom define a tensor product structure for the edge modes of the corresponding parent Hamiltonian, completely in line with how we identified the SPT edge modes in the one-dimensional case in \cref{sec:SPT1d}. When the PEPS parent Hamiltonian is subsequently perturbed, this induces an \emph{effective} Hamiltonian on these boundary degrees of freedom, potentially driving them through a phase transition~\cite{Yang2014}. Interestingly, in the case of PEPS representations of SPTs, these effective boundary Hamiltonians inherit the non-local MPO symmetries of the bulk. We refer to refs.~\cite{Williamson2016,Roose2018} for some examples and their numerical investigation. 

Provided an MPO representation of a group raises in turn the question what the phase diagram of a symmetric (edge) Hamiltonian can look like. As we will now demonstrate, the answer to this question depend strongly on whether the 3-cocycle associated to the MPO algebra is trivial or not. As demonstrated in ref.~\cite{Chen2011,Cirac2020} a very elegant and straightforward tensor network argument shows that whenever $\omega$ is cohomologically non-trivial, the Hamiltonians that commute with it must either break the symmetry spontaneously or be gapless. The cocycle $\omega$ is in that case said to represent the \emph{anomaly} of the symmetry. This observation closely parallels the celebrated Lieb–Schultz–Mattis theorems~\cite{Lieb1961,Oshikawa2000,Hastings2003}. The argument goes as follows.

In the argument we assume an \emph{injective} MPO representation of an arbitrary group $G$. Due to the injectivity we can now invoke the fundamental theorem to proof the existence of fusion tensors implementing the group multiplication locally, as in \cref{eq:GroupIntertwiner}. Assume now that the 3-cocycle defined as per \cref{eq:GroupMPOFusion} is non-trivial and let us assume the existence of a unique trivially gapped symmetric ground state, represented by an injective MPS. Graphically this boils down to:
\begin{equation}
    %
	\input{figs/GroupMPOMPS.tex}%
 \,\, = \,\,
    %
	\input{figs/GroupMPS.tex}%
 \,\, .
\end{equation}
An application of the fundamental theorem of MPS then reveals that there must exist a trivalent \emph{action tensor} (in gray) which satisfies:
\begin{equation}
    %
	\input{figs/GroupMPSIntertwiner_1.tex}%
 \,\, = \,\,
    %
	\input{figs/GroupMPSIntertwiner_2.tex}%
 \,\, .
\end{equation}
Similar as in the previous section, if we act with two symmetry MPOs on the state, the fusion of the MPO tensors followed by the action on the MPS is up to a phase equal to sequentially acting with the MPOs. Concretely:
\begin{equation}
    %
	\input{figs/GroupMPSIntertwinerRecoupling_1.tex}%
 \,\, = \,\, \lambda(g,h)
    %
	\input{figs/GroupMPSIntertwinerRecoupling_2.tex}%
 \,\, .
\end{equation}
If we now consider the action of three such MPOs on our MPS, and consider the two distinct ways in which we can locally recouple these, we can deduce the following condition
\begin{equation}
    \frac{\lambda(g,h)\lambda(gh,k)}{\lambda(h,k)\lambda(g,hk)}=\omega(g,h,k),
\end{equation}
but the left-hand side can be recognized as a coboundary, which would in turn imply that the 3-cocyle is cohomologically trivial. Since this is in contradiction with the assumption that $\omega$ is non-trivial, we must therefore conclude that there is no symmetric injective MPS, and that every MPO symmetric 1-D Hamiltonian must be symmetry broken or gapless. Note that this argument is true independent of whether the Hamiltonian encodes the edge modes of a 2-D SPT phase or of a genuine quantum spin chain.
\subsection{Topological order: quantum doubles \& the tube algebra}\label{sec:QuantumDoubles}
Intriguingly, beside SPT phases, there exist in spatial dimensions higher than one robust phases of quantum matter which are not protected by any symmetry. Such phases are said to exhibit \emph{intrinsic topological order}~\cite{Wen1989,Witten1988,Zeng2015,Wen2016} which is characterised by a ground state degeneracy that only depends on the topology of the underlying spatial manifold (sphere, torus,...), gapped excitations with \emph{anyonic} braiding statistics and, depending on its anyon content, a universal negative correction to the area law of entanglement entropy~\cite{Kitaev1997,Kitaev2005a,Levin2006}. Notably, intimately connected to the absence of symmetry, these phases are not characterised by a local order parameter,  and thus fall outside the scope of the traditional Landau paradigm of symmetry breaking~\cite{Landau1937,Wegner1971,Bravyi2006}. Nevertheless, the fact that there is a  ground state degeneracy clearly indicates that some symmetry is broken. It turns out that this symmetry is one that acts purely on the virtual degrees of freedom.

Recall that short-range entangled SPT order gets its stability from the existence of a topological index associated to the transformation properties of the entanglement degrees of freedom under the global symmetry (\cref{sec:SPT1d} and \ref{sec:SPT2d}). The robustness of intrinsic topological order on the other hand stems from its long-range entanglement~\cite{Chen2010}. In a sense, characterising topological orders thus boils down to the classification of equivalence classes of long-range entanglement. Similarly, phase boundaries between these topological phases can be understood from a radical change in the entanglement properties~\cite{Haegeman2015,Marien2016}. Having direct access to these entanglement degrees of freedom, as we have in PEPS, suggests that we could further reduce this problem to classifying \emph{local} PEPS tensors that give rise to \emph{global} topological states: the classification of intrinsically topologically ordered systems is equivalent to the construction of MPO-symmetric tensors. 

Indeed, in the language of PEPS, the presence of topological order can be traced back to the existence of \emph{virtual} MPO symmetries that commute with the ground state PEPS tensors as:
\begin{equation}\label{eq:PEPSPullingThrough}
    %
	\input{figs/PEPS_TopoPulling_1.tex}%
 \,=\, %
	\input{figs/PEPS_TopoPulling_2.tex}%
\,\,,
\end{equation}
Note that, as opposed to the SPT case, there is no action on the physical degrees of freedom. Consequently, these MPOs are completely deformable -- or thus \emph{topological} -- on the virtual level. This also implies that the underlying symmetry structure can be much richer than the ones described by groups. As we will see in lecture \ref{sec:GenSym}, the symmetry algebra is encoded by a \emph{fusion category}. Since the MPOs are blind to the physical degrees of freedom, such states are resilient to (small) local perturbations on the physical level~\cite{Marien2016}. However, \emph{anyon condensation} provides a mechanism to drive the system to a trivial phase under sufficiently strong perturbations~\cite{Bais2008,Burnell2017}.

It is conjectured that a non-chiral topological order is fully characterised by its anyon content, together with its fusion rules and braiding statistics. Mathematically, this data is encapsulated in a \emph{modular tensor category}, called the \emph{Drinfeld centre} $\mc Z(\mc D)$~\cite{Etingof2015,Kong2017,Bultinck2015,Kong2022}. The full extent of this statement, and the microscopic realisations of generic topological orders will be the topic of \cref{sec:SN}. In this section we will rather focus on the construction of a class of topological orders called (untwisted) \emph{quantum doubles}~\cite{Dijkgraaf1989,Roche1990,Kitaev1997}, as a more gentle introduction to PEPS representations of topological order and some of its properties.

\bigskip\noindent The quantum doubles as introduced by Kitaev in ref.~\cite{Kitaev1997} are a family of models constructed from a finite group $G$ by assigning $|G|$-dimensional qudits to the edges of a two-dimensional lattice, subject to a commuting projector Hamiltonian spelled out in the aforementioned reference. They are a lattice Hamiltonian realisation of untwisted Dijkgraaf-Witten topological field theory~\cite{Dijkgraaf1989}. For the group $\mbb Z_2$, the model reduces to the \emph{toric code}. The ground states of these quantum double models admit notably simple renormalisation fixed point PEPS representations called \emph{$G$-injective}, or more precisely \emph{$G$-isometric}, PEPS. The notion of $G$-isometry essentially captures PEPS tensors that are invertible only on a subspace of the virtual degrees of freedom which is invariant under an action of the group $G$, and is a particular incarnation of MPO injectivity.

Up to local physical transformations all $G$-isometric PEPS reduce to a universal form that we will now describe. In this section, we assume a square lattice on a two-dimensional torus. We write $L$ to denote the \emph{left regular representation} of $G$, acting as $L_g\ket{h} = \ket{gh}$. Define then the projector $1/|G|\sum_g L_g\otimes L_g\otimes L_g^\dagger\otimes L_g^\dagger$, which constitutes the ground state PEPS tensor, graphically:
\begin{equation}\label{eq:PEPSGTensor}
    \frac{1}{|G|}\sum_{g\in G}\,\,\, %
	\input{figs/PEPS_G_1.tex}%
\,\,.
\end{equation}
In this figure, the gray boxes represent $L_g$, and the bar denotes the inverse group element (recall that $L_g^\dagger = L_{\bar g}$). Note that the four innermost legs together form the physical level. As the sum in eq.~\ref{eq:PEPSGTensor} runs over all group elements,  this PEPS tensor satisfies the pulling through conditions~\cref{eq:PEPSPullingThrough} for the following MPOs labelled by group elements:
\begin{equation}\label{eq:MPOGroupRegular}
    %
	\input{figs/MPOGroupg.tex}%
:= \bigotimes_\msf i L_g,
\end{equation}
where the dotted line is used to emphasise that the bond dimension is 1, i.e. each MPO acts as a tensor product.

The PEPS tensor \cref{eq:PEPSGTensor} gives rise to only one of the ground states of the model. As demonstrated in~\cite{Schuch2010}, all the other ones can be obtained unitarily from this one by acting with operators that wrap around either or both of the non-contractible cycles of the torus. The extensive nature of these operators are reflective of the local indistinguishability of the ground states. Equivalently, one can obtain all the ground states by wrapping both cycles of the torus with MPOs on the virtual level and joining them on their intersection by a tensor $\mc P_Z$ that we need to solve for. In this, we require that the ground states are all translationally invariant and remain normalisable and orthogonal in the thermodynamic limit~\cite{Zhang2011}. Using the fusion tensors for the MPOs \cref{eq:MPOGroupRegular} we can expand such a state as follows:\footnote{Note that the fusion tensors in this case simply enforce the group multiplication, i.e. $\sum_{g,h,k}\ketbra{k}{g,h}=\sum_{g,h}\ketbra{gh}{g,h}$.}
\begin{equation}
    %
	\input{figs/PEPS_G_idempotent_1.tex}%
 \,\, = \sum_{g,h} \mc (\mc P_Z)_{g,h}\,\,%
	\input{figs/PEPS_G_idempotent_2.tex}%
\,\,.
\end{equation}
The translation invariance of these states is manifest by the pulling through condition. As demonstrated in refs.~\cite{Sahinoglu2014,Bultinck2015,Williamson2017} a basis of ground states in MPO injective PEPS is in one-to-one correspondence with \emph{minimal central idempotents} or -- equivalently -- irreducible representations of \emph{Ocneanu's tube algebra}~\cite{Ocneanu1993,Ocneanu2001}. In fact, the Drinfeld centre $\mc Z(\mc D)$ is in practice often computed via the tube algebra. For the particular case of quantum doubles, the relevant tube algebra is spanned by closed MPOs of the form:
\begin{equation}
    T_x^g := %
	\input{figs/PEPS_G_Tube.tex}%
\,\, ,
\end{equation}
thought of as acting from the outside indices to the inside. Their multiplication is defined by stacking them inside each other. Algebraically this boils down to:
\begin{equation}\label{eq:GTubesMult}
    T^g_x\circ T_y^h = \delta_{x,hyh^{-1}} T^{gh}_x.
\end{equation}
This multiplication defines an algebra often called the \emph{(untwisted) quantum double} $\mc D(G)$ of the group $G$~\cite{Roche1990}. Notably, the tube algebra is endowed with a \emph{star}, which is an antilinear involution that swaps the order of the product. For the case at hand: $\big( T_x^g\big)^\star := T^{g^{-1}}_{gxg^{-1}}$.
As such, the tube algebra is endowed with the structure of a \emph{$C^\star$-algebra}. Due to a famous result of Artin and Wedderburn, every $C^\star$-algebra can be decomposed in simple matrix algebras. The aforementioned minimal central idempotents then act as projectors on each of these simple blocks.

Let us compute them explicitly. To this end it is useful to focus first, for fixed $x\in G$, on all tubes $T_x^g$ where $g\in Z_x=\{g\in G|gx=xg\}\subseteq G$. These tubes form a representation of the stabiliser group $Z_x$ since
\begin{equation}
    T^{g_1}_x\circ T^{g_2}_x = T^{g_1g_2}_x, \quad \forall g_1,g_2\in Z_x,
\end{equation}
as follows directly from \cref{eq:GTubesMult}. Let $\mu_x$ denote an irreducible representation of $Z_x$ of dimension $d_{\mu_x}$ and with $\chi^{\mu_x}$ its corresponding character. The operators
\begin{equation}
    \mc P_{(x,\mu)} := \frac{d_{\mu_x}}{|G|} \sum_{k\in Z_x} \chi^{\mu_x}(k) T_x^k,
\end{equation}
form then hermitian projectors i.e.\footnote{Use Schur orthogonality of the characters $\sum_k \chi^{\mu_x}(k)\chi^{\nu_x}(k^{-1}m) = |Z_x|/d_\mu \chi^\mu(m)$.}
\begin{equation}
    \mc P_{(x,\mu_x)} \mc P_{(y,\nu_y)} = \delta_{x,y}\delta_{\mu_x,\nu_x} \mc P_{(x,\mu_x)}.
\end{equation}
These $\mc P_{(x,\mu)}$ are \emph{simple} idempotents. The central ones are then obtained by additionally summing over all $x$ in a conjugacy class $C_x = \{gxg^{-1}|g\in G\}$:
\begin{equation}
    \mc P_{(C_x,\mu_x)} \equiv \sum_{y\in C_x} \mc P_{(y,\mu_x)}.
\end{equation}
This reproduces the results of refs.~\cite{Kitaev2005a} that torus ground states of the quantum doubles are labelled by a conjugacy class $C_x$ and an irreducible representation of the stabiliser subgroup $Z_x$ of a representative $x$ of $C_x$.

Importantly, also anyonic excitations are labelled by minimal central idempotents of the tube algebra~\cite{Lan2013,Hu2015}. At the level of the PEPS, one can create pairs of anyonic excitations by replacing two of the ground state tensors by tensors whose support is contained in the image of the corresponding idempotent labelling the excitation, we refer to~\cite{Bultinck2017} for details.

The (untwisted) quantum doubles considered here are generalised in a number of ways. For one, one could consider a \emph{twisted} version of the model where the MPO symmetries form an anomalous representation of $G$ with the anomaly taking values in the third cohomology group $H^3(G,\rU(1))$, as described in \cref{sec:SPT1d}. For this, we refer to refs.~\cite{Buerschaper2013,Bultinck2016,Williamson2017}.

In \cref{sec:SN} we will discuss the PEPS representations of so-called string-net models constructed first by Levin and Wen in ref.~\cite{Levin2004}. These are a class of models that have a spherical fusion category as input, and realise all non-chiral topological orders $\mc Z(\mc D)$. One way of obtaining the torus ground states and anyonic excitations described by $\mc Z(\mc D)$ in these models is exactly via the tube algebra associated to $\mc D$, completely in line with the above construction. For the input $\mc D=\Vect_G^\omega$ the corresponding Levin-Wen model coincides with the twisted quantum doubles.
\subsection{Synopsis of lecture IV} 
Lecture IV introduced three new topics: symmetries in the form of MPOs which underlie the integrability of special classes of quantum spin chains; the higher dimensional generalisation of MPS to PEPS, and the corresponding manifold which parametrises low-energy states of gapped higher dimensional lattice models; and MPO symmetries that satisfy a so-called \emph{pulling through symmetry} which characterises the different gapped (topological) phases of such systems. Concerning integrability, we demonstrated that the Yang-Baxter equations are a consequence of the fundamental theorem of MPS. It would be wonderful if this point of view leads to more general classes of integrable models, such as the ones satisfying non-trivial categorical symmetries (see lecture V).

We showed that SPT phases in 2D lattice systems are characterised by non-trivial MPO symmetries whose associators encode 3-cocycles, leading to the classification of 2D SPT phases. However, the pulling through symmetries of PEPS open up completely new ways of realising topological phases of matter in 2D systems if no global symmetries are involved. We demonstrated the existence of intrinsic topologically ordered lattice systems in the form of quantum doubles, and showed that those systems exhibit symmetry breaking with respect to the entanglement degrees of freedom. This was the first hint that the Landau paradigm can be saved if you describe many-body phases of matter in terms of the symmetries of the underlying PEPS representation: different phases realise the symmetries of the tensors in a fundamentally different way. 

%% file: figs/MPOGroupgh.tex
\begin{tikzpicture}[baseline={([yshift=-.5ex]current bounding box.center)}]
	\def\l{.8};
    \def\h{.65};

	\draw[obj2] (-\l,0) -- (5*\l,0);
	
	\foreach \x in {0,1,2}{
		\draw[obj1] (2*\x*\l,\h) -- (2*\x*\l,-\h);
		\node[draw=black, line width=.7pt, fill=gray!20, rectangle, inner sep=3.8pt, rounded corners=2pt] at (2*\x*\l,0) {};
	}
    \node[above] at (3*\l,0) {${\sss gh}$};
\end{tikzpicture}

%% file: figs/GroupMPS.tex
\begin{tikzpicture}[baseline={([yshift=10pt]current bounding box.south)}]
	\def\l{.8};
    \def\h{.65};

	\draw[obj1] (-\l,0) -- (5*\l,0);
	
	\foreach \x in {0,1,2}{
		\draw[obj1] (2*\x*\l,\h) -- (2*\x*\l,0);
        \node[mpstensor] at (2*\x*\l,0) {};
	}
\end{tikzpicture}

%% file: figs/MPOGroupg.tex
\begin{tikzpicture}[baseline={([yshift=-.5ex]current bounding box.center)}]
	\def\l{1.6};
    \def\h{.65};

	\draw[obj2, dotted] (-1/2*\l,0) -- (2.5*\l,0);
	
	\foreach \x in {0,1,2}{
		\draw[obj1] (\x*\l,\h) -- (\x*\l,-\h);
		\node[mpotensor] at (\x*\l,0) {};
	}
	\node[above] at (1.5*\l,0) {${\sss g}$};
\end{tikzpicture}

%% file: _Lecture_5.tex
\section{Lecture V: Generalised symmetries}\label{sec:GenSym}
\subsection{Motivation}
In lecture~\ref{sec:PhasesOfMatter} we classified phase diagrams of one-dimensional quantum lattice models with a group symmetry. We also saw that in one-dimensional bosonic systems the presence of a symmetry is \emph{necessary} to have a non-trivial phase diagram. Recall from \cref{sec:Fermions} that the story was more subtle for the case of fermions, which can exhibit intrinsic topological order due to a superselection rule. In recent years, substantial research effort has been made to understand systems with symmetries beyond groups~\cite{Gaiotto2014,Thorngren2019,Aasen2020,Komargodski2020,McGreevy2022,Moradi2022,Freed2022,Shao2023,Schafer-Nameki2023,Moradi2023,Bhardwaj2023}. As we foreshadowed in the introduction to lecture~\ref{sec:PhasesOfMatter}, generalised symmetries described by fusion categories are encountered as the \emph{virtual symmetries} of non-chiral two-dimensional topological orders and therefore also arise as the symmetries of their boundary Hamiltonians or -- equivalently -- the symmetries of the effective theories describing their edge modes. Studying how those generalised symmetries influence the phase diagrams of systems has led to a \emph{generalised Landau paradigm}, where different phases of matter were found to be in correspondence with different ways to break a generalised symmetry~\cite{Thorngren2019,Moradi2022,Bhardwaj2023,Lootens2024,Chen2025}. With non-group symmetries, one cannot simply say that a symmetry is broken to a subgroup, so we should ask: how can a system (partially) break a generalised symmetry? Is there a map between systems exhibiting different ways of generalised symmetry breaking? These questions will all be expanded upon in this section.

Why are tensor networks a good description for such categorical symmetries? A crucial observation is that to encode symmetries beyond on-site, we need operators that act in a correlated way on neighbouring degrees of freedom. Additionally, we want to be able to define these operators on systems of any size, so we need to encode those correlated symmetries locally. MPOs are exactly the right language to use for this purpose. They are parametrised in terms of a single local tensor, can be defined on any number of sites, and share information between neighbouring physical degrees of freedom through their virtual degrees of freedom. Because of these considerations, we will often use the terms \emph{generalised}, \emph{categorical} and \emph{MPO symmetry} interchangeably.

Indeed, the study of general MPO symmetries revealed that, under mild conditions such as injectivity (discussed further in \cref{sec:MPORepresentationTheory}), their underlying structure is equivalent to that of fusion categories. In brief, fusion categories are semisimple associative structures of (simple) objects, very much like a group is an associative multiplication rule of its elements (its simple objects). However, unlike the group case, the multiplication of simple objects of a fusion category may lead to a direct sum of simple objects. A canonical example being the fusion category $\Rep(G)$ whose simple objects are the irreps of a group and whose multiplication rule is dictated by the direct sum decomposition of the tensor product of irreps, e.g. $1/2\otimes 1/2=0\oplus 1$. Fusion category theory is a very well studied topic in mathematics and mathematical physics, and  MPOs turn out to be the building blocks of \emph{representations} of those fusion categories on the lattice, much like matrices represent on-site group symmetries in spin systems.

In this final lecture we will first study the MPO representation theory of categorical symmetries and demonstrate how \emph{module categories} label the distinct representations. We will then provide three concrete applications. To start, we will see how these MPO symmetries arise in the Levin-Wen string-net models, which provide microscopic realisations of all non-chiral two-dimensional topological orders. We then provide two approaches to construct \emph{edge theories} of these string-net models. In \cref{sec:SC}, we will demonstrate how so-called \emph{strange correlators} of these PEPS representations lead to the construction of two-dimensional \emph{classical} partition functions of critical models~\cite{Vanhove2018}, generalizing a construction of Aasen, Mong and Fendley~\cite{Aasen2016}. The main merit of this strange correlator construction is that it makes the distinction between \emph{topological} and \emph{geometrical} or \emph{dynamical} features of the critical lattice model manifest: the (2+0)d partition functions inherit the MPO symmetries from the (2+1)d string-nets, and as such provides the lattice analogue of TFT/CFT correspondence. This holographic approach to the construction of edge theories with categorical symmetries can be interpreted as an explicit lattice realisation of the categorical construction of Kong, Wen and Zheng ~\cite{Kong2015} which provides a concrete foundation for what is now widely referred to as \emph{SymTFT} or \emph{topological holography}~\cite{Apruzzi2021,Chatterjee2022,Freed2022,Delcamp2024,Chen2025}. Finally, we will provide a systematic way to construct local (1+1)d MPO symmetric Hamiltonians and a generalised duality or gauging procedure for generalised symmetries. We conclude by discussing the anticipated generalised Landau paradigm for categorical symmetries, as well as some of its implications.
\subsection{MPO representation theory}
\label{sec:MPORepresentationTheory}
In this section we will follow a logic similar to the one in \cref{sec:YangBaxterFromMPO}. Assuming a collection of injective MPOs representing a fusion algebra, we deduce the consistency equations satisfied by the fusion tensors. We subsequently show how solutions of these equations can be used to construct representations of these MPOs and fusion tensors. It is demonstrated how distinct representations are constructed from the data encoded in \emph{module categories} over the fusion category describing the symmetry.

\bigskip\noindent Concretely, assume a finite collection of injective MPOs labelled by $a,b,c,\dots$ with periodic boundary conditions. We are interested in MPO algebras for which the structure constants $N_{ab}^c$ are independent of the number of sites in the periodic chain:
\begin{equation}
\label{eq:FusionRules}
    %
	\input{figs/MPOsStacked.tex}%
 = \sum_{c} N_{ab}^{c} %
	\input{figs/MPOc.tex}%
\,\, .
\end{equation}
Here and in what follows, the sums are over all distinct MPOs appearing in the algebra. Amongst them, there exists one labelled $\mbb 1$ for which $N_{a\mbb 1}^{a'}=N_{\mbb 1 a}^{a'}=\delta_{a,a'}$. The non-negative integer tensor $N$ defines a \emph{unital associative fusion} structure, meaning in particular that the structure constants satisfy $\sum_d N_{ab}^dN_{dc}^e = \sum_d N_{bc}^d N_{ad}^e$. Exploiting the fundamental theorem~\cref{eq:FundTh}, we find that \cref{eq:FusionRules} can be implemented locally by means of pulling the following \emph{fusion tensors} through the MPO tensors:
\begin{equation}
\label{eq:IntertwinerHorizontal}
    %
	\input{figs/MPOIntertwiner_1.tex}%
 \,= \,\,
    \raisebox{4.5pt}{%
	\input{figs/MPOIntertwiner_2.tex}%
}\,\,,
\end{equation}
where the multiplicity $i=1,2,\dots,N_{ab}^c$ labels the linearly independent fusion tensors. Similarly as in \cref{sec:SPT2d}, associativity of the MPO multiplication implies the existence of a collection of \emph{unitary} matrices called \emph{$F$-symbols} or \emph{associators} encoding the basis transformation
\begin{equation}\label{eq:MPOPentagon}
    \raisebox{-5.5pt}{%
	\input{figs/MPOPentagon_1.tex}%
} = \sum_{f,k,l}\big(F^{abc}_{d}\big)_{e,ij}^{f,kl}\,\,\raisebox{-5.5pt}{%
	\input{figs/MPOPentagon_2.tex}%
}.
\end{equation}
The requirement that the product of four MPOs -- which can be recoupled in two distinct ways -- is associative, reveals that these $F$-symbols obey a consistency condition colloquially known as the \emph{pentagon equation}. Schematically, the pentagon equation is a linear multivariate system of equations of the form
\begin{equation}\label{eq:pentagon}
    FF = \sum FFF,
\end{equation}
or, written out in components:
\begin{equation}\label{eq:pent}
    \sum_o \left(F^{fcd}_e\right)^{h,no}_{g,lm} \left(F^{abh}_e\right)^{i,pq}_{f,ko} = \sum_{j,rst} \left(F^{abc}_g\right)^{j,rs}_{f,kl} \left(F^{ajd}_e\right)^{i,tq}_{g,sm} \left(F^{bcd}_i\right)^{h,np}_{j,rt}.
\end{equation}
In essence, the fusion rules specified by the $N$-symbol, together with a unitary solution of the pentagon equation constitute the data encoded in a \emph{unitary fusion category} $\mc C$.\footnote{Note the resemblance of the pentagon equation with the 3-cocycle derived in \cref{eq:3cocycle} by an analogous reasoning applied to group symmetry. And indeed, a finite group together with a choice of 3-cocycle defines a fusion category called $\Vect_G^\omega$, as reviewed in \cref{sec:cat}, for which the $F$-symbol coincides with $\omega$.} Importantly, for any valid fusion structure, the number of solutions to the pentagon equation is finite, up to basis transformations on the multiplicity indices.

Given a solution to the pentagon equation, how can we find a corresponding MPO representation in the sense of \cref{eq:FusionRules}? Completely in line with the logic of \cref{sec:YangBaxterFromMPO}, we can find a representation by equating the $F$-symbols to (the components of) the fusion tensors defined in \cref{eq:IntertwinerHorizontal} as well as to the MPO tensors. In that case, both \cref{eq:MPOPentagon} as well as \cref{eq:IntertwinerHorizontal} are equivalent to the pentagon equation \eqref{eq:pent}~\cite{Sahinoglu2014,Bultinck2015,Williamson2017,Molnar2022}. Concretely, the assignment of non-zero components is as follows\footnote{We omit a conventional prefactor involving quantum dimensions~\cite{Bultinck2017,Lootens2020}.}:
\begin{equation}\label{eq:SingleLineLabels}
    %
	\input{figs/PEPS_C_single_labels.tex}%
 = \big(F^{abc}_d\big)_{e,ij}^{f,kl}= 
    %
	\input{figs/MPO_single_labels.tex}%
\,.
\end{equation}
Note that every index is labelled by a triple of objects $a,b,c$ and a multiplicity index taking values $1,2,\dots,N_{ab}^c$. In particular, all triples simultaneously have to satisfy the fusion rules in order for a component to be non-zero. Moreover, the structure of these tensors is such that some of the objects are shared between indices. This internal structure of the tensors invites us to depict these tensors as \emph{triple line tensors}:
\begin{equation}
    %
	\input{figs/PEPS_C_triple.tex}%
\equiv%
	\input{figs/PEPS_C_single_labels.tex}%
 ,\quad
    %
	\input{figs/MPO_Single_labels.tex}%
\equiv%
	\input{figs/MPO_triple.tex}%
\,.
\end{equation}
Contraction of these tensors takes place via identification of the simple objects on strands of concatenated tensors, pairing indices, and summing over them as detailed in e.g. refs.~\cite{Bultinck2017,Lootens2020}.

Due to the fusion rules, the MPOs act on a Hilbert space which is generically not a tensor product space, but rather a Hilbert space with local \emph{constraints} dictated by the fusion rules. In this sense, the MPOs constructed here and generalised below, form a representation and generalisation of the topological symmetries considered in the context of \emph{anyonic chains}~\cite{Feiguin2006,Trebst2008,Gils2013,Buican2017,Huang2021,Jones2024}. The models considered in the original reference~\cite{Feiguin2006} for instance are constructed from the Fibonacci fusion category $\Fib$ (see \cref{sec:cat}).

\bigskip\noindent The above provides only one MPO representation of $\mc C$. As demonstrated in ref.~\cite{Lootens2020}, any \emph{module category} over the given fusion category $\mc C$ provides a distinct way to realise that symmetry on the lattice. Succinctly, a $\mc C$-module category $\mc R$ consists of a finite set of simple objects $A,B,\dots$ together with an \emph{action} of $\mc C$ on $\mc R$, which can be expressed as $a\act A=\sum_{B} N_{aA}^{B}B$. Notably, this $N$-symbol is different from the $N$-symbol of $\mc C$ as it depends on two objects of $\mc R$ and one of $\mc C$. A module category also provides a collection of unitary \emph{$\mF$-symbols} or \emph{(left) module associators},which together with the $F$-symbols of $\mc C$ obey a \emph{mixed} pentagon equation which is of the form:
\begin{equation}\label{eq:ModulePentagon}
    \mF\mF = \sum F\mF\mF.
\end{equation}
Given a $\mc C$-module category and its corresponding module associators, we can construct fusion tensors by equating its components to the $\mF$-symbol -- completely in line with \cref{eq:SingleLineLabels} and as detailed below in \cref{eq:TripleLineTensors} -- so that the pulling through condition \cref{eq:IntertwinerHorizontal} is satisfied by virtue of \cref{eq:ModulePentagon}. Remark that every fusion category $\mc C$ defines a module category over itself, called the \emph{regular module category}. In that case, we recover the MPO representations of the previous paragraph. Before explaining what the MPO tensors evaluate to, let us establish some intuition for this framework by means of a simple example.

\paragraph{Example: $\mc C = \Vect_G$} Formally, a (non-anomalous) finite-group symmetry is described by the fusion category $\Vect_G$ of $G$-graded vector spaces. The simple objects correspond to group elements and the fusion rules are given by the group multiplication, i.e. $N_{g,h}^k = \delta_{gh,k}$. Module categories over $\Vect_G$ are in one-to-one correspondence with pairs $(H,\psi)$ where $H\subseteq G$ is a subgroup up to conjugacy and $\psi$ is a $\rU(1)$-valued 2-cocycle of the subgroup $H$. We refer to \cref{sec:cat} for details. Let us focus on the case where $H=G$ and $\psi$ is trivial. The corresponding module category, denoted $\Vect$, then results in MPOs
\begin{equation}\label{eq:MPOonsite}
    %
	\input{figs/MPOGroupg.tex}%
 \,\,= \bigotimes_\msf i u_\msf i(g).
\end{equation}
The $u_\msf i$ here are (not necessarily irreducible) unitary representations of $G$, and can be chosen independently on each site. The group structure of the MPOs follows from $u_\msf i(g)u_\msf i(h)=u_\msf i(gh)$, $\forall g,h\in G$, $\forall \msf i$. The dotted line is to emphasise that the bond dimension of the MPOs is $1$. Notably, the fact that the MPOs act on a tensor product Hilbert space is a direct consequence of the choice of module category $\Vect$. For the other allowed module categories over $\Vect_G$, the MPOs generically act on constrained Hilbert spaces as mentioned above.

\bigskip\noindent Note that in the case of on-site group symmetry \cref{eq:MPOonsite}, one can bring all MPO tensors $u_\msf i(g)$ -- indexed by the group elements $g\in G$ -- simultaneously in block-diagonal form. The blocks correspond to the irreducible representations contained in the representations $u_\msf i$. More generally, we could consider the collection of MPO tensors as operator-valued matrices labelled by the blocks $a,b,c,\dots$, and block-decompose the algebra spanned by them in the \emph{vertical} direction. The crucial insight is now that the labels of these injective blocks, and their fusion rules, i.e.\footnote{Note that these MPOs extend vertically -- as graphically indicated by the dots -- but not horizontally.}
\begin{equation}
    %
	\input{figs/MPOsStackedVertical_1.tex}%
 = \sum_\gamma N_{\alpha\beta}^\gamma \,\, %
	\input{figs/MPOsStackedVertical_2.tex}%
\,\,,
\end{equation}
are governed by another fusion category $\mc D$, which is generically \emph{distinct} from the symmetry category $\mc C$. Its simple objects will be labelled by Greek letters  $\alpha,\beta,\gamma,\dots$. Given $\mc C$ and $\mc R$, $\mc D$ is fully specified. In other words, fixing the symmetry category and a way to represent it on the lattice via the chosen module category $\mc R$, $\mc D$ is determined. Technically, $\mc D$ corresponds to the \emph{Morita dual} (see also \cref{sec:cat}) of $\mc C$ with respect to $\mc R$, denoted by $\mc D=\mc C_\mc R^\star$, and equivalently $\mc C =\mc D^\star_\mc R$. Morita equivalence of $\mc C$ and $\mc D$ guarantees in particular that the so-called \emph{Drinfeld centres} of $\mc C$ and $\mc D$ are equivalent: $\mc Z(\mc C)\simeq\mc Z(\mc D)$. In \cref{sec:SN} we will see that $\mc Z(\mc D)$ encodes all information about the ground state sectors and anyonic excitations of the string-net with topological virtual symmetries described by $\mc C$. In \cref{sec:SC} and \cref{sec:dualities}, elements of the Drinfeld centre will label the \emph{topological} or \emph{superselection sectors} of the string-net boundary theories.

Being a fusion category, $\mc D$ comes with its own set of $F$-symbols that we shall write as $\dF$. These satisfy a pentagon equation analogous to \cref{eq:pentagon}. The Morita dual $\mc D$ is constructed in such a way that $\mc R$ is also a module category over $\mc D$, with $\mc D$ now acting from the \emph{right} on $\mc R$. As a consequence there are once more corresponding $F$-symbols, denoted $\nF$, encoding the associativity of the $\mc D$ action on $\mc R$. These satisfy the fourth pentagon equation we encounter, now involving those of $\mc D$ as well:
\begin{equation}\label{eq:Pent2}
    \nF\nF = \sum \dF\nF\nF.
\end{equation}
Beyond being a module category for both $\mc C$ and $\mc D$, $\mc R$ is an (invertible) \emph{$(\mc C,\mc D)$-bimodule category} (see \cref{sec:cat}). Associated to its bimodule structure is a final collection of $F$-symbols denoted by $\rF$ obeying
\begin{align}
    \rF\mF &= \sum\mF\rF\rF, \label{eq:Pent3}\\
    \nF\rF &= \sum\rF\rF\nF \label{eq:Pent4}.
\end{align}
The components of the MPO tensors are equal to the $\rF$-symbols, while the $\nF$-symbols are exactly what the fusion tensors in the \emph{vertical} direction evaluate to. Gathering everything:
\begingroup
\allowdisplaybreaks
\begin{align}
    %
	\input{figs/PEPS_C_triple_mod.tex}%
&\equiv%
	\input{figs/PEPS_C_single_labels_mod.tex}%
 = \big(\mF^{abC}_A\big)_{c,ij}^{B,kl}, \notag\\
    %
	\input{figs/PEPS_D_triple.tex}%
&\equiv%
	\input{figs/PEPS_D_single_labels.tex}%
=\big(\nF^{A\alpha\beta}_B\big)_{C,ij}^{\gamma,kl},\label{eq:TripleLineTensors}\\
    %
	\input{figs/MPO_triple_mod.tex}%
&\equiv%
	\input{figs/MPO_single_labels_mod.tex}%
=\big(\rF^{aA\alpha}_D\big)_{C,ij}^{B,kl}.\notag
\end{align}
\endgroup
Note that in these triple line tensors, objects in the module category always live on the outer \emph{loops}. One can then verify that the vertical pulling through condition
\begin{equation}\label{eq:IntertwinerVertical}
    %
	\input{figs/MPOIntertwiner_3.tex}%
 =
    %
	\input{figs/MPOIntertwiner_4.tex}%

\end{equation}
is satisfied by virtue of \cref{eq:Pent4}, while \cref{eq:IntertwinerHorizontal} is equivalent to \cref{eq:Pent3}. These fusion tensors in turn are related via a rotated version of the diagram \cref{eq:MPOPentagon} involving these fusion tensors and the $\dF$-symbol, equivalent to \cref{eq:Pent2}.

In the example $\mc C=\Vect_G$, $\mc R=\Vect$ considered above, the fusion category $\mc D$ is $\Rep(G)$, whose simple objects are the irreps of the group $G$. The corresponding $\dF$-symbols  are equal to the 6j-symbols of $G$, while the  $\nF$-symbols are equal to  the \emph{Clebsch-Gordan coefficients}. Those Clebsch-Gordan coefficients decompose the MPO tensors $u_\msf i(g)$ in its irreducible blocks, and the data in these blocks is encoded in $\rF$, whose values are given by the irreps of $G$. In the general (non-group) case, the $\nF$-symbols will below be interpreted as \emph{generalised} Clebsch-Gordan coefficients for the MPO symmetries constructed here. At the same time, we will interpret \cref{eq:IntertwinerVertical} as a \emph{generalised Wigner-Eckart} theorem that states that MPO symmetric tensors are built from these generalised Clebsch-Gordan coefficients.

\bigskip\noindent In conclusion, the following picture has emerged. Given a symmetry specified by a fusion category $\mc C$, an MPO representation is constructed from an invertible $(\mc C,\mc D)$-bimodule category $\mc R$. This structure encodes associator data in the form of five F-symbols $(F,\mF,\rF,\nF,\dF)$ that solve six pentagon equations. The bimodule associator $\rF$ can be used to construct the injective MPO representation of the fusion category $\mc C$, whose tensors are locally fused horizontally \cref{eq:IntertwinerHorizontal} and vertically \cref{eq:IntertwinerVertical} via the tensors evaluating to $\mF$ and $\nF$ respectively. Both satisfy a recoupling identity involving $F$, respectively $\dF$.
\subsection{Application 1: string-net ground states \& intertwiners}\label{sec:SN}
\emph{String-net models} were first constructed by Levin and Wen in ref.~\cite{Levin2004}. They constitute a class of RG fixed point lattice models that takes as an input a (spherical) fusion category, denoted by $\mathcal{D}$, and realises the non-chiral topological order given by its Drinfeld centre $\mc Z(\mc D)$~\cite{Kirillov2011,Hu2015,Kawagoe2024}. By that, it is meant that the ground states and gapped excitations reproduce the topological sectors and anyonic statistics encoded in $\mc Z(\mc D)$.\footnote{Equivalently, they can be thought of as a Hamiltonian realisation of the Turaev--Viro--Barrett--Westbury topological field theory~\cite{Turaev1992,Barrett1993,Kirillov2011}.} The quantum doubles encountered in \cref{sec:QuantumDoubles} are reproduced for the choice $\mc D=\Vect_G$. The name \emph{string-net} stems from the interpretation of the model as describing a condensed phase of coloured strings, with the allowed string-types and their interactions encoded in $\mc D$. The Hamiltonian is a sum of local commuting projectors on each plaquette, allowing for a straightforward derivation of a PEPS representation of its ground states.

As shown in ref.~\cite{Lootens2020}, there exists a ground state PEPS representation for every module category $\mc R$ over the input $\mc D$. This observation generalises and unifies earlier references where implicitly the regular module category was assumed~\cite{Verstraete2006b,Buerschaper2009,Gu2009,Williamson2017,Bultinck2017}, or in the specific case of the quantum doubles, $\Vect$ as module category over $\Vect_G$~\cite{Schuch2010}. From this point of view, a string-net is thus a PEPS with a particular type of \emph{virtual} symmetry. The MPOs which form representations of the underlying Morita dual fusion category $\mathcal{C}$ are its symmetries, and they act purely on the virtual degrees of freedom of the tensor network. In accordance with the generalised Wigner-Eckart theorem discussed above, the PEPS tensors evaluate to $\nF$ as in \cref{eq:TripleLineTensors}, with the physical level taking values in $(\alpha\beta\gamma,k)$.

Akin to \cref{sec:QuantumDoubles}, the different torus ground state sectors and anyonic excitations, both labelled by elements in $\mc Z(\mc D)$, can be understood in terms of the \emph{tube algebra}. The tube algebra is constructed from the MPOs in $\mc C$, and their central idempotents are in one-to-one correspondence with elements in $\mc Z(\mc C)$~\cite{Williamson2017,Bultinck2017}. Recall that $\mc Z(\mc D)$ is guaranteed to be equivalent to $\mc Z(\mc C)$ due to the invertibility of $\mc R$. Indeed, if we consider a string-net on a torus, the different topological sectors can be obtained by winding the MPO symmetries of the previous section around its uncontractable loops. In their intersection, a tensor is placed whose components are determined by the central idempotents. The insertion of a central idempotent associated to the object $Z\in\mc Z(\mc C)$ in a string-net PEPS can be depicted as:
\begin{equation}
    \sum_{\substack{a_\alpha,b_\beta,c_\gamma\\i,j}} (\mc P_Z)_{a_\alpha,b_\beta,c_\gamma}^{i,j} \quad %
	\input{figs/String-net_PEPS.tex}%
 \,\, .
\end{equation}
Note that the purple tensors coincide with the fusion tensors of \cref{eq:IntertwinerVertical}, so that the exact location of the central idempotent is undetectable. Here, $a_\alpha$ stands for $a_\alpha\equiv(a,\alpha)$ where $a$ is a simple object in $\mc C$ and $\alpha$ labels the degeneracy of $a$ in $Z$.

Given a PEPS representation based on $\mc R$, one can locally change this to one based on $\mc R'$ via a new class of MPO \emph{intertwiners} that implement the change of module category. In particular, when $\mc R'=\mc D$, these MPO intertwiners are labelled by objects in $\mc R$ and are constructed from the module associator\footnote{\label{fn}In the general case, one requires the notion of \emph{($\mc D$-)module functors} between module categories, organised in a category denoted by $\Fun_\mc D(\mc R,\mc R')$. Notably, $\Fun_\mc D(\mc R,\mc R')$ can be endowed with the structure of an invertible $(\mc D_\mc R^\star,\mc D_{\mc R'}^\star)$-bimodule category.}. A change in module category hints at an equivalence between edge Hamiltonians for each of these distinct PEPS representations. Later, in \cref{sec:dualities}, we will see that the edge Hamiltonians corresponding to different representations $\mc R,\mc R'$ are related via \emph{duality transformations}, or -- equivalently -- via gauging (part of) its MPO symmetries. The aforementioned MPO intertwiners play the role of intertwiners between these dual models and can be promoted to isometries, if boundary conditions are adequately accounted for.

Furthermore, the tensor network picture can also be used to explicitly construct \emph{gapped boundaries} of string-net models. This is beyond the scope of these lecture notes and we refer to ref.~\cite{Lootens2020} for details, but let us mention that also in that case module categories show up as labelling different boundary conditions. In conclusion, the complete picture of string-net domain walls and boundaries, as developed by Kitaev and Kong in ref.~\cite{Kitaev2011}, can be recast in tensor network language, making these structures amenable to simulating them on a computer.
\subsection{Application 2: classical partition functions as strange correlators}\label{sec:SC}
With the tensor network representations of string-net ground states at our disposal, we can now discuss a second application of the general MPO formalism, the so-called \emph{strange correlator} construction. This construction appeared first in ref.~\cite{You2013} as a diagnosis to detect non-trivial SPT phases by considering the matrix element $C(\vec r,\vec r') = \mel{\Omega}{\phi(\vec r)\phi(\vec r')}{\Psi}$ where $\ket{\Psi}$ is the state under scrutiny, $\phi$ is a local observable, and $\ket{\Omega}$ denotes a well-chosen short-range entangled trivial state. It was shown for a number of one and two-dimensional states $\ket{\Psi}$ which exhibit SPT order that the quantity $C(\vec r,\vec r')$ either decays algebraically in the distance $|\vec r -\vec r'|$ or converges to a constant. Hence, this method provides an identification of SPT order in terms of the bulk ground state wave function.

A natural generalisation of this construction amounts to choosing a state $\ket{\Psi}$ which exhibits intrinsic topological order. Focussing on the case of non-chiral topological order in two dimensions, we can thus construct strange correlators from the overlap of string-net ground states of \cref{sec:SN} with a suitable state $\ket{\Omega}$. This approach was pioneered in ref.~\cite{Vanhove2018}. As we have seen above, the PEPS representations of string-net ground states have exact MPO symmetries acting purely on the virtual level. Consequently, strange correlators constructed from these states inherit these symmetries, regardless of the chosen state $\ket{\Omega}$. However, for many interesting examples, $\ket{\Omega}$ can simply be chosen as a product state. It is then natural to interpret this two-dimensional tensor network as the \emph{partition function} of a \emph{classical statistical model}\footnote{Note however, that depending on the chosen $\ket{\Omega}$ the classical model may have non-positive Boltzmann weights and requiring these to be positive generically requires fine-tuning the state $\ket{\Omega}$.} with topological defect lines/symmetries. Schematically for the case of $\ket{\Omega}$ being a (uniform) product state, the partition function thus takes the form
\begin{equation}
    Z(\{\beta_i\}) := \braket{\Omega[\{\beta_i\}]}{\Psi}= \,%
	\input{figs/SC_1.tex}%
= \,%
	\input{figs/SC_2.tex}%
\,\, ,
\end{equation}
where $\{\beta_i\}_i$ collectively denotes the parameters defining the state $\ket{\Omega}$, and thus indirectly determine the Boltzmann weights of the classical model. For appropriate values of the parameters $\{\beta_i\}_i$, the partition function can in fact describe a critical model. In that case, the MPO symmetries can be interpreted as lattice remnants of the continuum topological defects of the CFT. The strange correlator construction thus provides an explicit lattice representation of the correspondence between conformal field theories and topological field theories as envisioned first by Witten~\cite{Witten1988}, Moore and Seiberg~\cite{Moore1988} and subsequently developed further by Fr\"ohlich, Fuchs, Runkel and Schweigert in refs.~\cite{Frohlich2004,Frohlich2006,Fuchs2002,Fuchs2003,Fuchs2004}.

\paragraph{Example: hard hexagon model} Perhaps the simplest example is obtained from the string-net model based on the Fibonacci fusion category on the hexagonal lattice. This fusion category contains two simple objects denoted by $\{\mbb 1,\tau\}$, and the only non-trivial fusion rule reads $\tau\otimes\tau=\mbb 1\oplus\tau$. The strange correlator we will consider is a product state that projects all the physical degrees of freedom of the string-net onto the object $\tau$ -- or more precisely, in the notation of \cref{eq:TripleLineTensors}, to $\phi\cdot(\tau\tau\tau, 1)$, where $\phi=\frac{1+\sqrt 5}{2}$ is the golden ratio. The degrees of freedom on the loops then take on the role of classical binary variables prone to the constraint that no two neighbouring loops can be in the $\mbb 1$ state. It is suggestive to interpret this model as a classical gas of particles on the plaquettes such that a loop labelled by $\mbb 1$ signals the presence of a particle on the corresponding plaquette, and $\tau$ denotes its absence. This so-called \emph{hard hexagon} model was solved exactly in a tour de force by Baxter in 1980 and demonstrated to exhibit two phases~\cite{Baxter1980}. For a large value of the fugacity the model exhibits order characterised by a full occupancy of either of the three sublattices of the plaquettes of the hexagonal lattice, while for a low fugacity the gas is dilute and homogenous. These phases are separated by a second order phase transition for a critical value $z_c$ of the fugacity, whose scaling limit is described by the $c=4/5$ minimal Potts model~\cite{DiFrancesco1997}.

Focussing again on the strange correlator, the tensor from which the partition function is constructed has the following non-zero components:
\begin{equation}
    %
	\input{figs/SCFib_1.tex}%
 = -1, \qquad %
	\input{figs/SCFib_2.tex}%
 = \phi^{5/6}.
\end{equation}
With the above tensor at hand, it is then straightforward to derive that the partition function can be written as
\begin{equation}
    Z = \sum_{n=0}^{N/3} z_c^n \ g(n,N),
\end{equation}
where $N$ denotes the total number of plaquettes and $g(n,N)$ is a combinatorial factor equal to the number of ways to distribute $n$ particles over $N$ plaquettes subject to the aforementioned constraint~\cite{Fidkowski2009}. Straightforward algebra reveals that $z_c=\phi^{5}$, which is exactly the critical fugacity of the hard hexagon model. This simple strange correlator thus immediately yields the critical fugacity, without having to resort to the vastly more advanced techniques employed by Baxter.

From the transfer matrix one can also extract the finite-size CFT spectra. Above, we have highlighted already the significance of the role of the tube algebra to obtain the topological superselection sectors of quantum doubles and string-net models. As detailed in refs.~\cite{Petkova2000,Aasen2020,Vanhove2018} the transfer matrix of the strange correlator can be supplemented with symmetry-twisted boundary conditions labelled by objects of the symmetry fusion category. These boundary conditions are topological in the sense that they preserve lattice translation symmetry up to a unitary transformation. In their presence, the tube algebra can be used to decompose the spectrum of the transfer matrix in topological sectors. For the case of Fibonacci, there are 4 sectors labelled $0,\tau,\bar\tau,\tau\bar\tau$. Their corresponding spectra are shown in ref.~\cite{Vanhove2018}, where also the identification with the primaries of the Potts minimal model is explained.

\bigskip\noindent In the previous section we argued that different tensor network representations of string-net ground states correspond to module categories over the input fusion category. Coincidentally, we did not have this freedom for the case study in this section. It turns out that this choice of module category specifies the \emph{modular invariant} of the CFT on the torus as well as its partition function on higher genus surfaces~\cite{Moore1988,DiFrancesco1997,Fuchs2002}. The so-called \emph{diagonal case} corresponds to choosing the regular module category. Changing this module category is then referred to as \emph{orbifolding} or \emph{gauging}, numerically studied in ref.~\cite{Vanhove2021a}.

\bigskip\noindent There exists a vast literature on tensor network representations of critical partition functions using the strange correlator formalism. We refer to refs.~\cite{Freedman2011,Vanhove2018,Lootens2019,Brehm2021,Vanhove2021a,Vanhove2021b,Zeng2022,Ji2024,Shen2025,Hung2025} for additional case studies, some analytical results, and spectra on open boundary conditions and unoriented manifolds. Also, in recent work, systematic ways of constructing critical (integrable) strange correlators from purely topological data have been proposed~\cite{Fendley2020,Hung2025}. The appeal is clear: the strange correlator formalism provides a direct way of constructing critical lattice systems with exact categorical defects. The topological part (encoded in the string-net) is completely separated from the geometrical part (encoded in the overlap state $\ket{\Omega}$), thereby making the anomalies in the theory very explicit. One of the central challenges is now to construct critical lattice models corresponding to new CFTs -- the Haagerup critical lattice model \cite{Huang2021,Vanhove2021b,Hung2025} being the first possible example of such a theory.
\subsection{Application 3: dualities}\label{sec:dualities}
In the previous section we constructed two-dimensional classical partition functions with the MPO symmetries classified and constructed in \cref{sec:MPORepresentationTheory}. In this final application we will construct (1+1)D Hamiltonians commuting with those symmetries, and study their dualities. In a sense, these models can thus be viewed as \emph{edge Hamiltonians} of the string-nets of \cref{sec:SN}.

\bigskip\noindent For the case of on-site representations of groups, \emph{Clebsch-Gordan coefficients} are the building blocks of symmetric tensors. Indeed, the \emph{Wigner-Eckart theorem} dictates that symmetric tensors factorise as a product of Clebsch-Gordan coefficients and \emph{reduced matrix elements}. The latter notably only depend on the irrep labels, while the coupling of basis states is fully captured by the Clebsch-Gordan coefficients~\cite{Eckart1930,Wigner1960}. This result is not only central in the context of perturbation theory and spectroscopy in which they are often introduced, but also forms the cornerstone of symmetric tensor network algorithms~\cite{Sanz2009,Singh2009,Weichselbaum2012,Singh2013,Lootens2024,Devos2025}. Inspired by this perspective, we can deduce from \cref{eq:IntertwinerVertical} for the more general case of an MPO symmetry $\mc C$, that generic local operators are constructed from the fusion tensors as follows:
\begin{equation}\label{eq:LocalOp}
    h_{\msf i,\msf i+1}^\mc R = h(\alpha,\alpha',\beta,\beta',\gamma,i,i')%
	\input{figs/MPOLocalOp.tex}%
.
\end{equation}
Herein, the $h$ are complex numbers which can be chosen freely and may depend on the indicated labels. Importantly, these coefficients are independent of the internal virtual indices of the tensors, and can thus be interpreted as a generalisation of reduced matrix elements. Commutation of these operators with the MPO symmetries follows from
\begin{equation}
    %
	\input{figs/MPOLocalOp_1.tex}%
 =
    %
	\input{figs/MPOLocalOp_2.tex}%
 = 
    %
	\input{figs/MPOLocalOp_3.tex}%
 \,\, ,
\end{equation}
satisfied for all indicated simple objects of $\mc D$ and $\mc C$. From \cref{eq:LocalOp}, a local symmetric Hamiltonian can then be constructed as $H=\sum_{\msf i} h_{\msf i,\msf i+1}^\mc R$. Note that we restrict here to two-body operators, but symmetric operators acting on more sites can be obtained just as easily by taking products of~\cref{eq:LocalOp}. The construction of local operators \cref{eq:LocalOp} thus generalises the traditional Wigner-Eckart theorem to MPO symmetries~\cite{Bridgeman2022}.

So far, we have implicitly assumed infinite chains. Completely in line with \cref{sec:SC} we can also define these Hamiltonians in the presence of closed periodic and symmetry-twisted boundary conditions~\cite{Petkova2000,Aasen2020,Lootens2022}. For the latter, an extra degree of freedom is introduced on the bond between two sites, and the local Hamiltonian term across that bond is modified accordingly, we refer to ref.~\cite{Lootens2022} for details. The importance of including symmetry-twisted boundary conditions will become clear below.

\bigskip\noindent To illustrate the expression \cref{eq:LocalOp}, consider the spin 1/2 antiferromagnetic Heisenberg model encountered above in \cref{eq:Heisenberg}. By invoking the (traditional) Wigner-Eckart theorem, the local Hamiltonian term $\vec{S}_\msf i \cdot \vec{S}_{\msf i+1}$ can be constructed from the ${\rm SU}(2)$ Clebsch-Gordan coefficients. To rephrase this in the current framework, we need to choose $\mc D=\Rep({\rm SU}(2))$, the category of representations of ${\rm SU}(2)$, with $\mc R=\Vect$.\footnote{Technically speaking $\Rep({\rm SU}(2))$ is a tensor category, rather than a fusion category which is required to have a finite number of simple objects. Readers inclined to more mathematical rigour may wish to apply the formalism to a finite subgroup of ${\rm SU}(2)$ instead.} Spelled out explicitly:
\begin{equation}
    \vec{S}_\msf i \cdot \vec{S}_{\msf i+1} = -%
	\input{figs/MPOLocalHamSU2.tex}%
 = \sum_{\substack{i,j\\i'\!,j'}} \CC{{\bf 2}}{{\bf 2}}{{\bf 0}}{i'}{j'}{1} \CC{{\bf 2}}{{\bf 2}}{{\bf 0}}{i}{j}{1} \,\, |i',j'\rangle\langle i,j|,
\end{equation}
up to an additive constant. Here, ${\bf 2}$ and ${\bf 0}$ denote the ${\rm SU}(2)$ doublet and singlet respectively, and $\CC{{\bf 2}}{{\bf 2}}{{\bf 0}}{\bullet}{\bullet}{1}$ stands for the (real) Clebsch-Gordan coefficients associated to
\begin{equation}
    \CC{{\bf 2}}{{\bf 2}}{{\bf 0}}{\bullet}{\bullet}{1} : {\bf 2} \otimes {\bf 2} \rightarrow {\bf 0}: \ket{i}\otimes\ket{j} = \CC{{\bf 2}}{{\bf 2}}{{\bf 0}}{i}{j}{1} \ket{1},
\end{equation}
where $i,j,1$ denote basis vectors of the respective underlying representation spaces. This Hamiltonian term thus essentially projects two neighbouring spin 1/2's on the singlet representation, and the spin 1 representation appearing in the tensor product $\frac{1}{2}\otimes\frac{1}{2}$ is given zero weight by the chosen parameters $h$.

\bigskip\noindent Let us consider the set of all local symmetric operators \eqref{eq:LocalOp}, as well as their finite products and linear combinations. This constitutes an algebra -- sometimes referred to as \emph{bond algebra}~\cite{Cobanera2009,Cobanera2011,Moradi2022,Moradi2023,Jones2024} -- specified by the structure constants
\begin{equation}
    h_A^\mc R h_B^\mc R  = \sum_C f_{AB}^C(\dF) h_C^\mc R ,
\end{equation}
defined for the basis of operators $\{h_A^\mc R\}$. Crucially, the \emph{structure constants} $f_{AB}^C(\dF)$ depend on the associators $\dF$ of the fusion category $\mathcal{D}$ only, i.e. they are completely independent of the choice of module category $\mathcal{R}$. In particular, this observation implies that also the spectrum of the Hamiltonian is -- up to degeneracies -- independent of $\mc R$. For a fixed $\mc D$ and choice of coefficients $h$ in \cref{eq:LocalOp}, changing the module category $\mc R\rightarrow\mc R'$ boils down to a \emph{duality transformation}. As such, different module categories over $\mc D$, or equivalently the symmetry category $\mc C$, thus organise an entire web of dualities. Although this framework may appear abstract, most dualities of physical interest, including Kramers–Wannier duality, Kennedy–Tasaki duality, and the Jordan–Wigner transformation, naturally fit within it, as will be illustrated further below. Throughout, we will often refer to the (fictional) model of interest as the \emph{original} model, and compare it to its \emph{dual(s)} model(s). The dual model can look radically different from the original one in the following ways.

As mentioned above, the local symmetric operators \cref{eq:LocalOp} and their MPO symmetries act on a Hilbert space which is generically constrained if $\mc R$ has more than one simple object. As a consequence, the Hilbert space of one model is typically completely different from that of any of its dual models and their dimensions don't even have to match. This is closely related to the fact that dualities of the kind considered here can be realised by gauging (part of) the original MPO symmetry, so that the constrained Hilbert space arises as an \emph{effective} Hilbert space where (part of) the imposed Gauss constraints are resolved~\cite{Haegeman2014a,Bhardwaj2017,Seifnashri2025,Vancraeynest2025b}. Consequently, also the global symmetries of distinct dual models typically differ as they are encoded in $\mc D^\star_\mc R$ and $\mc D^\star_{\mc R'}$ respectively. In other words, given the initial symmetry and the duality transformation, the dual symmetry is fully determined. The existence of a duality between models implies also that every observable in one theory is equivalent to a \emph{dual} observable in the dual theory. However, only the \emph{symmetric}, local observables in one theory get mapped to dual local, symmetric operators in the dual theory, whereas charged operators, such as local order parameters, get mapped to non-local \emph{string order parameters}.

One of the upshots of the tensors network approach to dualities is that we can build explicit \emph{intertwiners} between dual models in terms of (non-invertible) matrix product operators~\cite{Lootens2021b}. At the level of the Hamiltonian terms they act according to:
\begin{equation}\label{eq:DualInter}
    h_{\msf i,\msf i+1}^{\mc R'} \circ D = D\circ h^{\mc R}_{\msf i,\msf i+1},
\end{equation}
from which it follows that acting with $D$ on an eigenstate of $H^\mc R$ results in an eigenstate of $H^{\mc R'}$ with exactly the same energy.

As demonstrated in ref.~\cite{Lootens2021b}, such an intertwiner only depends on the categories $\mc D,\mc R$ and $\mc R'$ and not on the specific model under scrutiny. In fact, these MPO intertwiners are precisely the MPO intertwiners mentioned in \cref{sec:SN} that transmute different PEPS representations of a $\mc D$ string-net model associated to the module categories $\mc R,\mc R'$ into each other. 

Another implication of \cref{eq:DualInter} is that given an intertwiner $D$, we can multiply $D$ from the \emph{right} with symmetry operators of the initial symmetry $\mc C=\mc D^\star_\mc R$ and from the \emph{left} with the dual symmetry operators in $\mc D^\star_\mc R$. As such, the different intertwiners are themselves organised in an invertible bimodule category -- that we will denote by $\mc M$ -- between the original and dual symmetry categories. The construction is consistent since $\mc D^\star_{\mc R'} =\mc C^\star_\mc M$.\footnote{This follows from the observation in \cref{fn} that duality operators are labelled by $\mc D$-module functors organised in $\Fun_\mc D(\mc R,\mc R')$.}

For a pair of dual models, we can collect these structures in a diagram as follows:
\begin{equation}
    %
	\input{figs/TriangleDiagram_1.tex}%
.
\end{equation}
In this diagram, every vertex stands for a fusion category. The one on top -- $\mc D$ -- encodes the bond algebra generated by local symmetric operators, whereas the other two -- $\mc D_\mc R^\star$ and $\mc D_{\mc R'}^\star$ -- encode the global symmetries. The connecting edges represent invertible bimodule categories. The two on the diagonals correspond to $\mc R$ and $\mc R'$, encoding the degrees of freedom of either model. The one on the bottom -- $\mc M$ -- encodes the MPO intertwiners.

Invertibility of $\mc R$ and $\mc R'$ guarantees in particular that the Drinfeld centres of the original symmetry $\mc D_\mc R^\star$ and the dual symmetry $\mc D_{\mc R'}^\star$ are equivalent. In this context, objects in the Drinfeld centre label \emph{topological sectors} of the theory. These sectors correspond to a boundary condition -- which breaks part of the symmetry -- together with a symmetry charge of the surviving symmetry. By carefully studying the interplay of the duality operators with these sectors, the non-invertible duality operators can be lifted to unitaries~\cite{Lootens2022}. Under a duality transformation the topological sectors are generically permuted, as exemplified in the Kramers-Wannier duality.

\paragraph{Example: Kramers-Wannier duality} Let us demonstrate how this formalism reproduces the famous Kramers-Wannier (KW) duality of the transverse field Ising model~\cref{eq:TFIM}~\cite{Kramers1941}. To this end, consider the Hamiltonian
\begin{equation}\label{eq:TFIM_per}
    H_{\rm TFIM}^{\Vect_{\mbb Z_2}} = -\sum_{\msf i=1}^{L} \sigma^Z_{\msf i}\sigma^Z_{\msf i+1} - \lambda\sum_{\msf i=1}^L \sigma^X_{\msf i},
\end{equation}
on periodic boundary conditions, i.e. $L+1\equiv 1$. The Hamiltonian clearly commutes with a $\mbb Z_2$ symmetry generated by $\prod_\msf i \sigma^X_\msf i$. The bond algebra corresponds to the fusion category $\mc D=\Vect_{\mbb Z_2}$ and the module category is $\mc R=\Vect_{\mbb Z_2}$. The Kramers-Wannier duality can be thought of as gauging the $\mbb Z_2$ symmetry, or thus more abstractly, changing the module category $\mc R=\Vect_{\mbb Z_2}\rightarrow\mc R'=\Vect$, corresponding to $\mc M=\Vect$. The (unique) duality MPO can be depicted as:
\begin{equation}\label{eq:KW_MPO_per}
    D_{\rm KW} = %
	\input{figs/KW_MPO_per.tex}%
\,\,,
\end{equation}
where we have explicitly indicated the periodic boundary conditions with the dotted lines, and here and below depict duality MPOs in {\color{BlueViolet}blue}. Note that the physical legs on either side are shifted by half a site. We emphasise again that this duality MPO is the same for every $\mbb Z_2$ symmetric model, and as such is not specific to the Ising model.

The MPO tensors are GHZ tensors in the $\sigma^Z$ and $\sigma^X$ eigenbasis and explicitly read 
\begin{equation}
    \raisebox{5pt}{%
	\input{figs/KW_tensor_1_a.tex}%
} = \ket{000} + \ket{111},\qquad \raisebox{5pt}{%
	\input{figs/KW_tensor_2.tex}%
} = \ket{+\!+\!+} + \ket{-\!-\!-}\,\,.
\end{equation}
All (global) properties of the intertwiner can be deduced from the symmetry properties of these local tensors. In particular, note that the first tensor is invariant under acting on any two legs with a $\sigma^Z$, and acting with $\sigma^X$ on all its legs simultaneously:
\begin{equation}
\begin{gathered}
    \raisebox{5pt}{%
	\input{figs/KW_tensor_1_a.tex}%
} = \raisebox{7pt}{%
	\input{figs/KW_tensor_1_c.tex}%
} =\\  %
	\input{figs/KW_tensor_1_e.tex}%
 = %
	\input{figs/KW_tensor_1_d.tex}%
 = %
	\input{figs/KW_tensor_1_b.tex}%

\end{gathered}\,\,.
\end{equation}
The second tensor has the same symmetries with $\sigma^Z$ and $\sigma^X$ interchanged. Hence, it follows that this MPO maps the local Hamiltonian terms according to
\begin{equation}\label{eq:KW_mapping}
    \sigma^Z_{\msf i}\sigma^Z_{\msf i+1} \rightarrow \sigma^Z_{\msf i+\frac{1}{2}}, \quad
    \sigma_\msf i^X \rightarrow \sigma^X_{\msf i-\frac{1}{2}}\sigma^X_{\msf i+\frac{1}{2}},
\end{equation}
so that the dual Hamiltonian reads
\begin{equation}\label{eq:TFIM_dual}
    H_{\rm TFIM}^{\Vect} = -\sum_{\msf i=1}^{L} \sigma^Z_{\msf i+\frac{1}{2}} - \lambda\sum_{\msf i=1}^L \sigma^X_{\msf i-\frac{1}{2}}\sigma^X_{\msf i+\frac{1}{2}}.
\end{equation}
This Hamiltonian has a symmetry generated by $\prod_\msf i \sigma^Z_{\msf i+1/2}$, which corresponds to the symmetry category $\Rep(\mbb Z_2)=(\Vect_{\mbb Z_2})_\Vect^\star$, in accordance with the prediction of the framework. The mapping \cref{eq:KW_mapping} also shows that the order parameter in the original model gets mapped to the disorder parameter in the dual and vice versa:
\begin{equation}
    \sigma^Z_\msf i \sigma^Z_\msf j \rightarrow \prod_{\msf k=\msf i}^{\msf j-1}\sigma^Z_{\msf k+\frac{1}{2}},\quad  \prod_{\msf k=\msf i}^{\msf j} \sigma^X_\msf k \rightarrow \sigma^X_{\msf i-\frac{1}{2}} \sigma^X_{\msf j+\frac{1}{2}}.
\end{equation}

The MPO \cref{eq:KW_MPO_per} has a null space which contains all $\mbb Z_2$ odd states: $\prod_\msf i\sigma_\msf i^X\ket{\Psi}=-\ket{\Psi}$. Similarly, its image consists of only states which are even under the dual symmetry. In order to access the non-trivial charge sectors in either model, we need to adapt the MPO to account for all symmetry-twisted boundary conditions. For the case at hand, this amounts to either periodic or anti-periodic boundary conditions. At the level of the Hamiltonian, changing the boundary condition boils down to flipping the sign of one of the nearest-neighbour interaction terms, conventionally the one between sites $L$ and $1$.

After having specified the boundary condition and the charge sector of the original model, the topological sector of the dual model is specified by the mapping in \cref{tab:KW}~\cite{BenTov2014,Radicevic2018,Li2023,Lootens2022}. Equivalently, after specifying the boundary conditions in both models, the accessible charge sectors are fully determined. Remark in particular the permutation of topological sectors: in essence the boundary condition and charge sector are interchanged under KW duality. For each of these sectors there is a corresponding MPO intertwiner of the form
\begin{equation}
    %
	\input{figs/KW_MPO_bc.tex}%
\,\,,
\end{equation}
where $\alpha$ is a placeholder for one of matrices $\{\mbb 1,\sigma^X,\sigma^Y,\sigma^Z\}$ as indicated in \cref{tab:KW} as well.
\begin{table}[htb]
    \centering
    \begin{tabular}{cc||cc||c}
        \shortstack{Boundary\\condition} & \shortstack{Charge\\sector} & \shortstack{Dual boundary\\condition} & \shortstack{Dual charge\\sector} & \shortstack{Boundary\\matrix} \\
        \hline
        P & $+1$ & P & $+1$ & $\mbb 1$\\
        P & $-1$& AP & $+1$ & $\sigma^Z$\\
        AP & $+1$ & P & $-1$ & $\sigma^X$\\
        AP & $-1$ & AP & $-1$ & $\sigma^Y$
    \end{tabular}.
    \caption{Mapping of topological sectors under the Kramers-Wannier duality. Here, \emph{(A)P} stands for \emph{(anti)periodic}.}
    \label{tab:KW}
\end{table}

\bigskip\noindent Note that the original model \cref{eq:TFIM_per} is in the symmetry-broken phase if $0\leq \lambda <1$ and in the symmetric phase if $\lambda>1$; under Kramers-Wannier duality these phases are interchanged in the dual model. Turning the logic around, for any value of the magnetic field strength $\lambda$, there is a model, either \cref{eq:TFIM_per} or \cref{eq:TFIM_dual}, that is (fully) symmetry-broken. We could conjecture that this observation holds generally: every phase can be mapped to a completely symmetry-broken phase -- with respect to the dual symmetry -- by means of a suitable duality mapping. Further evidence for this speculation is provided by the example of the Kennedy-Tasaki duality.
\paragraph{Example: Kennedy-Tasaki duality} The \emph{Kennedy-Tasaki} (KT) duality was originally devised to elucidate the ground state physics of the spin 1 antiferromagnetic Heisenberg model, which is now understood to exhibit non-trivial $\ZtZt$ SPT order. The duality was introduced as a non-unitary mapping on open chains between the Heisenberg model and a model that spontaneously breaks $\ZtZt$ symmetry~\cite{Kennedy1992,Oshikawa1992}. Recently, it has been reinterpreted in terms of twisted gauging of the $\ZtZt$ symmetry~\cite{Lootens2021b,Li2023,ParayilMana2024,Seifnashri2024,Vancraeynest2025a}. Equivalently, this is described as the change in module category $\mc R=\Vect\rightarrow\mc R'=\Vect^\psi$ over the input $\Vect_{\ZtZt}$, where $\psi$ is a non-trivial 2-cocycle of $\ZtZt$. As such, it is more precise to think about the symmetry as being labelled by characters in $\Rep(\ZtZt)$, which is equivalent to $\Vect_{\ZtZt}$ as a fusion category. It can be shown that $\mc M=\Rep^\psi(\ZtZt)$ so that given a $\psi$-projective representation $\pi$ of $\ZtZt$, the MPO is constructed from the tensor with components:
\begin{equation}\label{eq:KT_MPO}
    %
	\input{figs/KT_MPO.tex}%
 = \delta_{g,h} \pi(g)_{\alpha\beta}.
\end{equation}
For long it was unclear how to generalise KT from open to closed chains. The MPO framework provides a crisp answer: just take the trace of the MPO to define it on periodic boundary conditions. Also, the original KT transformation on the open chain is obtained by choosing suitable boundary conditions for the MPO \cref{eq:KT_MPO}. Note that the MPO tensors are diagonal in their physical leg, so that KT doesn't change the charge sector, but only permutes the boundary conditions.

Instead of applying it to the Heisenberg model, we focus here on the AKLT state \cref{eq:AKLTState}, also a non-trivial $\ZtZt$ SPT. Recall that the AKLT state can be represented as an MPS with matrices $A^i = \sigma^i/\sqrt 3$. If we take for concreteness the projective representation defined via $\pi((a,b))=(\sigma^X)^a(\sigma^Z)^b$, we obtain that acting with the duality on the AKLT state gives an MPS with matrices $A^i=\sigma^i\otimes\sigma^i$. This MPS is clearly reducible as all MPS matrices commute. Indeed, by making use the gauge transformation
\begin{equation}
    U = \frac{1}{\sqrt 2}
    \begin{pmatrix}
    1 & 0 & 0 & 1\\
    0 & 1 & 1 & 0 \\
    0 & 1 & -1 & 0 \\    
    1 & 0 & 0 & -1
    \end{pmatrix},
\end{equation}
this MPS can be reduced to a sum of four product states:
\begin{equation}
\begin{split}
    U(\sigma^X\otimes\sigma^X)U^\dagger &= {\rm diag}(1,1,-1,-1), \\
    U(\sigma^Y\otimes\sigma^Y)U^\dagger &= {\rm diag}(-1,1,-1,1), \\
    U(\sigma^Z\otimes\sigma^Z)U^\dagger &= {\rm diag}(1,-1,-1,1).
\end{split}
\end{equation}
This illustrates that the duality operator is fully responsible for the twofold degenerate entanglement spectrum of the AKLT state. Turning things around, by acting with the duality on the AKLT state -- or any other $\ZtZt$ SPT for that matter -- we can get rid of the degeneracies of the entanglement spectra. This observation can be used to our advantage as argued below.

In analogy with the example of KW, one can now define four symmetry-twisted boundary conditions, one for each element of $\ZtZt$. We relegate the reader to refs.~\cite{Li2023,Vancraeynest2025a} for a detailed discussion. As shown there, the charge sectors are mapped identically as anticipated, but depending on the charge sector the boundary conditions are permuted into each other.

\bigskip\noindent The observation that there seemingly always exists a unique duality transformation that maps a model to a completely symmetry-broken dual model suggests a one-to-one correspondence between module categories and categorically symmetric gapped phases of matter. And indeed, this is exactly the content of the \emph{generalised Landau paradigm}. Provided the symmetry $\mc C$, the gapped phases of matter are in one-to-one correspondence with $\mc C$-module categories $\mc P$~\cite{Thorngren2019,GarreRubio2022,Lootens2024}. This module category $\mc P$ encodes both the distinct ground states or injective blocks of the ground state MPS -- corresponding to its simple objects -- as well as the action of $\mc C$ on the ground state subspace -- via the module action. By making use of the parent Hamiltonian construction, the generalised Landau paradigm was made explicit in the MPS framework in ref.~\cite{GarreRubio2022}. In this context, the regular module category $\mc P=\mc C$ represents the completely symmetry-broken phase where the ground states are in bijection with the symmetry operators. As anticipated already in lecture \ref{sec:PhasesOfMatter}, this generalises the case of a group symmetry where both the module categories and phases of matter are characterised by pairs ($H\subseteq G,\psi$), $[\psi]\in H^2(H,\rU(1))$.

Combining the generalised Landau paradigm and the duality framework has far-reaching implications. For one, note that the symmetry-broken phase is the only gapped phase that is guaranteed to exist.\footnote{For some symmetry categories such as $\mc C=\Fib$ it is even the only gapped phase.} In fact, the existence of the completely symmetric phase is equivalent to the symmetry being anomaly-free and thus fully gaugeable. The corresponding duality operator that interchanges the two then acts by effectively averaging over the symmetry-broken ground states~\cite{Lootens2024,Vancraeynest2025b}. It is fascinating that, viewed this way, any continuous phase transition can be interpreted as a transition between a fully symmetry-broken phase on one side and a (partially) symmetric phase on the other.

Notably, the symmetry-broken dual model minimises the ground state entanglement entropy. This can be exploited in the DMRG algorithm, and for this reason the symmetry-broken model can be referred to as the \emph{optimal} (dual) model. Indeed, as demonstrated in \cref{sec:SPT1d} entanglement spectra can exhibit degeneracies, due to the fact that the entanglement degrees of freedom transform in multiplets of the (projective) virtual symmetry. If this is the case, there is redundancy in the MPS description of the ground state: computational resources are expended to fine-tune some variational parameters to reproduce these spectra. It is therefore beneficial to simulate the optimal dual model that maximally breaks the (dual) symmetry so that no degeneracies remain. In turn, the ground state of the original model is then obtained by acting with the duality operator which reinstates the degeneracies in the entanglement spectrum. A case study was carried out in ref.~\cite{Lootens2024}, and the benefit of this approach demonstrated.

Having identified the optimal dual model additionally provides a unifying perspective on the elementary excitations in any gapped phase $\mc P$. We now reserve the notation $\mc Q$ to denote the right module category over $\mc D$ corresponding to the optimal model. As argued in \cref{sec:Excitations}, the elementary excitations in this optimal and thus symmetry-broken phase $\mc D^\star_\mc Q$ are topological domain wall excitations. Recall that in the infinite-chain limit such excitations are created by acting on one half of the chain with a symmetry MPO in $\mc D^\star_\mc Q$, while its variational degrees of freedom are encoded in a tensor at its endpoint. The quasiparticle excitations in that case are organised in a fusion category, $\mc E$, which coincides with the symmetry category, i.e. $\mc E=\mc D_\mc Q^\star$. Provided this optimal model, its domain wall excitations can in turn be mapped back to the excitations of the original model, which are typically of a completely different nature, such as the magnon excitations in the Heisenberg model (see \cref{sec:magnon}). This situation is depicted as follows:
\begin{equation}\label{eq:duality_excit}
    %
	\input{figs/Duality_excit.tex}%
 \,\, .
\end{equation}
Herein, the blue MPO represents a duality operator -- coincidentally labelled by an object in $\mc P$ -- between the optimal model corresponding to $\mc Q$ and the model of interest $\mc R$ as indicated by the coupons. A domain wall excitation is then depicted on the bottom, and parametrised by the tensor $B$, similar as in \cref{sec:Excitations}. Note that this tensor carries a charge $e\in\mc E$ that terminates on the duality operator. This picture also reveals that the labels of $\mc Q$ capture the \emph{entanglement degrees of freedom} of the ground state of interest.

In this way, the domain wall excitations of the optimal dual model provide an invariant characterisation for excitations of all of its dual models. And the fundamental excitations in the any model can all be parametrised according to \cref{eq:duality_excit}.

Combined, these structures related to the quasiparticle excitations and the phase of a model are encapsulated in the following diagram, reproduced from ref.~\cite{Lootens2024}:
\begin{equation}
    %
	\input{figs/TriangleDiagram_2.tex}%
.
\end{equation}
\subsection{Synopsis of lecture V}
Lecture V can be seen as a symbiosis of the previous lectures: we encountered the fundamental theorem, MPO algebras, category theory, symmetry-breaking and string-nets models. Viewed from the point of view of tensor networks, topologically ordered systems exhibit symmetry breaking in their entanglement degrees of freedom. Those symmetries are described by MPO algebras which form representations of \emph{fusion categories}.

On the one hand, those symmetries describe the long-range entanglement structure in string-nets, which can themselves readily be expressed in terms of tensor networks. As an application, we introduced the concept of the \emph{strange correlator}. This prescription relates lattice representations of topological field theories to conformal field theories via an explicit holographic construction consisting of an overlap of a string-net with a featureless short-range entangled state. The MPO symmetries of the string-net then become the topological defects in the CFT, thereby providing an explicit lattice realisation of the Fr\"ohlich-Fuchs-Runkel-Schweigert characterisation of topological defects in CFTs.  

The same MPO algebras also describe the symmetries of the effective Hamiltonians that govern the physics of the edge modes of those 2+1D systems. In a fascinating twist, the different ways in which those symmetries can be represented -- technically described in terms of bimodule categories -- lead to a full comprehensive picture for defining dualities between all possible edge Hamiltonians in terms of a set of intertwining MPOs. One of the most fascinating physical consequences of this construction is the fact that every 1D phase of matter is dual to a phase that is completely symmetry broken (with respect to the dual symmetry). Notably, this is irrespective of whether this one-dimensional system is defined on a tensor product structure with on-site symmetries or on a constrained Hilbert space with categorical non-invertible symmetries. Among other things, this means that every possible second order phase transition for such 1D systems is equivalent to one to a completely symmetry broken phase.

In these lectures, we have not touched upon the higher-dimensional versions of MPO symmetries and MPO intertwiners. Interestingly, in higher dimensions, symmetries can act on freely-deformable submanifolds of various dimensionality of the lattice~\cite{Batista2004,Nussinov2006,Nussinov2009,Gaiotto2014,Freed2022,Lin2022,Roumpedakis2022,McGreevy2022}. These \emph{higher-form symmetries} are naturally represented in terms of tensor networks as well. In the case of (2+1)D, surface operators are implemented in terms of projected entangled-pair operators (PEPOs), while the 1-form symmetries, living on interfaces between PEPOs, are naturally described by MPOs. As a whole, this symmetry structure and its recoupling theory is encapsulated in a \emph{fusion 2-category}~\cite{Douglas2018,Decoppet2024}.  Similar to the one-dimensional case, distinct ways of representing a fusion 2-categorical symmetry on the lattice are classified by \emph{module 2-categories}~\cite{Decoppet2021,Delcamp2021}.

As pioneered in ref.~\cite{Haegeman2014a}, also dualities in 2+1D lattice systems can be implemented on the lattice by PEPOs. Based on this tensor network approach, significant progress in the understanding of higher-categorical symmetries and their dualities has been made~\cite{Delcamp2021,Delcamp2023,Inamura2023,Inamura2025,Eck2025,Vancraeynest2024,Vancraeynest2025a}. Despite these recent advances, plenty of interesting open research directions are still to be explored.

Further research directions include the incorporation of continuous symmetries within this categorical framework. As already touched upon in ref.~\cite{Lootens2024}, \emph{Schur-Weyl duality} is completely compatible with the MPO framework for describing dualities. It maps theories with continuous ${\rm SU}(N)$ symmetries to theories with discrete $\Rep({\rm SU}(N))$ symmetries. Since the Mermin-Wagner theorem no longer applies for the latter, the duality map can be used to circumvent it. As such, the Schur-Weyl duality maps interchange the symmetric and symmetry-broken phases, making the DMRG simulation of this dual theory much more efficient.

%% file: figs/MPOc.tex
\begin{tikzpicture}[baseline={(current bounding box.center)}]
	\def\l{1.6};
    \def\h{.65};

	\draw[obj2] (-.5*\l,0) -- (2.5*\l,0);
	
	\foreach \x in {0,1,2}{
		\draw[obj1] (\x*\l,\h) -- (\x*\l,-\h);
		\node[mpotensor] at (\x*\l,0) {};
	}
	\node[above] at (1.5*\l,0) {${\sss c}$};
\end{tikzpicture}

%% file: figs/MPO_single_labels.tex
\begin{tikzpicture}[baseline={([yshift=-.5ex]current bounding box.center)}]
	\def\l{.8};
    \def\h{.8};

	\draw[obj2] (-\l,0) node[text=black,anchor=east] {$\sss (eab,i)$} -- (\l,0) node[text=black,anchor=west] {$\sss (daf\!,l) $};
    \draw[obj1] (0,\h) node[anchor=south] {$\sss (ecd,j)$} -- (0,-\h) node[anchor=north] {$\sss (bcf\!,k)$};
    \node[mpotensor] at (0,0) {};
\end{tikzpicture}

%% file: figs/KW_tensor_1_a.tex
\begin{tikzpicture}[baseline=(current bounding box.center)]
	\def\l{.8};
    \def\h{.65};

    \draw[obj3] (-\l,0) -- (\l,0);
    \draw[obj1] (0,0) -- (0,-\h);
    
    \draw[obj1] (0,0) -- (0,-\h);
    \node[mpotensor, inner sep=3.6pt] at (0,0) {};
\end{tikzpicture}

%% file: figs/KW_tensor_2.tex
\begin{tikzpicture}[baseline=(current bounding box.center)]
	\def\l{.8};
    \def\h{.65};

    \draw[obj3] (-\l,0) -- (\l,0);
    \draw[obj1] (0,0) -- (0,\h);
    
    \node[mpotensor, inner sep=3.6pt] at (0,0) {};
    \node at (0,0) {$+$};
\end{tikzpicture}

%% file: figs/KT_MPO.tex
\begin{tikzpicture}[baseline={([yshift=-.5ex]current bounding box.center)}]
	\def\l{.8};
    \def\h{.8};

	\draw[obj3] (-\l,0) node[text=black,anchor=east] {$\sss \alpha$} -- (\l,0) node[text=black,anchor=west] {$\sss \beta$};
    \draw[obj1] (0,\h) node[anchor=south] {$\sss h$} -- (0,-\h) node[anchor=north] {$\sss g$};
    \node[mpotensor] at (0,0) {};
\end{tikzpicture}

%% file: _App_Cohomology.tex
\section{Group cohomology}\label{sec:GroupCohomology}
In this section we explain how cohomology groups $H^n(G,U(1)^\beta)$ for any finite group $G$ with a group action $\beta$ on $\rU(1)$ can be computed using simple linear algebra methods developed in \cite{Bulmash2020,Vancraeynest2022}. As a side product, we will also obtain representative cocycles for every cohomology class.

\bigskip\noindent  Let us first introduce our notation and relevant definitions. We denote the set of all functions $G^n\rightarrow \rU(1)$ by $C^n(G,\rU(1))$, these functions are called \emph{n-cochains}. The \emph{coboundary maps} $\partial^{n+1}:C^n(G,\rU(1))\rightarrow C^{n+1}(G,\rU(1))$ are then defined via:
\begin{equation}
\begin{split}
&(\partial^{n+1}\omega)(g_1,g_2,\dots,g_{n+1}) := \beta_{g_1}(\omega(g_2,g_3,\dots,g_{n+1})) \\
&\hfill\times \prod_{i=1}^n [\omega(g_1,g_2, \dots,g_{i-1},g_ig_{i+1},\dots,g_{n+1})]^{(-1)^i} \times  [\omega(g_1,g_2,\dots,g_n)]^{(-1)^{n + 1}} = 0.
\end{split}
\end{equation}
Here, $\beta$ denotes the group action of $G$ on (the $G$-module) $\rU(1)$. In the case of a symmetry group containing \emph{antiunitary} elements, such as time-reversal and spatial reflection encountered in the main text, $\beta_g(\bullet)$ acts via complex conjugation when $g$ is antiunitary and as the identity otherwise. It is easy to check that these maps are nilpotent, i.e. $\partial^{n+1}\circ \partial^n = 1$. We then define the abelian groups of \emph{n-cocycles} and \emph{n-coboundaries} as respectively:
\begin{align}
    Z^n(G,\rU(1)^\beta) &:= \ker(\partial^{n+1}),\\
    B^n(G,\rU(1)^\beta) &:= {\rm im}(\partial^n).
\end{align}
From the nilpotency of the coboundary maps, it follows that $B^n(G,\rU(1)^\beta)$ is a subgroup of $Z^n(G,\rU(1)^\beta)$, so that we can defined the \emph{n-th cohomology group} of $G$ as the quotient
\begin{equation}
    H^n(G,\rU(1)^\beta) = Z^n(G,\rU(1)^\beta)/B^n(G,\rU(1)^\beta).
\end{equation}

\bigskip\noindent The key insight in computing $H^n(G,\rU(1)^\beta)$ and finding representative cocycles is the following. By taking the logarithm of both sides of the \emph{cocycle condition} $\partial^{n+1}\omega=1$, and collecting the unknown phases $\omega(g_1,\dots,g_n)$ in a vector $\bm{\omega}$, the condition can be written as a linear system  that has to be satisfied modulo $2\pi$:
\begin{equation}
	\Omega \bm{\omega} = \vec{0}, \mod 2\pi.
	\label{eq:System}
\end{equation}
Here, the coefficient matrix $\Omega$ originates from the coboundary map $\partial^{n+1}$ and contains only integers. The dimension of $\bm{\omega}$ is $|G|^n$ and $\Omega$ is $|G|^{n+1}\times|G|^n$-dimensional. Since $\Omega$ only contains integers, we can write $\Omega$ in its \emph{Smith normal form}:
\begin{equation}
	\Omega = P\Lambda R.
\end{equation}
In this decomposition, $P$ and $R$ are respectively $\left|G\right|^{n+1}\times \left|G\right|^{n+1}$ and $\left|G\right|^n\times \left|G\right|^n$ matrices that only contain integers and have determinant one (and thus have integer-valued inverses). $\Lambda$ also only contains integers, is $\left|G\right|^{n+1}\times \left|G\right|^n$-dimensional and is of the form
\begin{equation}
	\Lambda
	=
	\begin{pmatrix}
		\begin{array}{c|c}
			\text{diag}\left(d_1,d_2,...,d_r\right) & 0_{r,n-r} \\
		      \hline
		      0_{n+1-r,r} & 0_{n+1-r,n-r}
		\end{array}
	\end{pmatrix}
	,
\end{equation}
in which the non-zero elements $d_1,...,d_r$ along the diagonal, some of which might be one, are in increasing order, $d_1\leq d_2\leq...$, and every element is a divisor of the next, $d_i | d_{i+1}$. The Smith normal form $\Lambda$ is unique. Inserting this decomposition in the system of equations (\ref{eq:System}) gives rise to the solution
\begin{equation}
	\bm{\omega} = 2\pi R^{-1}\Lambda^+\bm{\nu}.
	\label{eq:Sol2Cocyc}
\end{equation}
In this equation $\Lambda^+$ denotes the (unique) Moore-Penrose pseudoinverse of $\Lambda$ that satisfies $\Lambda\Lambda^+\Lambda = \Lambda$ and which is found to be
\begin{equation}
	\Lambda^+
	=
	\begin{pmatrix}
		\begin{array}{c|c}    \text{diag}\left(d_1^{-1},d_2^{-1},...,d_r^{-1}\right) & 0_{r,n-r} \\
		\hline
		0_{n+1-r,r} & 0_{n+1-r,n-r}
		\end{array}
	\end{pmatrix}.
\end{equation}
The vector $\bm{\nu}$ contains arbitrary integers. Writing the solution (\ref{eq:Sol2Cocyc}) in components yields
\begin{equation}
	\omega_i = \sum_{j=1}^r 2\pi\left(R^{-1}\right)_{ij}\frac{\nu_j}{d_j}.
	\label{eq:Sol2CocyBis}
\end{equation}
Because $\bm{\nu}$ can be chosen freely, one can choose subsequently $\nu_i = \delta_{1,i}, \delta_{2,i},...,\delta_{r,i}$ to obtain a basis of the solution space that can be written as
\begin{equation}
	(\bm{\omega}_j)_i = \frac{2\pi}{d_j}\left(R^{-1}\right)_{ij}, \quad \forall j\in\{1,...,r\}.
\end{equation}
Hence, the $n$-cocycles $\bm{\omega}$ are found to be the columns of $R^{-1}$. Since $R$ is full rank, all the solutions $\bm{\omega}$ are linearly independent. In particular, the non-trivial cocycles (below) can not be related by a coboundary, $\bm{\omega}'=\bm{\omega} + \Omega'\bm{\varphi}$, where $\Omega'$ denotes the coboundary map of degree $n-1$ and $\bm{\varphi}$ is a $|G|$-dimensional vector containing arbitrary real numbers. To classify all possible solutions, we now consider the diagonal entries of $\Lambda$.

From (\ref{eq:Sol2CocyBis}) and the fact that $R^{-1}$ contains only integers, it follows that for every diagonal entry $d_i=1$, a trivial solution $\omega_i=0 \mod 2\pi$ is obtained.
The non-trivial solutions are those that correspond to entries $d_i>1$. From (\ref{eq:Sol2CocyBis}) and the fact that the solution space is $\mathbb{Z}$-linear, it follows that the cocycle $\bm{\omega}_j$ corresponding to some $d_j$ generates a cyclic group. Now note that not all elements of $\bm{\omega}_j$ can be divisible by $d_j$ or any of its prime factors as this would be in contradiction with the fact that $R^{-1}$ has determinant one. Hence, the cyclic group generated by $\bm{\omega}_j$ is $\mathbb{Z}_{d_j}$.
Finally, the zero entries of $\Lambda$ can also be discarded in the cohomology as these correspond to trivial solutions of the cocycle equation that can be multiplied by arbitrary phases and thus correspond to coboundaries.

In conclusion following picture arises. Given some group $G$ one can write down the matrix representation of the coboundary map $\Omega$ that can be brought in Smith normal form $P\Lambda R$. The diagonal entries of $\Lambda$, $d_1,d_2,...,d_r$, determine the second cohomology group which is then of the form $\mathbb{Z}_{d_1}\times\mathbb{Z}_{d_2}\times ... \times \mathbb{Z}_{d_r}$, with the understanding that $\mathbb{Z}_1=\{e\}$ denotes the trivial group and that all zero diagonal entries can be discarded. The non-trivial n-cocycles in some arbitrary gauge correspond then to the columns of $R^{-1}$, weighted by the appropriate factor $2\pi/d_j$ as per \cref{eq:Sol2CocyBis}.

%% file: _App_Cat.tex
\section{Category theory}\label{sec:cat}
In this appendix we summarise relevant notions of fusion and (bi)module categories to the extent that they are used in the main text. We refer the reader to the standard references~\cite{Etingof2002,Etingof2015,Ostrik2001} and reviews~\cite{Kitaev2005a,Beer2018,Kong2022} for a more detailed and technically precise exposition.

\subsection{Fusion categories}
Succinctly, a (unitary) \emph{fusion category} $\mc D$ consists of
\begin{enumerate}
    \item A finite set of \emph{simple objects} (topological defects) $a,b,c,\dots$.
    \item Fusion rules $a\otimes b =\bigoplus_c N_{ab}^c c$, together with a unit for the tensor product denoted $\mbb 1$.
    \item Unitary maps called \emph{F-symbols} or \emph{associators} $F:(\bullet \otimes\bullet)\otimes\bullet\overset{\sim}{\rightarrow}\bullet\otimes(\bullet\otimes\bullet)$.
\end{enumerate}
The F-symbols adhere to a consistency condition, known as the \emph{pentagon equation}. Spelled out, the pentagon equation boils down to linear multivariate system of equations which schematically are of the form
\begin{equation}
    FF = \sum FFF,
\end{equation}
and written out in components read
\begin{equation}
    \sum_o \left(F^{fcd}_e\right)^{h,no}_{g,lm} \left(F^{abh}_e\right)^{i,pq}_{f,ko} = \sum_{j,rst} \left(F^{abc}_g\right)^{j,rs}_{f,kl} \left(F^{ajd}_e\right)^{i,tq}_{g,sm} \left(F^{bcd}_i\right)^{h,np}_{j,rt}.
\end{equation}
Notably, a fusion category is not necessarily endowed with a \emph{braiding}, i.e. unitary maps $R_{ab}:a\otimes b\overset{\sim}{\rightarrow}b\otimes a$. However, for every fusion category one can construct its \emph{Drinfeld centre} $\mc Z(\mc D)$, which is a \emph{modular} fusion category, meaning in particular that it is endowed with a \emph{non-degenerate} braiding. The centre can be physically realised in the string-net models of \cref{sec:SN} as the anyonic excitations of the model. Their exchange statistics are encoded in the braiding of $\mc Z(\mc D)$.
\paragraph{(Anomalous) Finite group symmetry} For any \emph{finite} group $G$, the fusion category $\Vect_G$ is the category of finite-dimensional $G$-graded vector spaces. In essence, the simple objects of $\Vect_G$ correspond to the group elements of $G$, and the fusion rules are dictated by the group multiplication. The $F$-symbols are all trivial. Whenever $G$ possesses a non-trivial third cohomology group (see \cref{sec:GroupCohomology}), any 3-cocycle $\omega$ can be used to construct $\Vect_G^\omega$. This fusion category is constructed analogous to $\Vect_G$ where now the non-zero components of the $F$-symbols evaluate to $\omega$. This cocycle encodes the \emph{('t Hooft) anomaly}, i.e. an obstruction to gauge to symmetry, or -- equivalently -- an obstruction for the symmetry to admit a trivial symmetric ground state as encountered in the main text in \cref{sec:SPT1d}.

\bigskip\noindent In this sense, a fusion category thus encodes a symmetry structure -- symmetry operators, their multiplication rules and junctions -- as well as their anomaly -- encoded in the F-symbol.
\paragraph{$\Rep(G)$ symmetry} The category $\Rep(G)$, for any finite group $G$, encodes as objects the representations of $G$. The irreducible representations then constitute the \emph{simple} objects. The fusion of those objects is then given by the usual tensor product of group representations. It can then be deduced that the $F$-symbols of $\Rep(G)$ are given by the 6j-symbols.
\paragraph{Fibonacci symmetry}
The Fibonacci fusion category has two simple objects $\{\mbb 1,\tau\}$ whose only non-trivial fusion rule reads $\tau\otimes\tau=\mbb 1\oplus\tau$. The F-symbols are spelled out in numerous places such as ref.~\cite{Beer2018}. The name \emph{Fibonacci} refers to the fact that the multiplicity of $\tau$ in subsequent tensor powers of $\tau$ generates the Fibonacci series. It can be shown that also the Fibonacci fusion category admits no local symmetric Hamiltonian with trivial symmetric ground state~\cite{GarreRubio2022}.
\subsection{Module categories}
Let $\mc C$ be a fusion category. A (unitary left) \emph{$\mc C$-module category $\mc M$} consists of
\begin{enumerate}
    \item A finite set of simple objects $A,B,C,\dots$.
    \item An \emph{action} $\act$ of $\mc C$ on $\mc M$: $a\act A =\bigoplus_{B} N_{aA}^{B}B$, for simple objects $B$ in $\mc M$.
    \item Unitary \emph{module $\mF$-symbols} or (left) \emph{module associators} $\mF:(\bullet \otimes\bullet)\act\bullet\overset{\sim}{\rightarrow}\bullet\act(\bullet\act\bullet)$.
\end{enumerate}
These $\mF$-symbols satisfy in turn a consistency equation involving both the $F$-symbols of $\mc C$ and the module $\mF$-symbols. Schematically:
\begin{equation}
    \mF\mF = \sum F\mF\mF.
\end{equation}
\paragraph{Module categories of $\Vect_G$}
Indecomposable module categories of $\Vect_G$ are labelled by pairs $(H,\psi)$, where $H\subseteq G$ is a subgroup (up to conjugacy) and $\psi$ is a 2-cocycle of $H$. In this case, the simple objects are labelled by cosets $r_1H,r_2H,\dots\in G/H$, and the action of $\Vect_G$ on them is defined as $g\act rH := (gr)H$. The module associator $\mF$ can be constructed from $\psi$ as detailed in e.g.~\cite{Naidu2006}.
\paragraph{Module categories of $\Rep(G)$} Also module categories over $\Rep(G)$ are classified by tuples $(H,\psi)$ due to the \emph{Morita equivalence} of $\Rep(G)$ and $\Vect_G$. The corresponding module category is $\Rep^\psi(H)$, the category of $\psi$-projective representations of $H$. The action of $\Rep(G)$ on $\Rep^\psi(H)$ is given via restriction of $G$-representations to the subgroup $H$ and subsequently taking the tensor product or (projective) $H$-representations.

\subsection{Bimodule categories}
Given a pair of fusion categories $\mc C,\mc D$, a \emph{$(\mc C,\mc D)$-bimodule category} $\mc M$ is a category which is both a left $\mc C$- and right $\mc D$-module category, together with additional unitary \emph{bimodule $\rF$-symbols} or  \emph{bimodule associators} $\rF:(\bullet \act \bullet)\cat\bullet \overset{\sim}{\rightarrow} \bullet\act(\bullet\cat\bullet)$.

\paragraph{$\Vect$} $\Vect$ can be interpreted as a $(\Vect_G,\Rep(G))$-bimodule category. In that case, the components of the $\rF$-symbols evaluate to the matrix elements of $G$-representations.

\bigskip\noindent Throughout this manuscript we exclusively deal with \emph{invertible} $(\mc C,\mc D)$-bimodule categories. In particular, this guarantees that $\mc C$ and $\mc D$ are \emph{Morita equivalent}, or -- equivalently -- that their Drinfeld centres $\mc Z(\mc C)$ and $\mc Z(\mc D)$ are equivalent modular fusion categories~\cite{Mueger2002}. In ref.~\cite{Bridgeman2022}, invertibility was shown to be equivalent to the notion of MPO injectivity, and a characterisation in terms of the bimodule associators derived.

Given a fusion category $\mc C$ and $\mc C$-module category $\mc M$, there is a unique fusion category, the \emph{Morita dual} $\mc C^\star_\mc M$, that renders $\mc M$ an invertible $(\mc C,\mc C^\star_\mc M)$-bimodule category. For example: $(\Vect_G)^\star_\Vect=\Rep(G)$. We refer to the relevant literature for a precise definition of the Morita dual fusion category.

%% file: main.bbl
\begin{thebibliography}{100}
\providecommand{\url}[1]{\texttt{#1}}
\providecommand{\urlprefix}{URL }
\expandafter\ifx\csname urlstyle\endcsname\relax
  \providecommand{\doi}[1]{doi:\discretionary{}{}{}#1}\else
  \providecommand{\doi}{doi:\discretionary{}{}{}\begingroup \urlstyle{rm}\Url}\fi
\providecommand{\eprint}[2][]{preprint doi:\href{#1}{#2}}

\bibitem{Verstraete2008}
F.~Verstraete, V.~Murg and J.~I. Cirac,
\newblock \emph{{Matrix product states, projected entangled pair states, and variational renormalization group methods for quantum spin systems}},
\newblock Adv. Phys. \textbf{57}(2), 143 (2008),
\newblock \doi{10.1080/14789940801912366}.

\bibitem{Schollwoeck2010}
U.~Schollwoeck,
\newblock \emph{{The density-matrix renormalization group in the age of matrix product states}},
\newblock Annals Phys. \textbf{326}, 96 (2011),
\newblock \doi{10.1016/j.aop.2010.09.012},
\newblock  \eprint[https://doi.org/10.48550/arXiv.1008.3477]{10.48550/arXiv.1008.3477}.

\bibitem{Bridgeman2016}
J.~C. Bridgeman and C.~T. Chubb,
\newblock \emph{{Hand-waving and Interpretive Dance: An Introductory Course on Tensor Networks}},
\newblock J. Phys. A \textbf{50}(22), 223001 (2017),
\newblock \doi{10.1088/1751-8121/aa6dc3},
\newblock  \eprint[https://doi.org/10.48550/arXiv.1603.03039]{10.48550/arXiv.1603.03039}.

\bibitem{Vanderstraeten2019}
L.~Vanderstraeten, J.~Haegeman and F.~Verstraete,
\newblock \emph{{Tangent-space methods for uniform matrix product states}},
\newblock SciPost Phys. Lect. Notes \textbf{7}, 1 (2019),
\newblock \doi{10.21468/SciPostPhysLectNotes.7},
\newblock  \eprint[https://doi.org/10.48550/arXiv.1810.07006]{10.48550/arXiv.1810.07006}.

\bibitem{Cirac2020}
J.~I. Cirac, D.~Perez-Garcia, N.~Schuch and F.~Verstraete,
\newblock \emph{{Matrix product states and projected entangled pair states: Concepts, symmetries, theorems}},
\newblock Rev. Mod. Phys. \textbf{93}(4), 045003 (2021),
\newblock \doi{10.1103/RevModPhys.93.045003},
\newblock  \eprint[https://doi.org/10.48550/arXiv.2011.12127]{10.48550/arXiv.2011.12127}.

\bibitem{Xiang2023}
T.~Xiang,
\newblock \emph{{Density Matrix and Tensor Network Renormalization}},
\newblock Cambridge University Press,
\newblock ISBN 978-1-009-39867-1, 978-1-009-39870-1,
\newblock \doi{10.1017/9781009398671} (2024).

\bibitem{Banuls2022}
M.~C. Ba{\~n}uls,
\newblock \emph{{Tensor Network Algorithms: A Route Map}},
\newblock Ann. Rev. Condensed Matter Phys. \textbf{14}, 173 (2023),
\newblock \doi{10.1146/annurev-conmatphys-040721-022705},
\newblock  \eprint[https://doi.org/10.48550/arXiv.2205.10345]{10.48550/arXiv.2205.10345}.

\bibitem{Verstraete2023}
F.~Verstraete, T.~Nishino, U.~Schollwöck, M.~C. Bañuls, G.~K. Chan and M.~E. Stoudenmire,
\newblock \emph{Density matrix renormalization group, 30 years on},
\newblock Nature Review Physics  (2023),
\newblock \doi{10.1038/s42254-023-00572-5}.

\bibitem{Chen2025}
X.~Chen,
\newblock \emph{{Essay: Generalized Landau Paradigm for Quantum Phases and Phase Transitions}},
\newblock Phys. Rev. Lett. \textbf{135}(25), 250001 (2025),
\newblock \doi{10.1103/tmvy-vsqd},
\newblock  \eprint[https://doi.org/10.48550/arXiv.2511.19793]{10.48550/arXiv.2511.19793}.

\bibitem{Poulin2011}
D.~Poulin, A.~Qarry, R.~Somma and F.~Verstraete,
\newblock \emph{{Quantum Simulation of Time-Dependent Hamiltonians and the Convenient Illusion of Hilbert Space}},
\newblock Phys. Rev. Lett. \textbf{106}, 170501 (2011),
\newblock \doi{10.1103/PhysRevLett.106.170501},
\newblock  \eprint[https://doi.org/10.48550/arXiv.1102.1360]{10.48550/arXiv.1102.1360}.

\bibitem{Slater1929}
J.~C. Slater,
\newblock \emph{{The Theory of Complex Spectra}},
\newblock Phys. Rev. \textbf{34}, 1293 (1929),
\newblock \doi{10.1103/PhysRev.34.1293}.

\bibitem{Slater1935}
J.~C. Slater,
\newblock \emph{The electronic structure of metals},
\newblock Rev. Mod. Phys. \textbf{6}, 209 (1934),
\newblock \doi{10.1103/RevModPhys.6.209}.

\bibitem{Hartree1947}
D.~R. Hartree,
\newblock \emph{The calculation of atomic structures},
\newblock Reports on Progress in Physics \textbf{11}(1), 113 (1947),
\newblock \doi{10.1088/0034-4885/11/1/305}.

\bibitem{GP}
F.~Dalfovo, S.~Giorgini, L.~P. Pitaevskii and S.~Stringari,
\newblock \emph{Theory of bose-einstein condensation in trapped gases},
\newblock Rev. Mod. Phys. \textbf{71}, 463 (1999),
\newblock \doi{10.1103/RevModPhys.71.463}.

\bibitem{CC}
R.~J. Bartlett and M.~Musia{\l},
\newblock \emph{Coupled-cluster theory in quantum chemistry},
\newblock Reviews of Modern Physics \textbf{79}, 291 (2007),
\newblock \doi{10.1103/RevModPhys.79.291}.

\bibitem{Kogut1979}
J.~B. Kogut,
\newblock \emph{{An Introduction to Lattice Gauge Theory and Spin Systems}},
\newblock Rev. Mod. Phys. \textbf{51}, 659 (1979),
\newblock \doi{10.1103/RevModPhys.51.659}.

\bibitem{Kogut1982}
J.~B. Kogut,
\newblock \emph{{A Review of the Lattice Gauge Theory Approach to Quantum Chromodynamics}},
\newblock Rev. Mod. Phys. \textbf{55}, 775 (1983),
\newblock \doi{10.1103/RevModPhys.55.775}.

\bibitem{Coffman1999}
V.~Coffman, J.~Kundu and W.~K. Wootters,
\newblock \emph{{Distributed entanglement}},
\newblock Phys. Rev. A \textbf{61}, 052306 (2000),
\newblock \doi{10.1103/PhysRevA.61.052306},
\newblock  \eprint[https://doi.org/10.48550/arXiv.quant-ph/9907047]{10.48550/arXiv.quant-ph/9907047}.

\bibitem{Osborne2006}
T.~J. Osborne and F.~Verstraete,
\newblock \emph{{General Monogamy Inequality for Bipartite Qubit Entanglement}},
\newblock Phys. Rev. Lett. \textbf{96}(22), 220503 (2006),
\newblock \doi{10.1103/PhysRevLett.96.220503},
\newblock  \eprint[https://doi.org/10.48550/arXiv.quant-ph/0502176]{10.48550/arXiv.quant-ph/0502176}.

\bibitem{Raggio1989}
G.~A. Raggio and R.~F. Werner,
\newblock \emph{{Quantum Statistical Mechanics of General Mean Field Systems}},
\newblock Helv. Phys. Acta \textbf{62}, 980 (1989).

\bibitem{Majumdar1969}
C.~K. Majumdar and D.~K. Ghosh,
\newblock \emph{{On Next-Nearest-Neighbor Interaction in Linear Chain. I}},
\newblock J. Math. Phys. \textbf{10}(8), 1388 (1969),
\newblock \doi{10.1063/1.1664978}.

\bibitem{Affleck1987}
I.~Affleck, T.~Kennedy, E.~H. Lieb and H.~Tasaki,
\newblock \emph{{Rigorous Results on Valence Bond Ground States in Antiferromagnets}},
\newblock Phys. Rev. Lett. \textbf{59}, 799 (1987),
\newblock \doi{10.1103/PhysRevLett.59.799}.

\bibitem{Fannes1990}
M.~Fannes, B.~Nachtergaele and R.~F. Werner,
\newblock \emph{{Finitely correlated states on quantum spin chains}},
\newblock Commun. Math. Phys. \textbf{144}, 443 (1992),
\newblock \doi{10.1007/BF02099178}.

\bibitem{Delgado2004}
F.~Verstraete, M.~A. Mart\'{\i}n-Delgado and J.~I. Cirac,
\newblock \emph{Diverging entanglement length in gapped quantum spin systems},
\newblock Phys. Rev. Lett. \textbf{92}, 087201 (2004),
\newblock \doi{10.1103/PhysRevLett.92.087201}.

\bibitem{Verstraete2006a}
F.~Verstraete and J.~I. Cirac,
\newblock \emph{{Matrix product states represent ground states faithfully}},
\newblock Phys. Rev. B \textbf{73}(9), 094423 (2006),
\newblock \doi{10.1103/PhysRevB.73.094423},
\newblock  \eprint[https://doi.org/10.48550/arXiv.cond-mat/0505140]{10.48550/arXiv.cond-mat/0505140}.

\bibitem{Hastings2007}
M.~B. Hastings,
\newblock \emph{{An area law for one-dimensional quantum systems}},
\newblock J. Stat. Mech. \textbf{0708}, P08024 (2007),
\newblock \doi{10.1088/1742-5468/2007/08/P08024},
\newblock  \eprint[https://doi.org/10.48550/arXiv.0705.2024]{10.48550/arXiv.0705.2024}.

\bibitem{Verstraete2004c}
F.~Verstraete, J.~J. Garc{\'\i}a-Ripoll and J.~I. Cirac,
\newblock \emph{{Matrix Product Density Operators: Simulation of Finite-Temperature and Dissipative Systems}},
\newblock Phys. Rev. Lett. \textbf{93}(20), 207204 (2004),
\newblock \doi{10.1103/PhysRevLett.93.207204},
\newblock  \eprint[https://doi.org/10.48550/arXiv.cond-mat/0406426]{10.48550/arXiv.cond-mat/0406426}.

\bibitem{Pirvu2010}
F.~Verstraete, J.~I. Cirac, V.~Murg and B.~Pirvu,
\newblock \emph{{Matrix product operator representations}},
\newblock New J. Phys. \textbf{12}(2), 025012 (2010),
\newblock \doi{10.1088/1367-2630/12/2/025012},
\newblock  \eprint[https://doi.org/10.48550/arXiv.0804.3976]{10.48550/arXiv.0804.3976}.

\bibitem{Haegeman2016}
J.~Haegeman and F.~Verstraete,
\newblock \emph{{Diagonalizing Transfer Matrices and Matrix Product Operators: A Medley of Exact and Computational Methods}},
\newblock Ann. Rev. Condensed Matter Phys. \textbf{8}, 355 (2017),
\newblock \doi{10.1146/annurev-conmatphys-031016-025507},
\newblock  \eprint[https://doi.org/10.48550/arXiv.1611.08519]{10.48550/arXiv.1611.08519}.

\bibitem{Haegeman2013a}
J.~Haegeman, T.~J. Osborne and F.~Verstraete,
\newblock \emph{{Post-matrix product state methods: To tangent space and beyond}},
\newblock Phys. Rev. B \textbf{88}(7), 075133 (2013),
\newblock \doi{10.1103/PhysRevB.88.075133},
\newblock  \eprint[https://doi.org/10.48550/arXiv.1305.1894]{10.48550/arXiv.1305.1894}.

\bibitem{Haegeman2014b}
J.~Haegeman, M.~Mari{\"e}n, T.~J. Osborne and F.~Verstraete,
\newblock \emph{{Geometry of matrix product states: Metric, parallel transport, and curvature}},
\newblock J. Math. Phys. \textbf{55}, 021902 (2014),
\newblock \doi{10.1063/1.4862851},
\newblock  \eprint[https://doi.org/10.48550/arXiv.1210.7710]{10.48550/arXiv.1210.7710}.

\bibitem{White1992}
S.~R. White,
\newblock \emph{Density matrix formulation for quantum renormalization groups},
\newblock Physical Review Letters \textbf{69}(19), 2863 (1992),
\newblock \doi{10.1103/PhysRevLett.69.2863}.

\bibitem{Haegeman2016b}
J.~Haegeman, C.~Lubich, I.~Oseledets, B.~Vandereycken and F.~Verstraete,
\newblock \emph{Unifying time evolution and optimization with matrix product states},
\newblock Phys. Rev. B \textbf{94}, 165116 (2016),
\newblock \doi{10.1103/PhysRevB.94.165116}.

\bibitem{Greenberger1989}
D.~M. Greenberger, M.~A. Horne and A.~Zeilinger,
\newblock \emph{{Going Beyond Bell{\textquoteright}s Theorem}},
\newblock Fundam. Theor. Phys. \textbf{37}, 69 (1989),
\newblock \doi{10.1007/978-94-017-0849-4_10},
\newblock  \eprint[https://doi.org/10.48550/arXiv.0712.0921]{10.48550/arXiv.0712.0921}.

\bibitem{PerezGarcia2006}
D.~Perez-Garcia, F.~Verstraete, M.~M. Wolf and J.~I. Cirac,
\newblock \emph{{Matrix product state representations}},
\newblock Quant. Inf. Comput. \textbf{7}(5-6), 401 (2007),
\newblock \doi{10.26421/QIC7.5-6-1},
\newblock  \eprint[https://doi.org/10.48550/arXiv.quant-ph/0608197]{10.48550/arXiv.quant-ph/0608197}.

\bibitem{Cadarso2012}
A.~Cadarso, M.~Sanz, M.~M. Wolf, J.~I. Cirac and D.~Perez-Garcia,
\newblock \emph{{Entanglement, fractional magnetization and long-range interactions}},
\newblock Phys. Rev. B \textbf{87}, 035114 (2013),
\newblock \doi{10.1103/PhysRevB.87.035114},
\newblock  \eprint[https://doi.org/10.48550/arXiv.1209.3898]{10.48550/arXiv.1209.3898}.

\bibitem{Schuch2011}
N.~Schuch, D.~P{\'e}rez-Garc{\'\i}a and I.~Cirac,
\newblock \emph{{Classifying quantum phases using matrix product states and projected entangled pair states}},
\newblock Phys. Rev. B \textbf{84}(16), 165139 (2011),
\newblock \doi{10.1103/PhysRevB.84.165139},
\newblock  \eprint[https://doi.org/10.48550/arXiv.1010.3732]{10.48550/arXiv.1010.3732}.

\bibitem{Pfeifer2014}
R.~N.~C. Pfeifer, J.~Haegeman and F.~Verstraete,
\newblock \emph{{Faster identification of optimal contraction sequences for tensor networks}},
\newblock Phys. Rev. E \textbf{90}, 033315 (2014),
\newblock \doi{10.1103/PhysRevE.90.033315},
\newblock  \eprint[https://doi.org/10.48550/arXiv.1304.6112]{10.48550/arXiv.1304.6112}.

\bibitem{Nishino1996}
T.~Nishino, K.~Okunishi and M.~Kikuchi,
\newblock \emph{Numerical renormalization group at criticality},
\newblock Physics Letters A \textbf{213}(1), 69 (1996),
\newblock \doi{10.1016/0375-9601(96)00128-4}.

\bibitem{Tagliacozzo2007}
L.~Tagliacozzo, T.~R. de~Oliveira, S.~Iblisdir and J.~I. Latorre,
\newblock \emph{{Scaling of entanglement support for Matrix Product States}},
\newblock Phys. Rev. B \textbf{78}, 024410 (2008),
\newblock \doi{10.1103/PhysRevB.78.024410},
\newblock  \eprint[https://doi.org/10.48550/arXiv.0712.1976]{10.48550/arXiv.0712.1976}.

\bibitem{Pollmann2009}
F.~Pollmann, S.~Mukerjee, A.~M. Turner and J.~E. Moore,
\newblock \emph{Theory of finite-entanglement scaling at one-dimensional quantum critical points},
\newblock Phys. Rev. Lett. \textbf{102}, 255701 (2009),
\newblock \doi{10.1103/PhysRevLett.102.255701}.

\bibitem{Pirvu2012}
B.~Pirvu, G.~Vidal, F.~Verstraete and L.~Tagliacozzo,
\newblock \emph{{Matrix product states for critical spin chains: Finite-size versus finite-entanglement scaling}},
\newblock Phys. Rev. B \textbf{86}(7), 075117 (2012),
\newblock \doi{10.1103/PhysRevB.86.075117},
\newblock  \eprint[https://doi.org/10.48550/arXiv.1204.3934]{10.48550/arXiv.1204.3934}.

\bibitem{Zauner2015}
V.~Zauner, D.~Draxler, L.~Vanderstraeten, M.~Degroote, J.~Haegeman, M.~M. Rams, V.~Stojevic, N.~Schuch and F.~Verstraete,
\newblock \emph{{Transfer Matrices and Excitations with Matrix Product States}},
\newblock New J. Phys. \textbf{17}(5), 053002 (2015),
\newblock \doi{10.1088/1367-2630/17/5/053002},
\newblock  \eprint[https://doi.org/10.48550/arXiv.1408.5140]{10.48550/arXiv.1408.5140}.

\bibitem{Rams2018}
M.~M. Rams, P.~Czarnik and L.~Cincio,
\newblock \emph{{Precise extrapolation of the correlation function asymptotics in uniform tensor network states with application to the Bose-Hubbard and XXZ models}},
\newblock Phys. Rev. X \textbf{8}, 041033 (2018),
\newblock \doi{10.1103/PhysRevX.8.041033},
\newblock  \eprint[https://doi.org/10.48550/arXiv.1801.08554]{10.48550/arXiv.1801.08554}.

\bibitem{Vanhecke2019}
B.~Vanhecke, J.~Haegeman, K.~Van~Acoleyen, L.~Vanderstraeten and F.~Verstraete,
\newblock \emph{{Scaling Hypothesis for Matrix Product States}},
\newblock Phys. Rev. Lett. \textbf{123}(25), 250604 (2019),
\newblock \doi{10.1103/PhysRevLett.123.250604},
\newblock  \eprint[https://doi.org/10.48550/arXiv.1907.08603]{10.48550/arXiv.1907.08603}.

\bibitem{Cardy}
J.~Cardy,
\newblock \emph{Finite-size scaling}, vol.~2,
\newblock Elsevier (2012).

\bibitem{Calabrese2004}
P.~Calabrese and J.~L. Cardy,
\newblock \emph{{Entanglement entropy and quantum field theory}},
\newblock J. Stat. Mech. \textbf{0406}, P06002 (2004),
\newblock \doi{10.1088/1742-5468/2004/06/P06002},
\newblock  \eprint[https://doi.org/10.48550/arXiv.hep-th/0405152]{10.48550/arXiv.hep-th/0405152}.

\bibitem{Verstraete2004a}
F.~Verstraete, D.~Porras and J.~I. Cirac,
\newblock \emph{{Density Matrix Renormalization Group and Periodic Boundary Conditions: A Quantum Information Perspective}},
\newblock Phys. Rev. Lett. \textbf{93}, 227205 (2004),
\newblock \doi{10.1103/PhysRevLett.93.227205},
\newblock  \eprint[https://doi.org/10.48550/arXiv.cond-mat/0404706]{10.48550/arXiv.cond-mat/0404706}.

\bibitem{Sanz2009}
M.~Sanz, M.~M. Wolf, D.~P{\'e}rez-Garc{\'\i}a and J.~I. Cirac,
\newblock \emph{{Matrix product states: Symmetries and two-body Hamiltonians}},
\newblock Phys. Rev. A \textbf{79}(4), 042308 (2009),
\newblock \doi{10.1103/PhysRevA.79.042308},
\newblock  \eprint[https://doi.org/10.48550/arXiv.0901.2223]{10.48550/arXiv.0901.2223}.

\bibitem{Singh2009}
S.~Singh, R.~N.~C. Pfeifer and G.~Vidal,
\newblock \emph{{Tensor network decompositions in the presence of a global symmetry}},
\newblock Phys. Rev. A \textbf{82}, 050301 (2010),
\newblock \doi{10.1103/PhysRevA.82.050301},
\newblock  \eprint[https://doi.org/10.48550/arXiv.0907.2994]{10.48550/arXiv.0907.2994}.

\bibitem{Weichselbaum2012}
A.~Weichselbaum,
\newblock \emph{{Non-abelian symmetries in tensor networks: A quantum symmetry space approach}},
\newblock Annals Phys. \textbf{327}(12), 2972 (2012),
\newblock \doi{10.1016/j.aop.2012.07.009},
\newblock  \eprint[https://doi.org/10.48550/arXiv.1202.5664]{10.48550/arXiv.1202.5664}.

\bibitem{Singh2013}
S.~Singh, R.~N.~C. Pfeifer, G.~Vidal and G.~K. Brennen,
\newblock \emph{{Matrix product states for anyonic systems and efficient simulation of dynamics}},
\newblock Phys. Rev. B \textbf{89}, 075112 (2014),
\newblock \doi{10.1103/PhysRevB.89.075112},
\newblock  \eprint[https://doi.org/10.48550/arXiv.1311.0967]{10.48550/arXiv.1311.0967}.

\bibitem{Lootens2024}
L.~Lootens, C.~Delcamp and F.~Verstraete,
\newblock \emph{{Entanglement and the density matrix renormalisation group in the generalised Landau paradigm}}  (2024),
\newblock \doi{10.1038/s41567-025-02961-2},
\newblock  \eprint[https://doi.org/10.48550/arXiv.2408.06334]{10.48550/arXiv.2408.06334}.

\bibitem{Devos2025}
L.~Devos and J.~Haegeman,
\newblock \emph{{TensorKit.jl: A Julia package for large-scale tensor computations, with a hint of category theory}}  (2025),
\newblock  \eprint[https://doi.org/10.48550/arXiv.2508.10076]{10.48550/arXiv.2508.10076}.

\bibitem{Zauner-Stauber2018a}
V.~Zauner-Stauber, L.~Vanderstraeten, M.~T. Fishman, F.~Verstraete and J.~Haegeman,
\newblock \emph{{Variational optimization algorithms for uniform matrix product states}},
\newblock Phys. Rev. B \textbf{97}(4), 045145 (2018),
\newblock \doi{10.1103/PhysRevB.97.045145},
\newblock  \eprint[https://doi.org/10.48550/arXiv.1701.07035]{10.48550/arXiv.1701.07035}.

\bibitem{McCulloch2008}
I.~P. McCulloch,
\newblock \emph{{Infinite size density matrix renormalization group, revisited}}  (2008),
\newblock  \eprint[https://doi.org/10.48550/arXiv.0804.2509]{10.48550/arXiv.0804.2509}.

\bibitem{Ostlund1995}
S.~\"Ostlund and S.~Rommer,
\newblock \emph{Thermodynamic limit of density matrix renormalization},
\newblock Phys. Rev. Lett. \textbf{75}, 3537 (1995),
\newblock \doi{10.1103/PhysRevLett.75.3537}.

\bibitem{Haegeman2011b}
J.~Haegeman, B.~Pirvu, D.~J. Weir, J.~I. Cirac, T.~J. Osborne, H.~Verschelde and F.~Verstraete,
\newblock \emph{{Variational matrix product ansatz for dispersion relations}},
\newblock Phys. Rev. B \textbf{85}(10), 100408 (2012),
\newblock \doi{10.1103/PhysRevB.85.100408},
\newblock  \eprint[https://doi.org/10.48550/arXiv.1103.2286]{10.48550/arXiv.1103.2286}.

\bibitem{Vanderstraeten2013}
L.~Vanderstraeten, J.~Haegeman, T.~J. Osborne and F.~Verstraete,
\newblock \emph{{S-matrix from matrix product states}},
\newblock Phys. Rev. Lett. \textbf{112}(25), 257202 (2014),
\newblock \doi{10.1103/PhysRevLett.112.257202},
\newblock  \eprint[https://doi.org/10.48550/arXiv.1312.6793]{10.48550/arXiv.1312.6793}.

\bibitem{Bijl1941}
A.~Bijl, J.~{de Boer} and A.~Michels,
\newblock \emph{{Properties of liquid helium II}},
\newblock Physica \textbf{8}(7), 655 (1941),
\newblock \doi{10.1016/S0031-8914(41)90422-6}.

\bibitem{Feynman1953}
R.~P. Feynman,
\newblock \emph{{Atomic Theory of Liquid Helium Near Absolute Zero}},
\newblock Phys. Rev. \textbf{91}, 1301 (1953),
\newblock \doi{10.1103/PhysRev.91.1301}.

\bibitem{Feynman1954}
R.~P. Feynman,
\newblock \emph{{Atomic Theory of the Two-Fluid Model of Liquid Helium}},
\newblock Phys. Rev. \textbf{94}, 262 (1954),
\newblock \doi{10.1103/PhysRev.94.262}.

\bibitem{Lieb1961}
E.~H. Lieb, T.~Schultz and D.~Mattis,
\newblock \emph{{Two soluble models of an antiferromagnetic chain}},
\newblock Annals Phys. \textbf{16}, 407 (1961),
\newblock \doi{10.1016/0003-4916(61)90115-4}.

\bibitem{Nachtergaele2006}
B.~Nachtergaele, Y.~Ogata and R.~Sims,
\newblock \emph{{Propagation of Correlations in Quantum Lattice Systems}},
\newblock J. Statist. Phys. \textbf{124}(1), 1 (2006),
\newblock \doi{10.1007/s10955-006-9143-6},
\newblock  \eprint[https://doi.org/10.48550/arXiv.math-ph/0603064]{10.48550/arXiv.math-ph/0603064}.

\bibitem{Hastings2010}
M.~B. Hastings,
\newblock \emph{{Locality in Quantum Systems}} (2010),  \eprint[https://doi.org/10.48550/arXiv.1008.5137]{10.48550/arXiv.1008.5137}.

\bibitem{Haegeman2013b}
J.~Haegeman, S.~Michalakis, B.~Nachtergaele, T.~J. Osborne, N.~Schuch and F.~Verstraete,
\newblock \emph{{Elementary Excitations in Gapped Quantum Spin Systems}},
\newblock Phys. Rev. Lett. \textbf{111}(8), 080401 (2013),
\newblock \doi{10.1103/PhysRevLett.111.080401},
\newblock  \eprint[https://doi.org/10.48550/arXiv.1305.2176]{10.48550/arXiv.1305.2176}.

\bibitem{Zauner-Stauber2018b}
V.~Zauner-Stauber, L.~Vanderstraeten, J.~Haegeman, I.~P. McCulloch and F.~Verstraete,
\newblock \emph{{Topological nature of spinons and holons: Elementary excitations from matrix product states with conserved symmetries}},
\newblock Phys. Rev. B \textbf{97}, 235155 (2018),
\newblock \doi{10.1103/PhysRevB.97.235155},
\newblock  \eprint[https://doi.org/10.48550/arXiv.1802.07197]{10.48550/arXiv.1802.07197}.

\bibitem{Haldane1982}
F.~D.~M. Haldane,
\newblock \emph{{Continuum dynamics of the 1-D Heisenberg antiferromagnetic identification with the O(3) nonlinear sigma model}},
\newblock Phys. Lett. A \textbf{93}, 464 (1983),
\newblock \doi{10.1016/0375-9601(83)90631-X}.

\bibitem{Haldane1983}
F.~D.~M. Haldane,
\newblock \emph{{Nonlinear field theory of large spin Heisenberg antiferromagnets. Semiclassically quantized solitons of the one-dimensional easy Axis Neel state}},
\newblock Phys. Rev. Lett. \textbf{50}, 1153 (1983),
\newblock \doi{10.1103/PhysRevLett.50.1153}.

\bibitem{VanDamme}
M.~Van~Damme, L.~Devos and J.~Haegeman,
\newblock \emph{{MPSKit}},
\newblock \doi{10.5281/zenodo.10654900} (2025).

\bibitem{Bravyi2006}
S.~Bravyi, M.~B. Hastings and F.~Verstraete,
\newblock \emph{{Lieb-Robinson Bounds and the Generation of Correlations and Topological Quantum Order}},
\newblock Phys. Rev. Lett. \textbf{97}, 050401 (2006),
\newblock \doi{10.1103/PhysRevLett.97.050401},
\newblock  \eprint[https://doi.org/10.48550/arXiv.quant-ph/0603121]{10.48550/arXiv.quant-ph/0603121}.

\bibitem{DiFrancesco1997}
P.~Di~Francesco, P.~Mathieu and D.~Senechal,
\newblock \emph{{Conformal Field Theory}},
\newblock Graduate Texts in Contemporary Physics. Springer-Verlag, New York,
\newblock ISBN 978-0-387-94785-3, 978-1-4612-7475-9,
\newblock \doi{10.1007/978-1-4612-2256-9} (1997).

\bibitem{Landau1937}
L.~D. Landau,
\newblock \emph{{On the theory of phase transitions}},
\newblock Zh. Eksp. Teor. Fiz. \textbf{7}, 19 (1937),
\newblock \doi{10.1016/B978-0-08-010586-4.50034-1}.

\bibitem{Kadanoff1970}
L.~P. Kadanoff and H.~Ceva,
\newblock \emph{{Determination of an opeator algebra for the two-dimensional Ising model}},
\newblock Phys. Rev. B \textbf{3}, 3918 (1971),
\newblock \doi{10.1103/PhysRevB.3.3918}.

\bibitem{Chen2010}
X.~Chen, Z.~C. Gu and X.~G. Wen,
\newblock \emph{{Local unitary transformation, long-range quantum entanglement, wave function renormalization, and topological order}},
\newblock Phys. Rev. B \textbf{82}, 155138 (2010),
\newblock \doi{10.1103/PhysRevB.82.155138},
\newblock  \eprint[https://doi.org/10.48550/arXiv.1004.3835]{10.48550/arXiv.1004.3835}.

\bibitem{Chen2011}
X.~Chen, Z.-X. Liu and X.-G. Wen,
\newblock \emph{{Two-dimensional symmetry-protected topological orders and their protected gapless edge excitations}},
\newblock Phys. Rev. B \textbf{84}(23), 235141 (2011),
\newblock \doi{10.1103/PhysRevB.84.235141},
\newblock  \eprint[https://doi.org/10.48550/arXiv.1106.4752]{10.48550/arXiv.1106.4752}.

\bibitem{GarreRubio2022}
J.~Garre-Rubio, L.~Lootens and A.~Moln{\'a}r,
\newblock \emph{{Classifying phases protected by matrix product operator symmetries using matrix product states}},
\newblock Quantum \textbf{7}, 927 (2023),
\newblock \doi{10.22331/q-2023-02-21-927},
\newblock  \eprint[https://doi.org/10.48550/arXiv.2203.12563]{10.48550/arXiv.2203.12563}.

\bibitem{Perez2008}
D.~P{\'e}rez-Garc{\'\i}a, M.~M. Wolf, M.~Sanz, F.~Verstraete and J.~I. Cirac,
\newblock \emph{{String Order and Symmetries in Quantum Spin Lattices}},
\newblock Phys. Rev. Lett. \textbf{100}(16), 167202 (2008),
\newblock \doi{10.1103/PhysRevLett.100.167202},
\newblock  \eprint[https://doi.org/10.48550/arXiv.0802.0447]{10.48550/arXiv.0802.0447}.

\bibitem{Cirac2016}
J.~I. Cirac, D.~Perez-Garcia, N.~Schuch and F.~Verstraete,
\newblock \emph{{Matrix Product Density Operators: Renormalization Fixed Points and Boundary Theories}},
\newblock Annals Phys. \textbf{378}, 100 (2017),
\newblock \doi{10.1016/j.aop.2016.12.030},
\newblock  \eprint[https://doi.org/10.48550/arXiv.1606.00608]{10.48550/arXiv.1606.00608}.

\bibitem{Vancraeynest2022}
B.~Vancraeynest-De~Cuiper, J.~C. Bridgeman, N.~Dewolf, J.~Haegeman and F.~Verstraete,
\newblock \emph{{One-dimensional symmetric phases protected by frieze symmetries}},
\newblock Phys. Rev. B \textbf{107}(11), 115123 (2023),
\newblock \doi{10.1103/PhysRevB.107.115123},
\newblock  \eprint[https://doi.org/10.48550/arXiv.2202.12880]{10.48550/arXiv.2202.12880}.

\bibitem{Wigner1960}
E.~P. Wigner,
\newblock \emph{{Group Theory and Its Application to the Quantum Mechanics of Atomic Spectra}},
\newblock American Journal of Physics \textbf{28}(4), 408 (1960),
\newblock \doi{10.1119/1.1935822}.

\bibitem{Bultinck2016}
N.~Bultinck, D.~J. Williamson, J.~Haegeman and F.~Verstraete,
\newblock \emph{{Fermionic matrix product states and one-dimensional topological phases}},
\newblock Phys. Rev. B \textbf{95}(7), 075108 (2017),
\newblock \doi{10.1103/PhysRevB.95.075108},
\newblock  \eprint[https://doi.org/10.48550/arXiv.1610.07849]{10.48550/arXiv.1610.07849}.

\bibitem{Kapustin2016}
A.~Kapustin, A.~Turzillo and M.~You,
\newblock \emph{{Spin Topological Field Theory and Fermionic Matrix Product States}},
\newblock Phys. Rev. B \textbf{98}(12), 125101 (2018),
\newblock \doi{10.1103/PhysRevB.98.125101},
\newblock  \eprint[https://doi.org/10.48550/arXiv.1610.10075]{10.48550/arXiv.1610.10075}.

\bibitem{Mortier2024}
Q.~Mortier, L.~Devos, L.~Burgelman, B.~Vanhecke, N.~Bultinck, F.~Verstraete, J.~Haegeman and L.~Vanderstraeten,
\newblock \emph{{Fermionic tensor network methods}},
\newblock SciPost Phys. \textbf{18}(1), 012 (2025),
\newblock \doi{10.21468/SciPostPhys.18.1.012},
\newblock  \eprint[https://doi.org/10.48550/arXiv.2404.14611]{10.48550/arXiv.2404.14611}.

\bibitem{Piroli2020}
L.~Piroli, A.~Turzillo, S.~K. Shukla and J.~I. Cirac,
\newblock \emph{{Fermionic quantum cellular automata and generalized matrix-product unitaries}},
\newblock J. Stat. Mech. \textbf{2101}, 013107 (2021),
\newblock \doi{10.1088/1742-5468/abd30f},
\newblock  \eprint[https://doi.org/10.48550/arXiv.2007.11905]{10.48550/arXiv.2007.11905}.

\bibitem{Kitaev2000}
A.~Kitaev,
\newblock \emph{{Unpaired Majorana fermions in quantum wires}},
\newblock Phys. Usp. \textbf{44}(10S), 131 (2001),
\newblock \doi{10.1070/1063-7869/44/10S/S29},
\newblock  \eprint[https://doi.org/10.48550/arXiv.cond-mat/0010440]{10.48550/arXiv.cond-mat/0010440}.

\bibitem{Fidkowski2010}
L.~Fidkowski and A.~Kitaev,
\newblock \emph{{Topological phases of fermions in one dimension}},
\newblock Phys. Rev. B \textbf{83}(7), 075103 (2011),
\newblock \doi{10.1103/PhysRevB.83.075103},
\newblock  \eprint[https://doi.org/10.48550/arXiv.1008.4138]{10.48550/arXiv.1008.4138}.

\bibitem{Gaudin2014}
M.~Gaudin,
\newblock \emph{The bethe wavefunction},
\newblock Cambridge University Press (2014).

\bibitem{Alcaraz2004}
F.~C. Alcaraz and M.~J. Lazo,
\newblock \emph{{The Bethe ansatz as a matrix product ansatz}},
\newblock J. Phys. A \textbf{37}, L1 (2004),
\newblock \doi{10.1088/0305-4470/37/1/L01},
\newblock  \eprint[https://doi.org/10.48550/arXiv.cond-mat/0304170]{10.48550/arXiv.cond-mat/0304170}.

\bibitem{Murg2012}
V.~Murg, V.~E. Korepin and F.~Verstraete,
\newblock \emph{{Numerical Contraction of the Tensor Network generated by the Algebraic Bethe Ansatz}},
\newblock Phys. Rev. B \textbf{86}, 045125 (2012),
\newblock \doi{10.1103/PhysRevB.86.045125},
\newblock  \eprint[https://doi.org/10.48550/arXiv.1201.5636]{10.48550/arXiv.1201.5636}.

\bibitem{Katsura2010}
H.~Katsura and I.~Maruyama,
\newblock \emph{{Derivation of Matrix Product Ansatz for the Heisenberg Chain from Algebraic Bethe Ansatz}},
\newblock J. Phys. A \textbf{43}, 175003 (2010),
\newblock \doi{10.1088/1751-8113/43/17/175003},
\newblock  \eprint[https://doi.org/10.48550/arXiv.0911.4215]{10.48550/arXiv.0911.4215}.

\bibitem{Verstraete2004b}
F.~Verstraete and J.~I. Cirac,
\newblock \emph{{Renormalization algorithms for quantum-many body systems in two and higher dimensions}}  (2004),
\newblock  \eprint[https://doi.org/10.48550/arXiv.cond-mat/0407066]{10.48550/arXiv.cond-mat/0407066}.

\bibitem{Vanderstraeten2016}
L.~Vanderstraeten, J.~Haegeman, P.~Corboz and F.~Verstraete,
\newblock \emph{{Gradient methods for variational optimization of projected entangled-pair states}},
\newblock Phys. Rev. B \textbf{94}(15), 155123 (2016),
\newblock \doi{10.1103/PhysRevB.94.155123},
\newblock  \eprint[https://doi.org/10.48550/arXiv.1606.09170]{10.48550/arXiv.1606.09170}.

\bibitem{Corboz2016}
P.~Corboz,
\newblock \emph{{Variational optimization with infinite projected entangled-pair states}},
\newblock Phys. Rev. B \textbf{94}(3), 035133 (2016),
\newblock \doi{10.1103/PhysRevB.94.035133},
\newblock  \eprint[https://doi.org/10.48550/arXiv.1605.03006]{10.48550/arXiv.1605.03006}.

\bibitem{Liao2019}
H.-J. Liao, J.-G. Liu, L.~Wang and T.~Xiang,
\newblock \emph{{Differentiable Programming Tensor Networks}},
\newblock Phys. Rev. X \textbf{9}(3), 031041 (2019),
\newblock \doi{10.1103/PhysRevX.9.031041},
\newblock  \eprint[https://doi.org/10.48550/arXiv.1903.09650]{10.48550/arXiv.1903.09650}.

\bibitem{Vanderstraeten2022}
L.~Vanderstraeten, L.~Burgelman, B.~Ponsioen, M.~Van~Damme, B.~Vanhecke, P.~Corboz, J.~Haegeman and F.~Verstraete,
\newblock \emph{{Variational methods for contracting projected entangled-pair states}},
\newblock Phys. Rev. B \textbf{105}(19), 195140 (2022),
\newblock \doi{10.1103/PhysRevB.105.195140},
\newblock  \eprint[https://doi.org/10.48550/arXiv.2110.12726]{10.48550/arXiv.2110.12726}.

\bibitem{Buerschaper2013}
O.~Buerschaper,
\newblock \emph{{Twisted injectivity in projected entangled pair states and the classification of quantum phases}},
\newblock Annals Phys. \textbf{351}, 447 (2014),
\newblock \doi{10.1016/j.aop.2014.09.007},
\newblock  \eprint[https://doi.org/10.48550/arXiv.1307.7763]{10.48550/arXiv.1307.7763}.

\bibitem{Williamson2016}
D.~J. Williamson, N.~Bultinck, M.~Mari{\"e}n, M.~B. {\c{S}}ahino{\u{g}}lu, J.~Haegeman and F.~Verstraete,
\newblock \emph{{Matrix product operators for symmetry-protected topological phases: Gauging and edge theories}},
\newblock Phys. Rev. B \textbf{94}(20), 205150 (2016),
\newblock \doi{10.1103/PhysRevB.94.205150},
\newblock  \eprint[https://doi.org/10.48550/arXiv.1412.5604]{10.48550/arXiv.1412.5604}.

\bibitem{Molnar2017}
A.~Molnar, Y.~Ge, N.~Schuch and J.~I. Cirac,
\newblock \emph{A generalization of the injectivity condition for projected entangled pair states},
\newblock Journal of Mathematical Physics \textbf{59}(2), 021902 (2018),
\newblock \doi{10.1063/1.5007017},
\newblock  \eprint[https://doi.org/10.48550/arXiv.1706.07329]{10.48550/arXiv.1706.07329}.

\bibitem{Liu2025}
Y.~Liu, A.~Molnar, X.-Q. Sun, F.~Verstraete, K.~Kato and L.~Lootens,
\newblock \emph{{Trading Mathematical for Physical Simplicity: Bialgebraic Structures in Matrix Product Operator Symmetries}}  (2025),
\newblock  \eprint[https://doi.org/10.48550/arXiv.2509.03600]{10.48550/arXiv.2509.03600}.

\bibitem{FloridoLlinas2025}
M.~Florido-Llin{\`a}s, {\'A}.~M. Alhambra, D.~P{\'e}rez-Garc{\'\i}a and J.~I. Cirac,
\newblock \emph{{Uniform matrix product states with a boundary}}  (2025),
\newblock  \eprint[https://doi.org/10.48550/arXiv.2512.11968]{10.48550/arXiv.2512.11968}.

\bibitem{Yang2014}
S.~Yang, L.~Lehman, D.~Poilblanc, K.~Van~Acoleyen, F.~Verstraete, J.~I. Cirac and N.~Schuch,
\newblock \emph{{Edge theories in Projected Entangled Pair State models}},
\newblock Phys. Rev. Lett. \textbf{112}, 036402 (2014),
\newblock \doi{10.1103/PhysRevLett.112.036402},
\newblock  \eprint[https://doi.org/10.48550/arXiv.1309.4596]{10.48550/arXiv.1309.4596}.

\bibitem{Roose2018}
G.~Roose, L.~Vanderstraeten, J.~Haegeman and N.~Bultinck,
\newblock \emph{{Anomalous domain wall condensation in a modified Ising chain}},
\newblock Phys. Rev. B \textbf{99}(19), 195132 (2019),
\newblock \doi{10.1103/PhysRevB.99.195132},
\newblock  \eprint[https://doi.org/10.48550/arXiv.1812.04656]{10.48550/arXiv.1812.04656}.

\bibitem{Oshikawa2000}
M.~Oshikawa,
\newblock \emph{{Commensurability, Excitation Gap, and Topology in Quantum Many-Particle Systems on a Periodic Lattice}},
\newblock Phys. Rev. Lett. \textbf{84}(7), 1535 (2000),
\newblock \doi{10.1103/PhysRevLett.84.1535}.

\bibitem{Hastings2003}
M.~B. Hastings,
\newblock \emph{{Lieb-Schultz-Mattis in higher dimensions}},
\newblock Phys. Rev. B \textbf{69}, 104431 (2004),
\newblock \doi{10.1103/PhysRevB.69.104431},
\newblock  \eprint[https://doi.org/10.48550/arXiv.cond-mat/0305505]{10.48550/arXiv.cond-mat/0305505}.

\bibitem{Wen1989}
X.~G. Wen,
\newblock \emph{{Topological Order in Rigid States}},
\newblock Int. J. Mod. Phys. B \textbf{4}, 239 (1990),
\newblock \doi{10.1142/S0217979290000139}.

\bibitem{Witten1988}
E.~Witten,
\newblock \emph{{Topological Quantum Field Theory}},
\newblock Commun. Math. Phys. \textbf{117}, 353 (1988),
\newblock \doi{10.1007/BF01223371}.

\bibitem{Zeng2015}
B.~Zeng, X.~Chen, D.-L. Zhou and X.-G. Wen,
\newblock \emph{{Quantum Information Meets Quantum Matter: From Quantum Entanglement to Topological Phases of Many-Body Systems}},
\newblock Quantum Science and Technology. Springer,
\newblock ISBN 978-1-4939-9082-5, 978-1-4939-9084-9,
\newblock \doi{10.1007/978-1-4939-9084-9} (2019),  \eprint[https://doi.org/10.48550/arXiv.1508.02595]{10.48550/arXiv.1508.02595}.

\bibitem{Wen2016}
X.-G. Wen,
\newblock \emph{{Zoo of quantum-topological phases of matter}},
\newblock Rev. Mod. Phys. \textbf{89}(4), 041004 (2017),
\newblock \doi{10.1103/RevModPhys.89.041004},
\newblock  \eprint[https://doi.org/10.48550/arXiv.1610.03911]{10.48550/arXiv.1610.03911}.

\bibitem{Kitaev1997}
A.~Y. Kitaev,
\newblock \emph{{Fault tolerant quantum computation by anyons}},
\newblock Annals Phys. \textbf{303}, 2 (2003),
\newblock \doi{10.1016/S0003-4916(02)00018-0},
\newblock  \eprint[https://doi.org/10.48550/arXiv.quant-ph/9707021]{10.48550/arXiv.quant-ph/9707021}.

\bibitem{Kitaev2005a}
A.~Kitaev,
\newblock \emph{{Anyons in an exactly solved model and beyond}},
\newblock Annals Phys. \textbf{321}(1), 2 (2006),
\newblock \doi{10.1016/j.aop.2005.10.005},
\newblock  \eprint[https://doi.org/10.48550/arXiv.cond-mat/0506438]{10.48550/arXiv.cond-mat/0506438}.

\bibitem{Levin2006}
M.~Levin and X.-G. Wen,
\newblock \emph{{Detecting Topological Order in a Ground State Wave Function}},
\newblock Phys. Rev. Lett. \textbf{96}, 110405 (2006),
\newblock \doi{10.1103/PhysRevLett.96.110405},
\newblock  \eprint[https://doi.org/10.48550/arXiv.cond-mat/0510613]{10.48550/arXiv.cond-mat/0510613}.

\bibitem{Wegner1971}
F.~J. Wegner,
\newblock \emph{{Duality in Generalized Ising Models and Phase Transitions Without Local Order Parameters}},
\newblock J. Math. Phys. \textbf{12}, 2259 (1971),
\newblock \doi{10.1063/1.1665530}.

\bibitem{Haegeman2015}
J.~Haegeman, V.~Zauner, N.~Schuch and F.~Verstraete,
\newblock \emph{{Shadows of Anyons}},
\newblock Nature Commun. \textbf{6}, 8284 (2015),
\newblock \doi{10.1038/ncomms9284},
\newblock  \eprint[https://doi.org/10.48550/arXiv.1410.5443]{10.48550/arXiv.1410.5443}.

\bibitem{Marien2016}
M.~Mari{\"e}n, J.~Haegeman, P.~Fendley and F.~Verstraete,
\newblock \emph{{Condensation-Driven Phase Transitions in Perturbed String Nets}},
\newblock Phys. Rev. B \textbf{96}, 155127 (2017),
\newblock \doi{10.1103/PhysRevB.96.155127},
\newblock  \eprint[https://doi.org/10.48550/arXiv.1607.05296]{10.48550/arXiv.1607.05296}.

\bibitem{Bais2008}
F.~A. Bais and J.~K. Slingerland,
\newblock \emph{{Condensate induced transitions between topologically ordered phases}},
\newblock Phys. Rev. B \textbf{79}, 045316 (2009),
\newblock \doi{10.1103/PhysRevB.79.045316},
\newblock  \eprint[https://doi.org/10.48550/arXiv.0808.0627]{10.48550/arXiv.0808.0627}.

\bibitem{Burnell2017}
F.~J. Burnell,
\newblock \emph{{Anyon condensation and its applications}},
\newblock Ann. Rev. Condensed Matter Phys. \textbf{9}, 307 (2018),
\newblock \doi{10.1146/annurev-conmatphys-033117-054154},
\newblock  \eprint[https://doi.org/10.48550/arXiv.1706.04940]{10.48550/arXiv.1706.04940}.

\bibitem{Etingof2015}
P.~Etingof, S.~Gelaki, D.~Nikshych and V.~Ostrik,
\newblock \emph{Tensor Categories},
\newblock American Mathematical Society (2015).

\bibitem{Kong2017}
L.~Kong and H.~Zheng,
\newblock \emph{{Gapless edges of 2d topological orders and enriched monoidal categories}},
\newblock Nucl. Phys. B \textbf{927}, 140 (2018),
\newblock \doi{10.1016/j.nuclphysb.2017.12.007},
\newblock  \eprint[https://doi.org/10.48550/arXiv.1705.01087]{10.48550/arXiv.1705.01087}.

\bibitem{Bultinck2015}
N.~Bultinck, M.~Mari{\"e}n, D.~J. Williamson, M.~B. {\c{S}}ahino{\u{g}}lu, J.~Haegeman and F.~Verstraete,
\newblock \emph{{Anyons and matrix product operator algebras}},
\newblock Annals Phys. \textbf{378}, 183 (2017),
\newblock \doi{10.1016/j.aop.2017.01.004},
\newblock  \eprint[https://doi.org/10.48550/arXiv.1511.08090]{10.48550/arXiv.1511.08090}.

\bibitem{Kong2022}
L.~Kong and Z.-H. Zhang,
\newblock \emph{{An invitation to topological orders and category theory}}  (2022),
\newblock  \eprint[https://doi.org/10.48550/arXiv.2205.05565]{10.48550/arXiv.2205.05565}.

\bibitem{Dijkgraaf1989}
R.~Dijkgraaf and E.~Witten,
\newblock \emph{{Topological Gauge Theories and Group Cohomology}},
\newblock Commun. Math. Phys. \textbf{129}, 393 (1990),
\newblock \doi{10.1007/BF02096988}.

\bibitem{Roche1990}
P.~Roche, V.~Pasquier and R.~Dijkgraaf,
\newblock \emph{{Quasi Hopf algebras, group cohomology and orbifold models}},
\newblock Nucl. Phys. B Proc. Suppl. \textbf{18}, 60 (1990),
\newblock \doi{10.1016/0920-5632(91)90123-V}.

\bibitem{Schuch2010}
N.~Schuch, I.~Cirac and D.~P{\'e}rez-Garc{\'\i}a,
\newblock \emph{{PEPS as ground states: Degeneracy and topology}},
\newblock Annals Phys. \textbf{325}, 2153 (2010),
\newblock \doi{10.1016/j.aop.2010.05.008},
\newblock  \eprint[https://doi.org/10.48550/arXiv.1001.3807]{10.48550/arXiv.1001.3807}.

\bibitem{Zhang2011}
Y.~Zhang, T.~Grover, A.~Turner, M.~Oshikawa and A.~Vishwanath,
\newblock \emph{{Quasi-particle Statistics and Braiding from Ground State Entanglement}},
\newblock Phys. Rev. B \textbf{85}, 235151 (2012),
\newblock \doi{10.1103/PhysRevB.85.235151},
\newblock  \eprint[https://doi.org/10.48550/arXiv.1111.2342]{10.48550/arXiv.1111.2342}.

\bibitem{Sahinoglu2014}
M.~B. {\c{S}}ahino{\u{g}}lu, D.~Williamson, N.~Bultinck, M.~Mari{\"e}n, J.~Haegeman, N.~Schuch and F.~Verstraete,
\newblock \emph{{Characterizing Topological Order with Matrix Product Operators}},
\newblock Annales Henri Poincare \textbf{22}(2), 563 (2021),
\newblock \doi{10.1007/s00023-020-00992-4},
\newblock  \eprint[https://doi.org/10.48550/arXiv.1409.2150]{10.48550/arXiv.1409.2150}.

\bibitem{Williamson2017}
D.~J. Williamson, N.~Bultinck and F.~Verstraete,
\newblock \emph{{Symmetry-enriched topological order in tensor networks: Defects, gauging and anyon condensation}}  (2017),
\newblock  \eprint[https://doi.org/10.48550/arXiv.1711.07982]{10.48550/arXiv.1711.07982}.

\bibitem{Ocneanu1993}
A.~Ocneanu,
\newblock \emph{Chirality for operator algebras},
\newblock In \emph{Taniguchi Symposium on Operator Algebras} (1993).

\bibitem{Ocneanu2001}
A.~Ocneanu,
\newblock \emph{Operator algebras, topology and subgroups of quantum symmetry},
\newblock In \emph{Advanced Studies in Pure Mathematics},
\newblock \doi{10.2969/aspm/03110235} (2001).

\bibitem{Lan2013}
T.~Lan and X.-G. Wen,
\newblock \emph{{Topological quasiparticles and the holographic bulk-edge relation in (2+1) -dimensional string-net models}},
\newblock Phys. Rev. B \textbf{90}(11), 115119 (2014),
\newblock \doi{10.1103/PhysRevB.90.115119},
\newblock  \eprint[https://doi.org/10.48550/arXiv.1311.1784]{10.48550/arXiv.1311.1784}.

\bibitem{Hu2015}
Y.~Hu, N.~Geer and Y.-S. Wu,
\newblock \emph{{Full dyon excitation spectrum in extended Levin-Wen models}},
\newblock Phys. Rev. B \textbf{97}(19), 195154 (2018),
\newblock \doi{10.1103/PhysRevB.97.195154},
\newblock  \eprint[https://doi.org/10.48550/arXiv.1502.03433]{10.48550/arXiv.1502.03433}.

\bibitem{Bultinck2017}
N.~Bultinck, R.~Vanhove, J.~Haegeman and F.~Verstraete,
\newblock \emph{{Global anomaly detection in two-dimensional symmetry-protected topological phases}},
\newblock Phys. Rev. Lett. \textbf{120}(15), 156601 (2018),
\newblock \doi{10.1103/PhysRevLett.120.156601},
\newblock  \eprint[https://doi.org/10.48550/arXiv.1710.02314]{10.48550/arXiv.1710.02314}.

\bibitem{Levin2004}
M.~A. Levin and X.-G. Wen,
\newblock \emph{{String net condensation: A Physical mechanism for topological phases}},
\newblock Phys. Rev. B \textbf{71}, 045110 (2005),
\newblock \doi{10.1103/PhysRevB.71.045110},
\newblock  \eprint[https://doi.org/10.48550/arXiv.cond-mat/0404617]{10.48550/arXiv.cond-mat/0404617}.

\bibitem{Gaiotto2014}
D.~Gaiotto, A.~Kapustin, N.~Seiberg and B.~Willett,
\newblock \emph{{Generalized Global Symmetries}},
\newblock JHEP \textbf{02}, 172 (2015),
\newblock \doi{10.1007/JHEP02(2015)172},
\newblock  \eprint[https://doi.org/10.48550/arXiv.1412.5148]{10.48550/arXiv.1412.5148}.

\bibitem{Thorngren2019}
R.~Thorngren and Y.~Wang,
\newblock \emph{{Fusion category symmetry. Part I. Anomaly in-flow and gapped phases}},
\newblock JHEP \textbf{04}, 132 (2024),
\newblock \doi{10.1007/JHEP04(2024)132},
\newblock  \eprint[https://doi.org/10.48550/arXiv.1912.02817]{10.48550/arXiv.1912.02817}.

\bibitem{Aasen2020}
D.~Aasen, P.~Fendley and R.~S.~K. Mong,
\newblock \emph{{Topological Defects on the Lattice: Dualities and Degeneracies}}  (2020),
\newblock  \eprint[https://doi.org/10.48550/arXiv.2008.08598]{10.48550/arXiv.2008.08598}.

\bibitem{Komargodski2020}
Z.~Komargodski, K.~Ohmori, K.~Roumpedakis and S.~Seifnashri,
\newblock \emph{{Symmetries and strings of adjoint QCD$_{2}$}},
\newblock JHEP \textbf{03}, 103 (2021),
\newblock \doi{10.1007/JHEP03(2021)103},
\newblock  \eprint[https://doi.org/10.48550/arXiv.2008.07567]{10.48550/arXiv.2008.07567}.

\bibitem{McGreevy2022}
J.~McGreevy,
\newblock \emph{{Generalized Symmetries in Condensed Matter}},
\newblock Ann. Rev. Condensed Matter Phys. \textbf{14}, 57 (2023),
\newblock \doi{10.1146/annurev-conmatphys-040721-021029},
\newblock  \eprint[https://doi.org/10.48550/arXiv.2204.03045]{10.48550/arXiv.2204.03045}.

\bibitem{Moradi2022}
H.~Moradi, S.~F. Moosavian and A.~Tiwari,
\newblock \emph{{Topological holography: Towards a unification of Landau and beyond-Landau physics}},
\newblock SciPost Phys. Core \textbf{6}, 066 (2023),
\newblock \doi{10.21468/SciPostPhysCore.6.4.066},
\newblock  \eprint[https://doi.org/10.48550/arXiv.2207.10712]{10.48550/arXiv.2207.10712}.

\bibitem{Freed2022}
D.~S. Freed, G.~W. Moore and C.~Teleman,
\newblock \emph{{Topological symmetry in quantum field theory}}  (2022),
\newblock  \eprint[https://doi.org/10.48550/arXiv.2209.07471]{10.48550/arXiv.2209.07471}.

\bibitem{Shao2023}
S.-H. Shao,
\newblock \emph{{What's Done Cannot Be Undone: TASI Lectures on Non-Invertible Symmetries}}  (2023),
\newblock  \eprint[https://doi.org/10.48550/arXiv.2308.00747]{10.48550/arXiv.2308.00747}.

\bibitem{Schafer-Nameki2023}
S.~Schafer-Nameki,
\newblock \emph{{ICTP lectures on (non-)invertible generalized symmetries}},
\newblock Phys. Rept. \textbf{1063}, 1 (2024),
\newblock \doi{10.1016/j.physrep.2024.01.007},
\newblock  \eprint[https://doi.org/10.48550/arXiv.2305.18296]{10.48550/arXiv.2305.18296}.

\bibitem{Moradi2023}
H.~Moradi, {\"O}.~M. Aksoy, J.~H. Bardarson and A.~Tiwari,
\newblock \emph{{Symmetry fractionalization, mixed-anomalies and dualities in quantum spin models with generalized symmetries}}  (2023),
\newblock  \eprint[https://doi.org/10.48550/arXiv.2307.01266]{10.48550/arXiv.2307.01266}.

\bibitem{Bhardwaj2023}
L.~Bhardwaj, L.~E. Bottini, D.~Pajer and S.~Schafer-Nameki,
\newblock \emph{{Categorical Landau Paradigm for Gapped Phases}},
\newblock Phys. Rev. Lett. \textbf{133}(16), 161601 (2024),
\newblock \doi{10.1103/PhysRevLett.133.161601},
\newblock  \eprint[https://doi.org/10.48550/arXiv.2310.03786]{10.48550/arXiv.2310.03786}.

\bibitem{Vanhove2018}
R.~Vanhove, M.~Bal, D.~J. Williamson, N.~Bultinck, J.~Haegeman and F.~Verstraete,
\newblock \emph{{Mapping topological to conformal field theories through strange correlators}},
\newblock Phys. Rev. Lett. \textbf{121}(17), 177203 (2018),
\newblock \doi{10.1103/PhysRevLett.121.177203},
\newblock  \eprint[https://doi.org/10.48550/arXiv.1801.05959]{10.48550/arXiv.1801.05959}.

\bibitem{Aasen2016}
D.~Aasen, R.~S.~K. Mong and P.~Fendley,
\newblock \emph{{Topological Defects on the Lattice I: The Ising model}},
\newblock J. Phys. A \textbf{49}(35), 354001 (2016),
\newblock \doi{10.1088/1751-8113/49/35/354001},
\newblock  \eprint[https://doi.org/10.48550/arXiv.1601.07185]{10.48550/arXiv.1601.07185}.

\bibitem{Kong2015}
L.~Kong, X.-G. Wen and H.~Zheng,
\newblock \emph{{Boundary-bulk relation for topological orders as the functor mapping higher categories to their centers}}  (2015),
\newblock  \eprint[https://doi.org/10.48550/arXiv.1502.01690]{10.48550/arXiv.1502.01690}.

\bibitem{Apruzzi2021}
F.~Apruzzi, F.~Bonetti, I.~Garc{\'\i}a~Etxebarria, S.~S. Hosseini and S.~Schafer-Nameki,
\newblock \emph{{Symmetry TFTs from String Theory}},
\newblock Commun. Math. Phys. \textbf{402}(1), 895 (2023),
\newblock \doi{10.1007/s00220-023-04737-2},
\newblock  \eprint[https://doi.org/10.48550/arXiv.2112.02092]{10.48550/arXiv.2112.02092}.

\bibitem{Chatterjee2022}
A.~Chatterjee and X.-G. Wen,
\newblock \emph{{Holographic theory for continuous phase transitions: Emergence and symmetry protection of gaplessness}},
\newblock Phys. Rev. B \textbf{108}(7), 075105 (2023),
\newblock \doi{10.1103/PhysRevB.108.075105},
\newblock  \eprint[https://doi.org/10.48550/arXiv.2205.06244]{10.48550/arXiv.2205.06244}.

\bibitem{Delcamp2024}
C.~Delcamp and N.~Ishtiaque,
\newblock \emph{{Symmetry topological field theory and non-abelian Kramers-Wannier dualities of generalised Ising models}},
\newblock Annales Henri Poincare  (2024),
\newblock \doi{10.1007/s00023-025-01591-x},
\newblock  \eprint[https://doi.org/10.48550/arXiv.2408.06074]{10.48550/arXiv.2408.06074}.

\bibitem{Molnar2022}
A.~Molnar, A.~R. de~Alarc{\'o}n, J.~Garre-Rubio, N.~Schuch, J.~I. Cirac and D.~P{\'e}rez-Garc{\'\i}a,
\newblock \emph{{Matrix product operator algebras I: representations of weak Hopf algebras and projected entangled pair states}}  (2022),
\newblock  \eprint[https://doi.org/10.48550/arXiv.2204.05940]{10.48550/arXiv.2204.05940}.

\bibitem{Lootens2020}
L.~Lootens, J.~Fuchs, J.~Haegeman, C.~Schweigert and F.~Verstraete,
\newblock \emph{{Matrix product operator symmetries and intertwiners in string-nets with domain walls}},
\newblock SciPost Phys. \textbf{10}(3), 053 (2021),
\newblock \doi{10.21468/SciPostPhys.10.3.053},
\newblock  \eprint[https://doi.org/10.48550/arXiv.2008.11187]{10.48550/arXiv.2008.11187}.

\bibitem{Feiguin2006}
A.~Feiguin, S.~Trebst, A.~W.~W. Ludwig, M.~Troyer, A.~Kitaev, Z.~Wang and M.~H. Freedman,
\newblock \emph{{Interacting anyons in topological quantum liquids: The golden chain}},
\newblock Phys. Rev. Lett. \textbf{98}, 160409 (2007),
\newblock \doi{10.1103/PhysRevLett.98.160409},
\newblock  \eprint[https://doi.org/10.48550/arXiv.cond-mat/0612341]{10.48550/arXiv.cond-mat/0612341}.

\bibitem{Trebst2008}
S.~Trebst, M.~Troyer, Z.~Wang and A.~W.~W. Ludwig,
\newblock \emph{{A Short Introduction to Fibonacci Anyon Models}},
\newblock Prog. Theor. Phys. Suppl. \textbf{176}, 384 (2008),
\newblock \doi{10.1143/ptps.176.384},
\newblock  \eprint[https://doi.org/10.48550/arXiv.0902.3275]{10.48550/arXiv.0902.3275}.

\bibitem{Gils2013}
C.~Gils, E.~Ardonne, S.~Trebst, D.~A. Huse, A.~W.~W. Ludwig, M.~Troyer and Z.~Wang,
\newblock \emph{{Anyonic quantum spin chains: Spin-1 generalizations and topological stability}},
\newblock Phys. Rev. B \textbf{87}(23), 235120 (2013),
\newblock \doi{10.1103/PhysRevB.87.235120},
\newblock  \eprint[https://doi.org/10.48550/arXiv.1303.4290]{10.48550/arXiv.1303.4290}.

\bibitem{Buican2017}
M.~Buican and A.~Gromov,
\newblock \emph{{Anyonic Chains, Topological Defects, and Conformal Field Theory}},
\newblock Commun. Math. Phys. \textbf{356}(3), 1017 (2017),
\newblock \doi{10.1007/s00220-017-2995-6},
\newblock  \eprint[https://doi.org/10.48550/arXiv.1701.02800]{10.48550/arXiv.1701.02800}.

\bibitem{Huang2021}
T.-C. Huang, Y.-H. Lin, K.~Ohmori, Y.~Tachikawa and M.~Tezuka,
\newblock \emph{{Numerical Evidence for a Haagerup Conformal Field Theory}},
\newblock Phys. Rev. Lett. \textbf{128}(23), 231603 (2022),
\newblock \doi{10.1103/PhysRevLett.128.231603},
\newblock  \eprint[https://doi.org/10.48550/arXiv.2110.03008]{10.48550/arXiv.2110.03008}.

\bibitem{Jones2024}
C.~Jones, K.~Schatz and D.~J. Williamson,
\newblock \emph{{Quantum cellular automata and categorical dualities of spin chains}}  (2024),
\newblock  \eprint[https://doi.org/10.48550/arXiv.2410.08884]{10.48550/arXiv.2410.08884}.

\bibitem{Kirillov2011}
A.~Kirillov, Jr,
\newblock \emph{{String-net model of Turaev-Viro invariants}}  (2011),
\newblock  \eprint[https://doi.org/10.48550/arXiv.1106.6033]{10.48550/arXiv.1106.6033}.

\bibitem{Kawagoe2024}
K.~Kawagoe, C.~Jones, S.~Sanford, D.~Green and D.~Penneys,
\newblock \emph{{Levin-Wen is a Gauge Theory: Entanglement from Topology}},
\newblock Commun. Math. Phys. \textbf{405}(11), 266 (2024),
\newblock \doi{10.1007/s00220-024-05144-x},
\newblock  \eprint[https://doi.org/10.48550/arXiv.2401.13838]{10.48550/arXiv.2401.13838}.

\bibitem{Turaev1992}
V.~G. Turaev and O.~Y. Viro,
\newblock \emph{{State sum invariants of 3-manifolds and quantum 6j-symbols}},
\newblock Topology \textbf{31}, 865 (1992),
\newblock \doi{10.1016/0040-9383(92)90015-A}.

\bibitem{Barrett1993}
J.~W. Barrett and B.~W. Westbury,
\newblock \emph{{Invariants of piecewise linear three manifolds}},
\newblock Trans. Am. Math. Soc. \textbf{348}, 3997 (1996),
\newblock \doi{10.1090/S0002-9947-96-01660-1},
\newblock  \eprint[https://doi.org/10.48550/arXiv.hep-th/9311155]{10.48550/arXiv.hep-th/9311155}.

\bibitem{Verstraete2006b}
F.~Verstraete, M.~M. Wolf, D.~Perez-Garcia and J.~I. Cirac,
\newblock \emph{{Criticality, the area law, and the computational power of PEPS}},
\newblock Phys. Rev. Lett. \textbf{96}, 220601 (2006),
\newblock \doi{10.1103/PhysRevLett.96.220601},
\newblock  \eprint[https://doi.org/10.48550/arXiv.quant-ph/0601075]{10.48550/arXiv.quant-ph/0601075}.

\bibitem{Buerschaper2009}
O.~Buerschaper, M.~Aguado and G.~Vidal,
\newblock \emph{{Explicit tensor network representation for the ground states of string-net models}},
\newblock Phys. Rev. B \textbf{79}(8), 085119 (2009),
\newblock \doi{10.1103/PhysRevB.79.085119}.

\bibitem{Gu2009}
Z.-C. Gu, M.~Levin, B.~Swingle and X.-G. Wen,
\newblock \emph{{Tensor-product representations for string-net condensed states}},
\newblock Phys. Rev. B \textbf{79}(8), 085118 (2009),
\newblock \doi{10.1103/PhysRevB.79.085118}.

\bibitem{Kitaev2011}
A.~Kitaev and L.~Kong,
\newblock \emph{{Models for Gapped Boundaries and Domain Walls}},
\newblock Commun. Math. Phys. \textbf{313}(2), 351 (2012),
\newblock \doi{10.1007/s00220-012-1500-5},
\newblock  \eprint[https://doi.org/10.48550/arXiv.1104.5047]{10.48550/arXiv.1104.5047}.

\bibitem{You2013}
Y.-Z. You, Z.~Bi, A.~Rasmussen, K.~Slagle and C.~Xu,
\newblock \emph{{Wave Function and Strange Correlator of Short Range Entangled states}},
\newblock Phys. Rev. Lett. \textbf{112}(24), 247202 (2014),
\newblock \doi{10.1103/PhysRevLett.112.247202},
\newblock  \eprint[https://doi.org/10.48550/arXiv.1312.0626]{10.48550/arXiv.1312.0626}.

\bibitem{Moore1988}
G.~W. Moore and N.~Seiberg,
\newblock \emph{{Classical and Quantum Conformal Field Theory}},
\newblock Commun. Math. Phys. \textbf{123}, 177 (1989),
\newblock \doi{10.1007/BF01238857}.

\bibitem{Frohlich2004}
J.~Frohlich, J.~Fuchs, I.~Runkel and C.~Schweigert,
\newblock \emph{{Kramers-Wannier duality from conformal defects}},
\newblock Phys. Rev. Lett. \textbf{93}, 070601 (2004),
\newblock \doi{10.1103/PhysRevLett.93.070601},
\newblock  \eprint[https://doi.org/10.48550/arXiv.cond-mat/0404051]{10.48550/arXiv.cond-mat/0404051}.

\bibitem{Frohlich2006}
J.~Frohlich, J.~Fuchs, I.~Runkel and C.~Schweigert,
\newblock \emph{{Duality and defects in rational conformal field theory}},
\newblock Nucl. Phys. B \textbf{763}, 354 (2007),
\newblock \doi{10.1016/j.nuclphysb.2006.11.017},
\newblock  \eprint[https://doi.org/10.48550/arXiv.hep-th/0607247]{10.48550/arXiv.hep-th/0607247}.

\bibitem{Fuchs2002}
J.~Fuchs, I.~Runkel and C.~Schweigert,
\newblock \emph{{TFT construction of RCFT correlators 1. Partition functions}},
\newblock Nucl. Phys. B \textbf{646}, 353 (2002),
\newblock \doi{10.1016/S0550-3213(02)00744-7},
\newblock  \eprint[https://doi.org/10.48550/arXiv.hep-th/0204148]{10.48550/arXiv.hep-th/0204148}.

\bibitem{Fuchs2003}
J.~Fuchs, I.~Runkel and C.~Schweigert,
\newblock \emph{{TFT construction of RCFT correlators. 2. Unoriented world sheets}},
\newblock Nucl. Phys. B \textbf{678}, 511 (2004),
\newblock \doi{10.1016/j.nuclphysb.2003.11.026},
\newblock  \eprint[https://doi.org/10.48550/arXiv.hep-th/0306164]{10.48550/arXiv.hep-th/0306164}.

\bibitem{Fuchs2004}
J.~Fuchs, I.~Runkel and C.~Schweigert,
\newblock \emph{{TFT construction of RCFT correlators. 3. Simple currents}},
\newblock Nucl. Phys. B \textbf{694}, 277 (2004),
\newblock \doi{10.1016/j.nuclphysb.2004.05.014},
\newblock  \eprint[https://doi.org/10.48550/arXiv.hep-th/0403157]{10.48550/arXiv.hep-th/0403157}.

\bibitem{Baxter1980}
R.~J. Baxter,
\newblock \emph{{Hard hexagons: exact solution}},
\newblock Journal of Physics A Mathematical General \textbf{13}(3), L61 (1980),
\newblock \doi{10.1088/0305-4470/13/3/007}.

\bibitem{Fidkowski2009}
L.~Fidkowski, M.~Freedman, C.~Nayak, K.~Walker and Z.~Wang,
\newblock \emph{{From string nets to nonabelions}},
\newblock Commun. Math. Phys. \textbf{287}, 805 (2009),
\newblock \doi{10.1007/s00220-009-0757-9}.

\bibitem{Petkova2000}
V.~B. Petkova and J.~B. Zuber,
\newblock \emph{{Generalized twisted partition functions}},
\newblock Phys. Lett. B \textbf{504}, 157 (2001),
\newblock \doi{10.1016/S0370-2693(01)00276-3},
\newblock  \eprint[https://doi.org/10.48550/arXiv.hep-th/0011021]{10.48550/arXiv.hep-th/0011021}.

\bibitem{Vanhove2021a}
R.~Vanhove, L.~Lootens, H.-H. Tu and F.~Verstraete,
\newblock \emph{{Topological aspects of the critical three-state Potts model}},
\newblock J. Phys. A \textbf{55}(23), 235002 (2022),
\newblock \doi{10.1088/1751-8121/ac68b1},
\newblock  \eprint[https://doi.org/10.48550/arXiv.2107.11177]{10.48550/arXiv.2107.11177}.

\bibitem{Freedman2011}
M.~H. Freedman, J.~Gukelberger, M.~B. Hastings, S.~Trebst, M.~Troyer and Z.~Wang,
\newblock \emph{{Galois Conjugates of Topological Phases}},
\newblock Phys. Rev. B \textbf{85}, 045414 (2012),
\newblock \doi{10.1103/PhysRevB.85.045414},
\newblock  \eprint[https://doi.org/10.48550/arXiv.1106.3267]{10.48550/arXiv.1106.3267}.

\bibitem{Lootens2019}
L.~Lootens, R.~Vanhove, J.~Haegeman and F.~Verstraete,
\newblock \emph{{Galois Conjugated Tensor Fusion Categories and Nonunitary Conformal Field Theory}},
\newblock Phys. Rev. Lett. \textbf{124}(12), 120601 (2020),
\newblock \doi{10.1103/PhysRevLett.124.120601},
\newblock  \eprint[https://doi.org/10.48550/arXiv.1902.11241]{10.48550/arXiv.1902.11241}.

\bibitem{Brehm2021}
E.~M. Brehm and I.~Runkel,
\newblock \emph{{Lattice models from CFT on surfaces with holes: I. Torus partition function via two lattice cells}},
\newblock J. Phys. A \textbf{55}(23), 235001 (2022),
\newblock \doi{10.1088/1751-8121/ac6a91},
\newblock  \eprint[https://doi.org/10.48550/arXiv.2112.01563]{10.48550/arXiv.2112.01563}.

\bibitem{Vanhove2021b}
R.~Vanhove, L.~Lootens, M.~Van~Damme, R.~Wolf, T.~J. Osborne, J.~Haegeman and F.~Verstraete,
\newblock \emph{{Critical Lattice Model for a Haagerup Conformal Field Theory}},
\newblock Phys. Rev. Lett. \textbf{128}(23), 231602 (2022),
\newblock \doi{10.1103/PhysRevLett.128.231602},
\newblock  \eprint[https://doi.org/10.48550/arXiv.2110.03532]{10.48550/arXiv.2110.03532}.

\bibitem{Zeng2022}
X.~Zeng, R.~Wang, C.~Shen and L.-Y. Hung,
\newblock \emph{{Virasoro generators in the Fibonacci model tensor network: Tackling finite-size effects}},
\newblock Phys. Rev. B \textbf{107}(24), 245146 (2023),
\newblock \doi{10.1103/PhysRevB.107.245146},
\newblock  \eprint[https://doi.org/10.48550/arXiv.2212.02937]{10.48550/arXiv.2212.02937}.

\bibitem{Ji2024}
K.~Ji, L.~Chen, L.-P. Yang and L.-Y. Hung,
\newblock \emph{{Simplex tensor network renormalization group for boundary theory of 3+1D symTFT}}  (2024),
\newblock  \eprint[https://doi.org/10.48550/arXiv.2412.08374]{10.48550/arXiv.2412.08374}.

\bibitem{Shen2025}
C.~Shen,
\newblock \emph{{Exploring the phase diagram of $SU(2)_4$ strange correlator}}  (2025),
\newblock  \eprint[https://doi.org/10.48550/arXiv.2502.14556]{10.48550/arXiv.2502.14556}.

\bibitem{Hung2025}
L.-Y. Hung, K.~Ji, C.~Shen, Y.~Wan and Y.~Zhao,
\newblock \emph{{A 2D-CFT Factory: Critical Lattice Models from Competing Anyon Condensation Processes in SymTO/SymTFT}}  (2025),
\newblock  \eprint[https://doi.org/10.48550/arXiv.2506.05324]{10.48550/arXiv.2506.05324}.

\bibitem{Fendley2020}
P.~Fendley,
\newblock \emph{{Integrability and braided tensor categories}},
\newblock J. Statist. Phys. \textbf{182}, 43 (2021),
\newblock \doi{10.1007/s10955-021-02712-6},
\newblock  \eprint[https://doi.org/10.48550/arXiv.2008.02292]{10.48550/arXiv.2008.02292}.

\bibitem{Eckart1930}
C.~Eckart,
\newblock \emph{{The Application of Group Theory to the Quantum Dynamics of Monatomic Systems}},
\newblock Rev. Mod. Phys. \textbf{2}, 305 (1930),
\newblock \doi{10.1103/RevModPhys.2.305}.

\bibitem{Bridgeman2022}
J.~C. Bridgeman, L.~Lootens and F.~Verstraete,
\newblock \emph{{Invertible Bimodule Categories and Generalized Schur Orthogonality}},
\newblock Commun. Math. Phys. \textbf{402}(3), 2691 (2023),
\newblock \doi{10.1007/s00220-023-04781-y},
\newblock  \eprint[https://doi.org/10.48550/arXiv.2211.01947]{10.48550/arXiv.2211.01947}.

\bibitem{Lootens2022}
L.~Lootens, C.~Delcamp and F.~Verstraete,
\newblock \emph{{Dualities in One-Dimensional Quantum Lattice Models: Topological Sectors}},
\newblock PRX Quantum \textbf{5}(1), 010338 (2024),
\newblock \doi{10.1103/PRXQuantum.5.010338},
\newblock  \eprint[https://doi.org/10.48550/arXiv.2211.03777]{10.48550/arXiv.2211.03777}.

\bibitem{Cobanera2009}
E.~Cobanera, G.~Ortiz and Z.~Nussinov,
\newblock \emph{{Unified approach to Quantum and Classical Dualities}},
\newblock Phys. Rev. Lett. \textbf{104}, 020402 (2010),
\newblock \doi{10.1103/PhysRevLett.104.020402},
\newblock  \eprint[https://doi.org/10.48550/arXiv.0907.0733]{10.48550/arXiv.0907.0733}.

\bibitem{Cobanera2011}
E.~Cobanera, G.~Ortiz and Z.~Nussinov,
\newblock \emph{{The Bond-Algebraic Approach to Dualities}},
\newblock Adv. Phys. \textbf{60}, 679 (2011),
\newblock \doi{10.1080/00018732.2011.619814},
\newblock  \eprint[https://doi.org/10.48550/arXiv.1103.2776]{10.48550/arXiv.1103.2776}.

\bibitem{Haegeman2014a}
J.~Haegeman, K.~Van~Acoleyen, N.~Schuch, J.~I. Cirac and F.~Verstraete,
\newblock \emph{{Gauging quantum states: from global to local symmetries in many-body systems}},
\newblock Phys. Rev. X \textbf{5}(1), 011024 (2015),
\newblock \doi{10.1103/PhysRevX.5.011024},
\newblock  \eprint[https://doi.org/10.48550/arXiv.1407.1025]{10.48550/arXiv.1407.1025}.

\bibitem{Bhardwaj2017}
L.~Bhardwaj and Y.~Tachikawa,
\newblock \emph{{On finite symmetries and their gauging in two dimensions}},
\newblock JHEP \textbf{03}, 189 (2018),
\newblock \doi{10.1007/JHEP03(2018)189},
\newblock  \eprint[https://doi.org/10.48550/arXiv.1704.02330]{10.48550/arXiv.1704.02330}.

\bibitem{Seifnashri2025}
S.~Seifnashri, S.-H. Shao and X.~Yang,
\newblock \emph{{Gauging non-invertible symmetries on the lattice}},
\newblock SciPost Phys. \textbf{19}(2), 063 (2025),
\newblock \doi{10.21468/SciPostPhys.19.2.063},
\newblock  \eprint[https://doi.org/10.48550/arXiv.2503.02925]{10.48550/arXiv.2503.02925}.

\bibitem{Vancraeynest2025b}
B.~Vancraeynest-De~Cuiper, J.~Garre-Rubio, F.~Verstraete, K.~Vervoort, D.~J. Williamson and L.~Lootens,
\newblock \emph{{From gauging to duality in one-dimensional quantum lattice models}}  (2025),
\newblock \doi{10.48550/arXiv.2509.22051},
\newblock  \eprint[https://doi.org/10.48550/arXiv.2509.22051]{10.48550/arXiv.2509.22051}.

\bibitem{Lootens2021b}
L.~Lootens, C.~Delcamp, G.~Ortiz and F.~Verstraete,
\newblock \emph{{Dualities in One-Dimensional Quantum Lattice Models: Symmetric Hamiltonians and Matrix Product Operator Intertwiners}},
\newblock PRX Quantum \textbf{4}(2), 020357 (2023),
\newblock \doi{10.1103/PRXQuantum.4.020357},
\newblock  \eprint[https://doi.org/10.48550/arXiv.2112.09091]{10.48550/arXiv.2112.09091}.

\bibitem{Kramers1941}
H.~A. Kramers and G.~H. Wannier,
\newblock \emph{{Statistics of the two-dimensional ferromagnet. Part 1.}},
\newblock Phys. Rev. \textbf{60}, 252 (1941),
\newblock \doi{10.1103/PhysRev.60.252}.

\bibitem{BenTov2014}
Y.~BenTov,
\newblock \emph{{Fermion masses without symmetry breaking in two spacetime dimensions}},
\newblock JHEP \textbf{07}, 034 (2015),
\newblock \doi{10.1007/JHEP07(2015)034},
\newblock  \eprint[https://doi.org/10.48550/arXiv.1412.0154]{10.48550/arXiv.1412.0154}.

\bibitem{Radicevic2018}
D.~Radicevic,
\newblock \emph{{Spin Structures and Exact Dualities in Low Dimensions}}  (2018),
\newblock  \eprint[https://doi.org/10.48550/arXiv.1809.07757]{10.48550/arXiv.1809.07757}.

\bibitem{Li2023}
L.~Li, M.~Oshikawa and Y.~Zheng,
\newblock \emph{{Noninvertible duality transformation between symmetry-protected topological and spontaneous symmetry breaking phases}},
\newblock Phys. Rev. B \textbf{108}(21), 214429 (2023),
\newblock \doi{10.1103/PhysRevB.108.214429},
\newblock  \eprint[https://doi.org/10.48550/arXiv.2301.07899]{10.48550/arXiv.2301.07899}.

\bibitem{Kennedy1992}
T.~Kennedy and H.~Tasaki,
\newblock \emph{{Hidden Z2\texttimes{}Z2 symmetry breaking in Haldane-gap antiferromagnets}},
\newblock Phys. Rev. B \textbf{45}(1), 304 (1992),
\newblock \doi{10.1103/PhysRevB.45.304}.

\bibitem{Oshikawa1992}
M.~Oshikawa,
\newblock \emph{{Hidden $\mathbb Z_2 \times \mathbb Z_2$ symmetry in quantum spin chains with arbitrary integer spin}},
\newblock Journal of Physics: Condensed Matter \textbf{4}(36), 7469 (1992),
\newblock \doi{10.1088/0953-8984/4/36/019}.

\bibitem{ParayilMana2024}
A.~Parayil~Mana, Y.~Li, H.~Sukeno and T.-C. Wei,
\newblock \emph{{Kennedy-Tasaki transformation and noninvertible symmetry in lattice models beyond one dimension}},
\newblock Phys. Rev. B \textbf{109}(24), 245129 (2024),
\newblock \doi{10.1103/PhysRevB.109.245129},
\newblock  \eprint[https://doi.org/10.48550/arXiv.2402.09520]{10.48550/arXiv.2402.09520}.

\bibitem{Seifnashri2024}
S.~Seifnashri and S.-H. Shao,
\newblock \emph{{Cluster State as a Noninvertible Symmetry-Protected Topological Phase}},
\newblock Phys. Rev. Lett. \textbf{133}(11), 116601 (2024),
\newblock \doi{10.1103/PhysRevLett.133.116601},
\newblock  \eprint[https://doi.org/10.48550/arXiv.2404.01369]{10.48550/arXiv.2404.01369}.

\bibitem{Vancraeynest2025a}
B.~Vancraeynest-De~Cuiper and C.~Delcamp,
\newblock \emph{{Twisted gauging and topological sectors in (2+1)d Abelian lattice gauge theories}},
\newblock SciPost Phys. \textbf{19}(2), 054 (2025),
\newblock \doi{10.21468/SciPostPhys.19.2.054},
\newblock  \eprint[https://doi.org/10.48550/arXiv.2501.16301]{10.48550/arXiv.2501.16301}.

\bibitem{Batista2004}
C.~D. Batista and Z.~Nussinov,
\newblock \emph{{Generalized Elitzur's theorem and dimensional reduction}},
\newblock Phys. Rev. B \textbf{72}, 045137 (2005),
\newblock \doi{10.1103/PhysRevB.72.045137},
\newblock  \eprint[https://doi.org/10.48550/arXiv.cond-mat/0410599]{10.48550/arXiv.cond-mat/0410599}.

\bibitem{Nussinov2006}
Z.~Nussinov and G.~Ortiz,
\newblock \emph{{Sufficient symmetry conditions for Topological Quantum Order}},
\newblock Proc. Nat. Acad. Sci. \textbf{106}, 16944 (2009),
\newblock \doi{10.1073/pnas.0803726105},
\newblock  \eprint[https://doi.org/10.48550/arXiv.cond-mat/0605316]{10.48550/arXiv.cond-mat/0605316}.

\bibitem{Nussinov2009}
Z.~Nussinov and G.~Ortiz,
\newblock \emph{{A symmetry principle for topological quantum order}},
\newblock Annals Phys. \textbf{324}, 977 (2009),
\newblock \doi{10.1016/j.aop.2008.11.002},
\newblock  \eprint[https://doi.org/10.48550/arXiv.cond-mat/0702377]{10.48550/arXiv.cond-mat/0702377}.

\bibitem{Lin2022}
L.~Lin, D.~G. Robbins and E.~Sharpe,
\newblock \emph{{Decomposition, Condensation Defects, and Fusion}},
\newblock Fortsch. Phys. \textbf{70}(11), 2200130 (2022),
\newblock \doi{10.1002/prop.202200130},
\newblock  \eprint[https://doi.org/10.48550/arXiv.2208.05982]{10.48550/arXiv.2208.05982}.

\bibitem{Roumpedakis2022}
K.~Roumpedakis, S.~Seifnashri and S.-H. Shao,
\newblock \emph{{Higher Gauging and Non-invertible Condensation Defects}},
\newblock Commun. Math. Phys. \textbf{401}(3), 3043 (2023),
\newblock \doi{10.1007/s00220-023-04706-9},
\newblock  \eprint[https://doi.org/10.48550/arXiv.2204.02407]{10.48550/arXiv.2204.02407}.

\bibitem{Douglas2018}
C.~L. Douglas and D.~J. Reutter,
\newblock \emph{{Fusion 2-categories and a state-sum invariant for 4-manifolds}}  (2018),
\newblock  \eprint[https://doi.org/10.48550/arXiv.1812.11933]{10.48550/arXiv.1812.11933}.

\bibitem{Decoppet2024}
T.~D. D{\'e}coppet, P.~Huston, T.~Johnson-Freyd, D.~Nikshych, D.~Penneys, J.~Plavnik, D.~Reutter and M.~Yu,
\newblock \emph{{The Classification of Fusion 2-Categories}}  (2024),
\newblock  \eprint[https://doi.org/10.48550/arXiv.2411.05907]{10.48550/arXiv.2411.05907}.

\bibitem{Decoppet2021}
T.~D. D{\'e}coppet,
\newblock \emph{{Finite semisimple module 2-categories}},
\newblock Selecta Math. \textbf{31}(1), 5 (2025),
\newblock \doi{10.1007/s00029-024-00998-4},
\newblock  \eprint[https://doi.org/10.48550/arXiv.2107.11037]{10.48550/arXiv.2107.11037}.

\bibitem{Delcamp2021}
C.~Delcamp,
\newblock \emph{{Tensor network approach to electromagnetic duality in (3+1)d topological gauge models}},
\newblock JHEP \textbf{08}, 149 (2022),
\newblock \doi{10.1007/JHEP08(2022)149},
\newblock  \eprint[https://doi.org/10.48550/arXiv.2112.08324]{10.48550/arXiv.2112.08324}.

\bibitem{Delcamp2023}
C.~Delcamp and A.~Tiwari,
\newblock \emph{{Higher categorical symmetries and gauging in two-dimensional spin systems}},
\newblock SciPost Phys. \textbf{16}(4), 110 (2024),
\newblock \doi{10.21468/SciPostPhys.16.4.110},
\newblock  \eprint[https://doi.org/10.48550/arXiv.2301.01259]{10.48550/arXiv.2301.01259}.

\bibitem{Inamura2023}
K.~Inamura and K.~Ohmori,
\newblock \emph{{Fusion surface models: 2+1d lattice models from fusion 2-categories}},
\newblock SciPost Phys. \textbf{16}, 143 (2024),
\newblock \doi{10.21468/SciPostPhys.16.6.143},
\newblock  \eprint[https://doi.org/10.48550/arXiv.2305.05774]{10.48550/arXiv.2305.05774}.

\bibitem{Inamura2025}
K.~Inamura, S.-J. Huang, A.~Tiwari and S.~Schafer-Nameki,
\newblock \emph{{(2+1)d Lattice Models and Tensor Networks for Gapped Phases with Categorical Symmetry}}  (2025),
\newblock  \eprint[https://doi.org/10.48550/arXiv.2506.09177]{10.48550/arXiv.2506.09177}.

\bibitem{Eck2025}
L.~Eck,
\newblock \emph{{Dualities between 2+1d fusion surface models from braided fusion categories}},
\newblock SciPost Phys. \textbf{19}, 157 (2025),
\newblock \doi{10.21468/SciPostPhys.19.6.157},
\newblock  \eprint[https://doi.org/10.48550/arXiv.2501.14722]{10.48550/arXiv.2501.14722}.

\bibitem{Vancraeynest2024}
B.~V.-D. Cuiper and J.~Garre-Rubio,
\newblock \emph{{Systematic construction of stabilizer codes via gauging abelian boundary symmetries}},
\newblock Quantum \textbf{9}, 1852 (2025),
\newblock \doi{10.22331/q-2025-09-08-1852},
\newblock  \eprint[https://doi.org/10.48550/arXiv.2410.09044]{10.48550/arXiv.2410.09044}.

\bibitem{Bulmash2020}
D.~Bulmash and M.~Barkeshli,
\newblock \emph{{Absolute anomalies in (2+1)D symmetry-enriched topological states and exact (3+1)D constructions}},
\newblock Phys. Rev. Res. \textbf{2}(4), 043033 (2020),
\newblock \doi{10.1103/PhysRevResearch.2.043033},
\newblock  \eprint[https://doi.org/10.48550/arXiv.2003.11553]{10.48550/arXiv.2003.11553}.

\bibitem{Etingof2002}
P.~Etingof, D.~Nikshych and V.~Ostrik,
\newblock \emph{{On fusion categories}}  (2002),
\newblock  \eprint[https://doi.org/10.48550/arXiv.math/0203060]{10.48550/arXiv.math/0203060}.

\bibitem{Ostrik2001}
V.~Ostrik,
\newblock \emph{{Module categories, weak Hopf algebras and modular invariants}},
\newblock Transform. Groups \textbf{8}(2), 177 (2003),
\newblock \doi{10.1007/s00031-003-0515-6},
\newblock  \eprint[https://doi.org/10.48550/arXiv.math/0111139]{10.48550/arXiv.math/0111139}.

\bibitem{Beer2018}
K.~Beer \emph{et~al.},
\newblock \emph{{From categories to anyons: a travelogue}}  (2018),
\newblock  \eprint[https://doi.org/10.48550/arXiv.1811.06670]{10.48550/arXiv.1811.06670}.

\bibitem{Naidu2006}
D.~Naidu,
\newblock \emph{Categorical morita equivalence for group-theoretical categories},
\newblock Communications in Algebra  (2006),
\newblock \doi{10.1080/00927870701511996}.

\bibitem{Mueger2002}
M.~Mueger,
\newblock \emph{{From Subfactors to Categories and Topology I. Frobenius algebras in and Morita equivalence of tensor categories}},
\newblock Journal of Pure and Applied Algebra  (2002),
\newblock \doi{10.1016/S0022-4049(02)00247-5},
\newblock  \eprint[https://doi.org/10.48550/arXiv.math/0111204]{10.48550/arXiv.math/0111204}.

\end{thebibliography}
